\documentclass[aps,showpacs,pra,twocolumn]{revtex4-1}

\usepackage{exscale}
\usepackage{graphicx}
\usepackage{amsmath}
\usepackage{latexsym}
\usepackage[caption=false]{subfig}
\usepackage{amsfonts}
\usepackage{amssymb}
\usepackage{bbm}
\usepackage{times}
\usepackage[T1]{fontenc}
\usepackage{lipsum}
\usepackage{amsthm}
\usepackage{fancyhdr,txfonts,bbm}

\usepackage{tikz}
\usetikzlibrary{matrix,calc}
\usetikzlibrary{shapes.misc}

\usepackage{bbold}
\usepackage{bm}
\usepackage{subfig}
\usepackage{color}
\usepackage{orcidlink}
\usepackage{academicons}
\definecolor{orcidlogocol}{HTML}{A6CE39}

\usepackage{epsfig,pstricks}
\usepackage{changes}
\definechangesauthor[name={Markus}, color=red]{M}
\definechangesauthor[name={Dario}, color=magenta]{D}

\newcommand{{\Cd}}{{\mathbb{C}^d}}
\newcommand{{\C}}{{\mathbb{C}}}

\DeclareMathOperator{\Tr}{Tr}

\begin{document}

\title{Lorentz invariant polynomials as entanglement indicators for Dirac particles}

\author{Markus Johansson}
\affiliation{Barcelona (Barcelona), Spain}
\date{\today}
\begin{abstract}
The spinorial degrees of freedom of two or more spacelike separated Dirac particles are considered and a method for constructing mixed polynomials that are invariant under the spinor representations of the local proper orthochronous Lorentz groups is described. The method is an extension of the method for constructing homogeneous polynomials introduced in [Phys. Rev. A {\bf 105}, 032402 (2022), arXiv:2103.07784] and [Ann. Phys. (N. Y.) {\bf 457}, 169410 (2023), arXiv:2105.07503]. The mixed polynomials constructed by this method are identically zero for all product states. 
Therefore they are considered indicators of the spinor entanglement of Dirac particles.
Mixed polynomials can be constructed to indicate spinor entanglement that involves all the particles, or alternatively to indicate spinor entanglement that involves only a proper subset of the particles.
It is shown that the mixed polynomials can indicate some types of spinor entanglement that involves all the particles but cannot be indicated by any homogeneous locally Lorentz invariant polynomial.
For the case of two Dirac particles mixed polynomials of bidegree (2,2) and bidegree (3,1) are constructed.
For the case of three Dirac particles mixed polynomials of bidegree (2,2), bidegree (3,1) and bidegree (3,3) are constructed.
The relations of the polynomials constructed for two and three Dirac particles to the polynomial spin entanglement indicators for two and three non-relativistic spin-$\frac{1}{2}$ particles are described.
Moreover, the constructed polynomial indicators of spinor entanglement are in general not invariant under local time evolutions of the particles but evolve dynamically and we discuss how to describe this dynamical evolution.

\end{abstract}
\maketitle
\section{Introduction}
The Dirac equation \cite{dirac2,dirac} was first introduced as a description of the relativistic electron. As such it is used for example in relativistic quantum mechanics \cite{bjorken} and in relativistic quantum chemistry \cite{pykk}. 
Subsequently the Dirac equation has been used to describe also other relativistic spin-$\frac{1}{2}$ particles, collectively referred to here as Dirac particles.
For example, it is used in the Standard Model of particle physics to describe quarks and leptons \cite{schwartz} and in the Yukawa model of hadrons to describe baryons \cite{yukawa}. 
Moreover, for the case of zero mass particles the Dirac equation has solutions with definite chirality, so called Weyl particles \cite{weyl}. In addition to the original equation several modifications of the Dirac equation with extra or alternative terms, such as a Yukawa pseudoscalar coupling, have been introduced \cite{thaller}.
The intrinsic spinorial degree of freedom of a particle described by the Dirac equation, or one of its modifications, is represented by a four component Dirac spinor.

Quantum entanglement is a feature of quantum mechanics that permits nonlocal causation between spacelike separated events \cite{epr,bell,chsh,bell2}, sometimes called action at a distance.
A system with two or more spacelike separated particles is entangled if it is in a superposition where some physical property of one particle is conditioned on the physical properties of one or more other particles. In this case the state of the system cannot be fully described by only local variables specifying the properties of the individual particles \cite{bell,chsh,bell2}.
Some physical phenomena are impossible without nonlocal causation, i.e., impossible without the presence of entanglement. These phenomena include the violation of a Bell inequality \cite{bell,chsh,svet}, quantum steering \cite{steer,wise} and quantum teleportation \cite{bennett}.
 
A function on the state space of a system that takes non-zero value for some entangled states but not for any product state is here called an {\it entanglement indicator}. In general multiple independent such functions may exist that are sensitive to different ways in which the system can be entangled. One way to construct entanglement indicators is as polynomials in the state coefficients or as polynomials in the state coefficients and their complex conjugates, here called {\it mixed polynomials} following Ref. \cite{oka}. Homogeneous polynomials in the state coefficients have been considered as entanglement indicators for non-relativistic spin-$\frac{1}{2}$ particles in multiple works \cite{grassl,wootters,wootters2,popescu,carteret,carsud,coffman,toni,sud,wong,tarrach,moor,luque}. In addition to the homogeneous polynomials, the mixed polynomials have also been considered as they can indicate some types of entanglement not indicated by the homogeneous polynomials \cite{grassl,popescu,lindpop,carsud,carteret,coffman,toni,kempe,sud,tarrach,toumazet}.

The quantum entanglement of Dirac particles has been considered in multiple works \cite{czachor,alsing,terno,adami,pachos,ahn,terno2,tera,tera2,mano,won,caban3,caban,geng,leon,delgado,moradi,caban2,tessier,spinorent,multispinor}. 
The tools and methods used to describe the spin entanglement of non-relativistic spin-$\frac{1}{2}$ particles are in general not well suited for systems of relativistic spin-$\frac{1}{2}$ particles.
Therefore, several tools and methods for describing the entanglement of Dirac particles in a relativistic setting have been suggested and discussed \cite{czachor,alsing,terno,adami,pachos,ahn,terno2,tera,tera2,mano,won,caban3,caban,geng,leon,delgado,moradi,caban2,tessier,spinorent,multispinor}.
One such tool is locally Lorentz invariant polynomials.
The characterization of the entanglement of spinorial degrees of freedom in a system of two or more Dirac particles with definite momenta using locally Lorentz invariant homogeneous polynomials has been considered in Refs. \cite{spinorent,multispinor}.

In this work we consider the question of constructing locally Lorentz invariant polynomial spinor entanglement indicators for two or more spacelike separated Dirac particles. In particular we describe a method for constructing such polynomials in the state coefficients and their complex conjugates
of different bidegrees that extends the method for constructing homogeneous polynomials described in Refs. \cite{spinorent,multispinor}. The homogeneous locally Lorentz invariant polynomials constructed in Refs. \cite{spinorent,multispinor} are all spinor entanglement indicators. However there exist spinor entangled states that are not indicated by any homogeneous polynomials but are still indicated by mixed polynomials. One such state is the threepartite entangled so called W-state \cite{dur}.
Furthermore, unlike the homogeneous polynomials the mixed polynomials can be used to indicate spinor entanglement involving only a proper subset of the particles.
Moreover, since the spinor entanglement indicators are polynomials in the state coefficients and their complex conjugates they follow dynamical equations that can be derived from the Dirac equation. We therefore briefly discuss how to describe their local time evolution.

This work is organized as follows. Sections \ref{dir}--\ref{invariants} review the background material, discuss the physical assumptions made and describe the tools used to construct the locally Lorentz invariant polynomials.
In particular, section \ref{dir} reviews the description of Dirac particles and discusses the fundamental assumptions made in this work. In section \ref{rep} we review the spinor representation of the Lorentz group and the charge conjugation transformation. Section \ref{invariants}  describes how to construct skew-symmetric bilinear forms and sesquilinear forms that are invariant under the spinor representation of the proper orthochronous Lorentz group. 
Sections \ref{dynnn}--\ref{tpo} contain the results.
In particular, section \ref{dynnn} describes the dynamical evolution of the bilinear and sesquilinear forms.
Section \ref{consrt} describes the method for constructing polynomial spinor entanglement indicators for two Dirac particles and give the constructed mixed polynomials.
Section \ref{ent} describes the method for constructing polynomial entanglement indicators for the case of multiple Dirac particles. 
In section \ref{three} the case of three Dirac particles is considered and a selection of constructed polynomials is given. Section \ref{tpo} 
briefly discusses the role of the mixed polynomial spinor entanglement indicators in relation to the homogeneous polynomials.
Section \ref{diss} is the discussion and conclusions.

\section{Dirac particles}\label{dir}

The Dirac equation was introduced in Ref. \cite{dirac2} as a relativistic wave equation describing a spin-$\frac{1}{2}$ particle, or Dirac particle. For a Dirac particle with mass $m$ and electromagnetic charge $q$ coupled to a four-potential $A_{\mu}(x)$ it can be given, with natural units $\hbar=c=1$, on the form
\begin{eqnarray}
\left[\sum_{\mu}\gamma^\mu(i\partial_{\mu}-qA_{\mu}(x)) -m\right]\psi(x)=0.
\end{eqnarray}
Here $\psi(x)$ is a four component Dirac spinor 
\begin{eqnarray}\label{spinor}
\psi(x)\equiv
\begin{pmatrix}
\psi_0(x) \\
\psi_1(x)\\
\psi_2(x) \\
\psi_3(x) \\
\end{pmatrix},
\end{eqnarray}
where each component is a complex valued function of the four-vector $x$, and $\gamma^0,\gamma^1,\gamma^2,\gamma^3$ are $4\times 4$ matrices
defined by the relations
\begin{eqnarray}\label{anti}
\gamma^\mu\gamma^\nu+\gamma^\nu\gamma^\mu=2g^{\mu\nu}I,
\end{eqnarray}
where $g^{\mu\nu}$ is the Minkowski metric with signature $(+---)$. The relations in Eq. (\ref{anti}) do not uniquely define the matrices $\gamma^0,\gamma^1,\gamma^2,\gamma^3$ and they can be chosen in different physically equivalent ways.
A common choice that we use here is the so called Dirac matrices or {\it gamma matrices} given by
\begin{align}
\gamma^0&=
\begin{pmatrix}
I & 0 \\
0 & -I  \\
\end{pmatrix},
&\gamma^1=
\begin{pmatrix}
0  & \sigma^1 \\
-\sigma^1 &  0 \\
\end{pmatrix},\nonumber\\
\gamma^2&=
\begin{pmatrix}
0  & \sigma^2 \\
-\sigma^2 &  0 \\
\end{pmatrix},
&\gamma^3=
\begin{pmatrix}
0  & \sigma^3 \\
-\sigma^3 &  0 \\
\end{pmatrix},
\end{align}
where $I$ is the $2\times 2$ identity matrix and $\sigma^1,\sigma^2,\sigma^3$ are the Pauli matrices
\begin{eqnarray}
I=
\begin{pmatrix}
1 & 0 \\
0 & 1  \\
\end{pmatrix},\phantom{o}
\sigma^1=
\begin{pmatrix}
0  & 1 \\
1 &  0 \\
\end{pmatrix},\phantom{o}
\sigma^2=
\begin{pmatrix}
0  & -i \\
i &  0 \\
\end{pmatrix},\phantom{o}
\sigma^3=
\begin{pmatrix}
1  & 0 \\
0 &  -1 \\
\end{pmatrix}.
\end{eqnarray}
For a derivation of the Dirac equation and the properties of the gamma matrices see e.g. Ref. \cite{dirac2} or Ref. \cite{dirac} Ch. XI. 

The matrix $\gamma^0$ is Hermitian whereas the other gamma matrices are anti-Hermitian.
As a consequence of this the anticommutator relations in Eq. (\ref{anti}) implies that for each gamma matrix $\gamma^\mu$ and its conjugate transpose $\gamma^{\mu \dagger}$ we have that
\begin{eqnarray}\label{uumb}
 \gamma^0\gamma^\mu \gamma^0=\gamma^{\mu \dagger}.
\end{eqnarray}
This is a property that is utilized in the following. Note also that $\gamma^0$ is its own transpose and its own inverse, i.e., $\gamma^0=(\gamma^{0})^{\dagger}=(\gamma^{0})^{T}=(\gamma^{0})^{-1}$.
We can identify two other matrices in the algebra of gamma matrices that also have useful properties.
One is the matrix
\begin{eqnarray}
C\equiv i\gamma^1\gamma^3=
\begin{pmatrix}
-\sigma^2 & 0 \\
0 & -\sigma^2  \\
\end{pmatrix},
\end{eqnarray}
which is such that for each gamma matrix $\gamma^\mu$ and its transpose $\gamma^{\mu T}$ we have that
\begin{eqnarray}\label{uub}
 C\gamma^\mu C=\gamma^{\mu T}.
\end{eqnarray}
Moreover, $C$ is Hermitian and also its own inverse, i.e., $C=C^\dagger=C^{-1}$.
The other matrix is 
\begin{eqnarray}
\gamma^5\equiv i\gamma^0\gamma^1\gamma^2\gamma^3=
\begin{pmatrix}
0 & I \\
I & 0  \\

\end{pmatrix},
\end{eqnarray}
which anticommutes with each of the gamma matrices $\gamma^\mu$
\begin{eqnarray}
\gamma^5\gamma^\mu+\gamma^\mu\gamma^5=0.
\end{eqnarray}
We can see that $\gamma^5$ is real Hermitian and its own inverse, i.e., $\gamma^5=(\gamma^{5})^{\dagger}=(\gamma^{5})^{T}=(\gamma^{5})^{-1}$.

For a free Dirac-particle at rest, i.e., for zero momentum and $A_1=A_2=A_3=A_0=0$ the Dirac equation reduces to $(i\gamma^0\partial_{t} -m)\psi(x)=0$. The solutions to the Dirac equation for this case is a space spanned by the four orthogonal spinors
\begin{eqnarray}\label{spinorz}
\begin{pmatrix}
1 \\
0\\
0 \\
0 \\
\end{pmatrix}e^{-imt},\begin{pmatrix}
0 \\
1\\
0 \\
0 \\
\end{pmatrix}e^{-imt},
\begin{pmatrix}
0 \\
0\\
1 \\
0 \\
\end{pmatrix}e^{imt},
\begin{pmatrix}
0 \\
0\\
0 \\
1 \\
\end{pmatrix}e^{imt}.
\end{eqnarray}
The first two spinors are solutions to the Dirac equation with positive energy $m$ while the latter two are are solutions with negative energy $-m$. The positive energy solutions are often interpreted as states of a non-relativistic free spin-$\frac{1}{2}$ particle while the negative energy solutions are interpreted as states of a non-relativistic free spin-$\frac{1}{2}$ antiparticle (See e.g. Ref. \cite{peskin} Ch. 3.5.).

For the case of zero particle momentum and zero electromagnetic four-potential we can see that there is an invariant positive energy subspace defined by the projector $P_+=\frac{1}{2}(I+\gamma^0)$ and an invariant negative energy subspace defined by the projector $P_-=\frac{1}{2}(I-\gamma^0)$. The spinors $\psi_+$ in the image of $P_+$ and the spinors $\psi_-$ in the image of $P_-$ are of the form
\begin{eqnarray}\label{weyn}
\psi_+=
\begin{pmatrix}
{\psi}_0 \\
{\psi}_1\\
0 \\
0 \\
\end{pmatrix},\phantom{o}
\psi_-=
\begin{pmatrix}
0 \\
0\\
{\psi}_2 \\
{\psi}_3 \\
\end{pmatrix},
\end{eqnarray}
respectively. We can see in Eq. (\ref{weyn}) that both the spinors in the image of $P_+$ and the spinors in the image of $P_-$ have only two nonzero spinor components.

The Dirac equation for the case of a zero mass particle was considered by Weyl in Ref. \cite{weyl}
\begin{eqnarray}\label{weyl}
\sum_{\mu}\gamma^\mu(i\partial_{\mu}-qA_{\mu}(x))\psi(x)=0.
\end{eqnarray}
For this equation there is an invariant subspace that is the image of the projector $P_L=\frac{1}{2}(1-\gamma^5)$ called the left-handed chiral subspace, and an invariant subspace that is the image of the projector $P_R=\frac{1}{2}(1+\gamma^5)$ called the right-handed chiral subspace. Solutions to Eq. (\ref{weyl}) that belong to the image of $P_R$ are called right-handed Weyl particles $\psi_R$ and solutions that belong to the image of $P_L$ are called left-handed Weyl particles $\psi_L$. These have the form
\begin{eqnarray}\label{wey}
\psi_R=
\begin{pmatrix}
{\psi}_0 \\
{\psi}_1\\
{\psi}_0 \\
{\psi}_1 \\
\end{pmatrix},\phantom{o}
\psi_L=
\begin{pmatrix}
{\psi}_0 \\
{\psi}_1\\
-{\psi}_0 \\
-{\psi}_1 \\
\end{pmatrix}.
\end{eqnarray}
We can see in Eq. (\ref{wey}) that both the left-handed spinors and the right-handed spinors have only two independent spinor components.

Any solution to the Dirac equation can be expanded using a set of basis spinors $\phi_j$ as
\begin{eqnarray}\label{bass}
\psi(x)=\sum_{j}\psi_{j}(x)\phi_j,
\end{eqnarray}
where the $\psi_{j}(x)$ are complex valued functions of $x$, and the $\phi_j$ are the four spinors
\begin{eqnarray}\label{basis}
{\phi_0}=
\begin{pmatrix}
1 \\
0   \\
0   \\
0   \\
\end{pmatrix},\phantom{o}
{\phi_1}=
\begin{pmatrix}
0 \\
1   \\
0   \\
0   \\
\end{pmatrix},\phantom{o}
{\phi_2}=
\begin{pmatrix}
0 \\
0   \\
1   \\
0   \\
\end{pmatrix},\phantom{o}
{\phi_3}=
\begin{pmatrix}
0 \\
0   \\
0   \\
1   \\
\end{pmatrix}.
\end{eqnarray}
In the following we frequently use this spinor basis to represent states of Dirac particles.

In this work we consider a scenario with two or more spacelike separated Dirac particles.
As in Refs. \cite{spinorent,multispinor} we introduce a number of laboratories and assume that each laboratory contains only one Dirac particle.
Furthermore, as was done in Refs. \cite{alsing,pachos,moradi,caban2,caban3,spinorent,multispinor} we assume that for any spacelike separated particles that have not previously interacted the state can be described as a tensor product  $\psi_A(x_A)\otimes \varphi_B(x_B)\otimes \zeta_C(x_C)\otimes\dots$ of single particle states. We also assume that the tensor products of the elements of the single particle spinor bases $\phi_{j_A}\otimes \phi_{k_B}\otimes \phi_{l_C}\otimes\dots$ is a basis for the multi-particle states.

The assumption that the state of a system of spacelike separated particles can be described by a tensor product structure is
often made, but is not trivial.
The reason for making this assumption is that the operations on any particle in such a system can be made jointly with the operations on any of the other particles, i.e., the operations on different spacelike separated particles commute.
However, it is not known if a description in terms of commuting operator algebras is equivalent to a description in terms of tensor product spaces in general \cite{navascues,tsirelson,werner} and this open question is called Tsirelson's Problem \cite{tsirelson}. Nevertheless, if for each observer the algebra of operations is finite dimensional it has been shown in References \cite{tsirelson,werner} that the two descriptions are equivalent. In particular this is the case if the Hilbert space of the shared system has finite dimension. In any experiment a Hilbert space with sufficiently large finite dimension can be constructed that operationally describes the system (See Appendix \ref{opp} for a discussion). Therefore we assume that it is operationally motivated to use a tensor product structure.

As in Refs. \cite{spinorent,multispinor} we describe each different Dirac particle as belonging to a different Minkowski space. These different Minkowski spaces should be understood as the respective local descriptions of spacetime used by the different laboratories holding the particles. 
If the laboratories are in a flat spacetime these different Minkowski spaces are the spacelike separated  observers different descriptions of the same Minkowski space they all inhabit. Alternatively, if the laboratories are in a curved spacetime described by General Relativity (See e.g. Ref. \cite{wald}) the different Minkowski spaces are the Minkowski tangent-spaces of the different spacelike separated points of the observers.

\section{The spinor representation of the Lorentz group and the charge conjugation}\label{rep}

In General Relativity a spacetime is described by a four-dimensional Lorentzian manifold that in general has nonzero curvature.
At every non-singular point of such a spacetime one can define a four-dimensional tangent vector space that is isomorphic to the Minkowski space (See e.g. Ref. \cite{wald}).
As in References \cite{spinorent,multispinor} we make the assumption that the local curvature of spacetime is small enough that it is physically motivated to neglect it. We then treat the Dirac particle as being in the Minkowski tangent space of a point rather than in the spacetime manifold itself. If we do not make the assumption that the curvature can be neglected the Dirac equation must be replaced by some curved spacetime counterpart such as that introduced by Weyl and Fock \cite{weyl,fock}.

A change of inertial reference frame, i.e., a Lorentz transformation is a coordinate transformation on the local Minkowski tangent space to a spacetime point. However, a Lorentz transformation also induces an action on the Dirac spinor in the point given by the spinor representation of the Lorentz transformation. Let us consider a Lorentz transformation $\Lambda$ and its spinor representation $S(\Lambda)$. The spinor then transforms as $\psi(x)\to \psi'(x')=S(\Lambda)\psi(x)$ where $x'=\Lambda x$ (See e.g. Ref. \cite{zuber}), and the Dirac equation transforms as
\begin{eqnarray}
&&\left[\sum_{\mu}\gamma^\mu(i\partial_{\mu}-qA_{\mu}) -m\right]\psi(x)=0\nonumber\\
\to&&\left[\sum_{\mu,\nu}\gamma^\mu(\Lambda^{-1})^{\nu}_{\mu}(i\partial_{\nu}-qA_{\nu}) -m\right]S(\Lambda)\psi(x)=0.
\end{eqnarray}
The invariance of the Dirac equation implies that
\begin{eqnarray}
S^{-1}(\Lambda)\gamma^\mu S(\Lambda)=\sum_\nu\Lambda^{\mu}_{\nu}\gamma^\nu.
\end{eqnarray}

The group of all Lorentz transformations is the Lorentz group.
This group is a six-dimensional Lie group with four connected components. The connected component of the Lorentz group that contains the identity element is called the proper orthochronous Lorentz group.
This subgroup consists of the Lorentz transformations that preserve both the orientation of space and the direction of time. 
The spinor representation of the Lorentz group is also a six-dimensional Lie group with four connected components. The connected component of this group that contains the identity element is the spinor representation of the proper orthochronous Lorentz group. This subgroup can be generated by the exponentials of its Lie algebra. 
The Lie algebra of the spinor representation of the Lorentz group has six generators $S^{\rho\sigma}$ defined by
\begin{eqnarray}\label{gene}
S^{\rho\sigma}=\frac{1}{4}[\gamma^\rho,\gamma^\sigma]=\frac{1}{2}\gamma^\rho\gamma^\sigma-\frac{1}{2}g^{\rho\sigma}I,
\end{eqnarray}
where $g^{\rho\sigma}$ is again the Minkowski metric with signature $(+---)$.
The spinor representations of the spatial rotations have three generators $S^{12},S^{13}$, and $S^{23}$ while the spinor representations of the Lorentz boosts have three generators $S^{01},S^{02}$, and $S^{03}$.
By taking the matrix exponential of an element $\frac{1}{2}\sum_{\rho,\sigma} \omega_{\rho\sigma}S^{\rho\sigma}$ of the Lie algebra, where the $\omega_{\rho\sigma}$ are real numbers, we obtain a finite transformation
 \begin{eqnarray}
S(\Lambda)=\exp\left(\frac{1}{2}\sum_{\rho,\sigma} \omega_{\rho\sigma}S^{\rho\sigma}\right).
\end{eqnarray}
Any spinor representation of a proper orthochronous Lorentz transformation
is a product of such finite transformations. See e.g. Ref. \cite{zuber}.

The proper orthochronous Lorentz group is related to the other three connected components of the Lorentz group by the parity inversion P, the time reversal T and the combined PT transformation, respectively. Likewise, the spinor representation of the proper orthochronous Lorentz group is related to the other three connected components of the spinor representation of the Lorentz group by the spinor representations of the parity inversion P, the time reversal T, and the PT transformation, respectively. 
The spinor representations of the P and T transformations are only defined up to a multiplicative U(1) factor and can therefore be chosen in different physically equivalent ways.
Here the spinor representation of the parity transformation P is chosen as
\begin{eqnarray}
S(\textrm{P})=\gamma^0.
\end{eqnarray}
The spinor representation of the time reversal transformation T involves the matrix $C$ and the complex conjugation of the spinor and is chosen here as $\psi \to C\psi^*$. 

Other than the Lorentz group we can consider also the charge conjugation transformation C, the charge parity CP transformation, the charge time CT transformation, and the charge parity time CPT transformation. Similarly to the spinor representations of P and T, the spinor representation of C is only defined up to a multiplicative U(1) factor that can be chosen. The spinor representation of the charge conjugation C involves complex conjugation of the spinor and is chosen here as $\psi \to i\gamma^2\psi^*$. With this choice it follows that the spinor representation of the CP transformation is $\psi \to -i\gamma^0\gamma^2\psi^*= iC\gamma^5\psi^*$ and the spinor representation of the CT transformation is given by the matrix $-i\gamma^0\gamma^5$
\begin{eqnarray}
S(\textrm{CT})=-i\gamma^0\gamma^5.
\end{eqnarray}
Finally, it follows that the spinor representation of the CPT transformation is given by the matrix $-i\gamma^5$
\begin{eqnarray}
S(\textrm{CPT})=-i\gamma^5.
\end{eqnarray}
 See e.g. Ref. \cite{bjorken} Ch. 5. 
We use these choices of the spinor representations of the P, T and C transformations in the following.

\section{Bilinear and sesquilinear forms invariant under the spinor representation of the proper orthochronous Lorentz group}
\label{invariants}

A physical quantity that is invariant under some representation of the Lorentz group is called a Lorentz invariant.
Here we will refer to a quantity that is invariant under the action of the spinor representation of the proper orthochronous Lorentz group as a Lorentz invariant for convenience even if it is not invariant under the spinor representation of the full Lorentz group.
Lorentz invariants of this kind can be constructed as bilinear or sesquilinear forms on the Dirac spinors (See e.g. Ref. \cite{pauli}). 

We first consider the Lorentz invariant bilinear forms.
From the properties of the matrix $C$ described in Eq. (\ref{uub}) and the definition of the generators $S^{\rho\sigma}$ of the spinor representation of the proper orthochronous Lorentz group in Eq. (\ref{gene}) it follows that
\begin{eqnarray}
S^{\rho\sigma T}=\frac{1}{4}[\gamma^{\sigma T},\gamma^{\rho T}]=-\frac{1}{4}C[\gamma^{\rho},\gamma^{\sigma}]C=-CS^{\rho\sigma}C.
\end{eqnarray}
This in turn implies that for any finite transformation $S(\Lambda)$ it holds that $S(\Lambda)^TC=CS(\Lambda)^{-1}$.
Therefore we can construct a Lorentz invariant bilinear form as  
\begin{eqnarray}
\psi^TC\varphi,
\end{eqnarray}
where $\psi$ and $\varphi$ are Dirac spinors.
For any spinor representation $S(\Lambda)$ of a proper orthochronous Lorentz transformation this bilinear form transforms as $\psi^TS(\Lambda)^TCS(\Lambda)\varphi=\psi^TCS(\Lambda)^{-1}S(\Lambda)\varphi=\psi^TC\varphi$. Moreover $\psi^TC\varphi$ is invariant under the spinor representation of the parity transformation P.
This can be seen from the relations $\gamma^0=(\gamma^{0})^{T}=(\gamma^{0})^{-1}$ and $\gamma^0C=C\gamma^0$ which imply that $\psi^TS(\textrm{P})^TCS(\textrm{P})\varphi=\psi^T\gamma^0C\gamma^0\varphi=\psi^TC\varphi$.
However, $\psi^TC\varphi$ is not invariant under the spinor representation of the CPT transformation but changes sign.
This can be seen from the definition $S(\textrm{CPT})=-i\gamma^{5}$ and the properties $\gamma^5C=C\gamma^5$ and $\gamma^5=(\gamma^{5})^{T}=(\gamma^{5})^{-1}$ which imply that $\psi^TS(\textrm{CPT})^TCS(\textrm{CPT})\varphi=-\psi^T\gamma^5C\gamma^5\varphi=-\psi^TC\varphi$.

Next we can see that since the matrix $\gamma^5$ anti-commutes with all $\gamma^{\mu}$ it commutes with any generator $S^{\rho\sigma}$
\begin{eqnarray}
[S^{\rho\sigma},\gamma^5]=\frac{1}{4}[\gamma^{\rho},\gamma^{\sigma}]\gamma^5-\frac{1}{4}\gamma^5[\gamma^{\rho},\gamma^{\sigma}]=0,
\end{eqnarray}
and it follows that $\gamma^5$ commutes with any spinor representation of a proper orthochronous Lorentz transformation $S(\Lambda)\gamma^5=\gamma^5S(\Lambda)$.
This allows us to construct another Lorentz invariant bilinear form as 
\begin{eqnarray}
\psi^TC\gamma^5\varphi.
\end{eqnarray}
This bilinear form transforms under the spinor representation of a Lorentz transformation as $\psi^TS(\Lambda)^TC\gamma^5S(\Lambda)\varphi=\psi^TCS(\Lambda)^{-1}\gamma^5S(\Lambda)\varphi=\psi^TC\gamma^5S(\Lambda)^{-1}S(\Lambda)\varphi=\psi^TC\gamma^5\varphi$.
Moreover, unlike $\psi^TC\varphi$ the bilinear form $\psi^TC\gamma^5\varphi$ is not invariant under the spinor representation of the parity transformation but changes sign.
This follows from the anticommutation $\gamma^0\gamma^5=-\gamma^5\gamma^0$ which implies $\psi^TS(\textrm{P})^TC\gamma^5S(\textrm{P})\varphi=\psi^T\gamma^0C\gamma^5\gamma^0\varphi=-\psi^TC\gamma^5\varphi$.
Furthermore, $\psi^TC\gamma^5\varphi$ is not invariant under the spinor representation of the CPT transformation but changes sign. 
This follows from the definition $S(\textrm{CPT})=-i\gamma^{5}$ which gives that $\psi^TS(\textrm{CPT})^TC\gamma^5S(\textrm{CPT})\varphi=-\psi^T\gamma^5C\gamma^5\gamma^5\varphi=-\psi^TC\gamma^5\varphi$. 

Next we consider the Lorentz invariant sesquilinear forms.
We can see from the properties of the matrix $\gamma^0$ described in Eq. (\ref{uumb}) and the form of the generators $S^{\rho\sigma}$ of the spinor representation of the proper orthochronous Lorentz group in Eq. (\ref{gene}) that
\begin{eqnarray}
S^{\rho\sigma \dagger}=\frac{1}{4}[\gamma^{\sigma \dagger},\gamma^{\rho \dagger}]=-\frac{1}{4}\gamma^0[\gamma^{\rho},\gamma^{\sigma}]\gamma^0=-\gamma^0S^{\rho\sigma}\gamma^0.
\end{eqnarray}
Thus it holds for any finite transformation $S(\Lambda)$ that $S(\Lambda)^\dagger \gamma^0=\gamma^0S(\Lambda)^{-1}$.
This allows us to construct a Lorentz invariant sesquilinear form as  
\begin{eqnarray}
\psi^\dagger \gamma^0\varphi.
\end{eqnarray}
This sesquilinear form transforms as $\psi^\dagger S(\Lambda)^\dagger \gamma^0S(\Lambda)\varphi=\psi^\dagger \gamma^0S(\Lambda)^{-1}S(\Lambda)\varphi=\psi^\dagger \gamma^0\varphi$ for any spinor representation $S(\Lambda)$ of a proper orthochronous Lorentz transformation. Moreover $\psi^\dagger \gamma^0\varphi$ is invariant under the spinor representation of the parity transformation P.
This follows since $\gamma^0=(\gamma^{0})^{\dagger}=(\gamma^{0})^{-1}$ and thus $\psi^\dagger S(\textrm{P})^\dagger\gamma^0S(\textrm{P})\varphi=\psi^\dagger\gamma^0\gamma^0\gamma^0\varphi=\psi^\dagger \gamma^0\varphi$.
The sesquilinear form $\psi^\dagger\gamma^0\varphi$ is not invariant under the spinor representation of the CPT transformation but changes sign.
This follows since $S(\textrm{CPT})=-i\gamma^{5}$ and $\gamma^5\gamma^0=-\gamma^0\gamma^5$ and $\gamma^5=(\gamma^{5})^{\dagger}=(\gamma^{5})^{-1}$ and thus $\psi^\dagger S(\textrm{CPT})^\dagger\gamma^0S(\textrm{CPT})\varphi=\psi^\dagger\gamma^5\gamma^0\gamma^5\varphi=-\psi^\dagger \gamma^0\varphi$.

Using again that $\gamma^5$ commutes with any $S(\Lambda)$ we can construct another Lorentz invariant sesquilinear form as
\begin{eqnarray}
\psi^\dagger \gamma^0\gamma^5\varphi,
\end{eqnarray}
This sesquilinear form transforms as $\psi^\dagger S(\Lambda)^\dagger \gamma^0\gamma^5S(\Lambda)\varphi=\psi^\dagger \gamma^0\gamma^5S(\Lambda)^{-1}S(\Lambda)\varphi=\psi^\dagger \gamma^0\gamma^5\varphi$ for any spinor representation $S(\Lambda)$ of a proper orthochronous Lorentz transformation. Moreover, unlike $\psi^\dagger \gamma^0\varphi$ the sesquilinear form $\psi^\dagger \gamma^0\gamma^5\varphi$ is not invariant under the spinor representation of the parity transformation P but changes sign.
This follows since $\gamma^0$ anti-commutes with $\gamma^5$ and thus $\psi^\dagger S(\textrm{P})^\dagger\gamma^0\gamma^5S(\textrm{P})\varphi=\psi^\dagger\gamma^0\gamma^0\gamma^5\gamma^0\varphi=-\psi^\dagger \gamma^0\gamma^5\varphi$. The sesquilinear form $\psi^\dagger\gamma^0\gamma^5\varphi$ is not invariant under the spinor representation of the CPT transformation but changes sign.
This follows since $S(\textrm{CPT})=-i\gamma^{5}$ and $\gamma^5\gamma^0=-\gamma^0\gamma^5$ and $\gamma^5=(\gamma^{5})^{\dagger}=(\gamma^{5})^{-1}$ and thus $\psi^\dagger S(\textrm{CPT})^\dagger\gamma^0\gamma^5S(\textrm{CPT})\varphi=\psi^\dagger\gamma^5\gamma^0\gamma^5\gamma^5\varphi=-\psi^\dagger \gamma^0\gamma^5\varphi$.

The bilinear forms $\psi^TC\varphi$ and $\psi^TC\gamma^5\varphi$ are both skew-symmetric, i.e., $\psi^TC\varphi=-\varphi^TC\psi$ and $\psi^TC\gamma^5\varphi=-\varphi^TC\gamma^5\psi$ due to the antisymmetry of the matrices $C$ and $C\gamma^5$ respectively. Thus in particular we have that $\psi^TC\psi=0$ and $\psi^TC\gamma^5\psi=0$. The sesquilinear form $\psi^\dagger \gamma^0\varphi$ is Hermitian while $\psi^\dagger \gamma^0\gamma^5\varphi$ is skew-Hermitian, i.e., $\psi^\dagger \gamma^0\varphi=(\varphi^\dagger \gamma^0\psi)^*$ and $\psi^\dagger \gamma^0\gamma^5\varphi=-(\varphi^\dagger \gamma^0\gamma^5\psi)^*$. In contrast to the skew-symmetric bilinear forms $\psi^TC\psi$ and $\psi^TC\gamma^5\psi$ the sesquilinear forms  $\psi^\dagger \gamma^0\psi$ and $\psi^\dagger \gamma^0\gamma^5\psi$ are in general non-zero. Moreover, $\psi^\dagger \gamma^0\psi$ is real valued since it is Hermitian and $\psi^\dagger \gamma^0\gamma^5\psi$ is pure imaginary valued since it is skew-Hermitian.

We recall that the U(1) phase factors of the spinor representations of the P, T and C transformations have been arbitrarily chosen. Therefore the U(1) phases acquired by the two bilinear forms $\psi^TC\varphi$ and $\psi^TC\gamma^5\varphi$ under the $S(\textrm{P})$ transformation and the phases acquired under the $S(\textrm{CPT})$ transformation are also arbitrary. However, the difference by a factor of $-1$ between the phase acquired by $\psi^TC\gamma^5\varphi$, and the phase acquired by $\psi^TC\varphi$ under the $S(\textrm{P})$ transformation, does not depend of the choices of U(1) phase factors. The phases acquired by the two sesquilinear forms $\psi^\dagger \gamma^0\varphi$ and $\psi^\dagger \gamma^0\gamma^5\varphi$ under the $S(\textrm{P})$ transformation or under the $S(\textrm{CPT})$ transformation are independent of the choices of U(1) phase factors.

Each of the bilinear forms $\psi^TC\varphi$ and $\psi^TC\gamma^5\varphi$ as well as each of the sesquilinear forms $\psi^\dagger \gamma^0\varphi$ and $\psi^\dagger \gamma^0\gamma^5\varphi$ is invariant under a larger connected Lie group than the spinor representation of the proper orthochronous Lorentz group (See Appendix \ref{lie} for a description of these groups).
However, the largest connected matrix Lie group that preserve both the bilinear forms $\psi^TC\varphi$ and $\psi^TC\gamma^5\varphi$ as well as both the sesquilinear forms $\psi^\dagger\gamma^0\varphi$ and $\psi^\dagger\gamma^0\gamma^5\varphi$ is the spinor representation of the proper orthochronous Lorentz group.

\section{Dynamical evolution of the bilinear and sesquilinear forms}
\label{dynnn}
The two bilinear forms $\psi^T(x)C\varphi(x)$ and $\psi^T(x)C\gamma^5\varphi(x)$ and the two sesquilinear forms $\psi^\dagger(x)\gamma^0\varphi(x)$ and $\psi^\dagger(x)\gamma^0\gamma^5\varphi(x)$ described in Section \ref{invariants} are constructed from Dirac spinors. Therefore they obey dynamical equations that can be derived from the Dirac equation. Here we describe a few such dynamical equations and in particular consider the time evolution of the bilinear and sesquilinear forms.

We first consider again the Dirac equation and include a pseudoscalar coupling term $g\gamma^5\phi(x)$ where $g$ is a real valued constant and $\phi(x)$ is a real valued function (See e.g. Ref. \cite{thaller})
\begin{eqnarray}
\sum_{\mu}\gamma^\mu\partial_{\mu}\psi(x) =(-iq\sum_{\mu}\gamma^\mu A_{\mu}(x) -im+g\gamma^5\phi(x))\psi(x).\nonumber\\
\end{eqnarray}
The transpose of this Dirac equation is then given by
\begin{eqnarray}
\sum_{\mu}\partial_{\mu}\psi(x)^{T}\gamma^{\mu T} =\psi(x)^{T}(-iq\sum_{\mu}\gamma^{\mu T} A_{\mu}(x) -im+g\gamma^5\phi(x)),\nonumber\\
\end{eqnarray}
and the conjugate transpose of this Dirac equation is given by
\begin{eqnarray}
\sum_{\mu}\partial_{\mu}\psi(x)^{\dagger}\gamma^{\mu \dagger} =\psi(x)^{\dagger}(+iq\sum_{\mu}\gamma^{\mu \dagger} A_{\mu}(x) +im+g\gamma^5\phi(x)).\nonumber\\
\end{eqnarray}

For the bilinear form $\psi^T(x)C\varphi(x)$ we can use the Dirac equation and its transpose to derive the following equation describing the dynamical evolution of $\psi^T(x)C\varphi(x)$,
\begin{eqnarray}\label{hugc}
&&\psi^T(x)C\sum_{\mu}\gamma^0\gamma^\mu(\partial_{\mu}\varphi(x)) + \sum_{\mu}(\partial_{\mu}\psi^T(x))\gamma^{\mu T}\gamma^0 C\varphi(x)\nonumber\\
=&&\psi^T(x)C(-iq\sum_{\mu}\gamma^0\gamma^\mu A_{\mu}(x) -im\gamma^0 +g\gamma^0\gamma^5\phi(x))\varphi(x)\nonumber\\&&+\psi^T(x)(-iq\sum_{\mu}\gamma^{\mu T}\gamma^0 A_{\mu}(x) -im\gamma^0+g\gamma^5\gamma^0\phi(x))C\varphi(x)\nonumber\\
=&&2\psi^T(x)C(-iq A_{0}(x) -im \gamma^0)\varphi(x).
\end{eqnarray}
In particular, Eq. (\ref{hugc}) implies that the time derivative of $\psi^T(x)C\varphi(x)$ is given by
\begin{eqnarray}\label{hugc2}
&&\partial_{0}(\psi^T(x)C\varphi(x))\nonumber\\
&=&2\psi^T(x)C(-iq A_{0}(x) -im \gamma^0)\varphi(x)\nonumber\\
&&+\sum_{\mu=1,2,3}\left[ (\partial_{\mu}\psi^T(x))C\gamma^0\gamma^{\mu} \varphi(x)-\psi^T(x)C\gamma^0\gamma^\mu(\partial_{\mu}\varphi(x))\right].\nonumber\\
\end{eqnarray}
We can see from  Eq. (\ref{hugc2}) that if $\psi(x)$ and $\varphi(x)$ are momentum eigenstates with the same definite momentum and the mass $m$ equals zero, we have the first order linear ordinary differential equation $\partial_{0}(\psi^T(x)C\varphi(x))=-2iq A_{0}(x)\psi^T(x)C\varphi(x)$. It follows that $\psi^T(t,\mathbf{x})C\varphi(t,\mathbf{x})=K\exp(-2iq\int_{0}^t A_{0}(s,\mathbf{x})ds)$ where $K$ is a constant.
Thus in this case $\psi^T(x)C\varphi(x)$ as a function of time is invariant up to a U(1) phase as described in Ref. \cite{spinorent}.

Similarly, for the bilinear form $\psi^T(x)C\gamma^5\varphi(x)$ we can use the Dirac equation and its transpose to derive the following equation describing the dynamical evolution of $\psi^T(x)C\gamma^5\varphi(x)$,
\begin{eqnarray}\label{hugd}
&&\psi^T(x)C\gamma^5\sum_{\mu}\gamma^0\gamma^\mu(\partial_{\mu}\varphi(x)) + \sum_{\mu}(\partial_{\mu}\psi^T(x))\gamma^{\mu T}\gamma^0 C\gamma^5\varphi(x)\nonumber\\
&=&\psi^T(x)C\gamma^5(-iq\sum_{\mu}\gamma^0\gamma^\mu A_{\mu}(x) -im\gamma^0 +g\gamma^0\gamma^5\phi(x))\varphi(x)\nonumber\\&&+\psi^T(x)(-iq\sum_{\mu}\gamma^{\mu T}\gamma^0 A_{\mu}(x) -im\gamma^0+g\gamma^5\gamma^0\phi(x))C\gamma^5\varphi(x)\nonumber\\
&=&2\psi^T(x)C\gamma^5(-iq A_{0}(x) +g\gamma^0\gamma^5\phi(x))\varphi(x).
\end{eqnarray}
In particular Eq. (\ref{hugd}) implies that the time derivative of $\psi^T(x)C\gamma^5\varphi(x)$ is given by
\begin{eqnarray}\label{hugd2}
&&\partial_{0}(\psi^T(x)C\gamma^5\varphi(x))\nonumber\\
&=&2\psi^T(x)C\gamma^5(-iq A_{0}(x) +g\gamma^0\gamma^5\phi(x))\varphi(x)\nonumber\\
&&+\sum_{\mu=1,2,3}\left[ (\partial_{\mu}\psi^T(x))C\gamma^5\gamma^0\gamma^{\mu} \varphi(x)-\psi^T(x)C\gamma^5\gamma^0\gamma^\mu(\partial_{\mu}\varphi(x))\right].\nonumber\\
\end{eqnarray}
From  Eq. (\ref{hugd2}) we can see that if $\psi(x)$ and $\varphi(x)$ are momentum eigenstates with the same definite momentum and the pseudoscalar coupling $g$ equals zero, we have that $\partial_{0}(\psi^T(x)C\gamma^5\varphi(x))=-2iq A_{0}(x)\psi^T(x)C\gamma^5\varphi(x)$. Thus in this case $\psi^T(x)C\gamma^5\varphi(x)$ as a function of time is invariant up to a U(1) phase as described in Ref. \cite{spinorent}.

For the sesquilinear form $\psi^\dagger(x)\gamma^0\varphi(x)$ we can use the Dirac equation and its conjugate transpose to derive the following equation describing the dynamical evolution of $\psi^\dagger(x)\gamma^0\varphi(x)$,
\begin{eqnarray}\label{huge}
&&\psi^\dagger(x)\gamma^0\sum_{\mu}\gamma^0\gamma^\mu(\partial_{\mu}\varphi(x)) + \sum_{\mu}(\partial_{\mu}\psi^\dagger(x))\gamma^{\mu \dagger}\gamma^0\gamma^0\varphi(x)\nonumber\\
&=&\psi^\dagger(x)\gamma^0(-iq\sum_{\mu}\gamma^0\gamma^\mu A_{\mu}(x) -im\gamma^0 +g\gamma^0\gamma^5\phi(x))\varphi(x)\nonumber\\&&+\psi^\dagger(x)(iq\sum_{\mu}\gamma^{\mu \dagger}\gamma^0 A_{\mu}(x) +im\gamma^0+g\gamma^5\gamma^0\phi(x))\gamma^0\varphi(x)\nonumber\\
&=&2\psi^\dagger(x)(-iq\sum_{\mu=1,2,3}\gamma^\mu A_{\mu}(x) +g\gamma^5\phi(x))\varphi(x).
\end{eqnarray}
In particular Eq. (\ref{huge}) implies that the time derivative of $\psi^\dagger(x)\gamma^0\varphi(x)$ is given by
\begin{eqnarray}\label{huge2}
&&\partial_{0}(\psi^\dagger(x)\gamma^0\varphi(x))\nonumber\\
&=&2\psi^\dagger(x)(-iq\sum_{\mu=1,2,3}\gamma^\mu A_{\mu}(x) +g\gamma^5\phi(x))\varphi(x)\nonumber\\
&&+\sum_{\mu=1,2,3}\left[ (\partial_{\mu}\psi^\dagger(x))\gamma^{\mu} \varphi(x)-\psi^\dagger(x)\gamma^\mu(\partial_{\mu}\varphi(x))\right].
\end{eqnarray}

Similarly, for the sesquilinear form $\psi^\dagger(x)\gamma^0\gamma^5\varphi(x)$ we can use the Dirac equation and its conjugate transpose to derive the following equation describing the dynamical evolution of $\psi^\dagger(x)\gamma^0\gamma^5\varphi(x)$,
\begin{eqnarray}\label{hugf}
&&\psi^\dagger(x)\gamma^0\gamma^5\sum_{\mu}\gamma^0\gamma^\mu(\partial_{\mu}\varphi(x)) + \sum_{\mu}(\partial_{\mu}\psi^\dagger(x))\gamma^{\mu \dagger}\gamma^0\gamma^0\gamma^5\varphi(x)\nonumber\\
&=&\psi^\dagger(x)\gamma^0\gamma^5(-iq\sum_{\mu}\gamma^0\gamma^\mu A_{\mu}(x) -im\gamma^0 +g\gamma^0\gamma^5\phi)\varphi(x)\nonumber\\&&+\psi^\dagger(x)(iq\sum_{\mu}\gamma^{\mu \dagger}\gamma^0 A_{\mu}(x) +im\gamma^0+g\gamma^5\gamma^0\phi)\gamma^0\gamma^5\varphi(x)\nonumber\\
&=&2\psi^\dagger(x)\gamma^5(iq\sum_{\mu=1,2,3}\gamma^\mu A_{\mu}(x)  +im)\varphi(x).
\end{eqnarray}
In particular Eq. (\ref{hugf}) implies that the time derivative of $\psi^\dagger(x)\gamma^0\gamma^5\varphi(x)$ is given by
\begin{eqnarray}\label{hugf2}
&&\partial_{0}(\psi^\dagger(x)\gamma^0\gamma^5\varphi(x))\nonumber\\
&=&2\psi^\dagger(x)\gamma^5(iq\sum_{\mu=1,2,3}\gamma^\mu A_{\mu}(x)  +im)\varphi(x)\nonumber\\&&+
\sum_{\mu=1,2,3}[\psi^\dagger(x)\gamma^5\gamma^\mu(\partial_{\mu}\varphi(x)) - (\partial_{\mu}\psi^\dagger(x))\gamma^5\gamma^{\mu}\varphi(x)].
\end{eqnarray}

Note that unlike the two bilinear forms neither of the two sesquilinear forms are invariant up to a U(1) phase as a function of time when the spinors are momentum eigenstates with the same definite momentum, regardless of whether $m$ or $g$ are zero. Thus they cannot be used to construct the kind of invariants of the time evolution as have been considered in Ref. \cite{spinorent} and Ref. \cite{multispinor}.

\section{Constructing polynomial spinor entanglement indicators for two Dirac particles}
\label{consrt}
Here we describe a method for constructing locally Lorentz invariant polynomials that are indicators of spinor entanglement for two spacelike separated Dirac particles. The method is an extension of the method introduced in Reference \cite{spinorent}.

Consider two spacelike separated observers, here named Alice and Bob, that each has their own laboratory containing a Dirac particle. Let the two particles be in a joint state and assume that operations on Alice's particle  can be made jointly with operations on Bob's particle, i.e., assume that Alice's operations commute with Bob's operations. We assume that a tensor product structure can be used to describe the shared two-particle system and use the tensor products $\phi_{j_A}\otimes \phi_{k_B}$ of local basis spinors as a basis. Let $x_A$ be the spacetime coordinate in Alice's Minkowski space and let $x_B$ be the spacetime coordinate in Bob's Minkowski space.
Then we can expand the state in this basis as
\begin{eqnarray}\label{bilbo}
\psi_{AB}(x_A,x_B)=\sum_{j_A,k_B}\psi_{j_A,k_B}(x_A,x_B)\phi_{j_A}\otimes \phi_{k_B},
\end{eqnarray}
where $\psi_{j_A,k_B}(x_A,x_B)$ are complex valued functions of the coordinates $x_A$ and $x_B$.

Any state of a system of two Dirac particles that can be created using only local resources, i.e., any product state, is by definition not entangled while all other states are entangled. Any product state can be completely factorized as $\psi(x_A)\otimes\varphi(x_B)$ for some $\psi(x_A)$ and $\varphi(x_B)$. Thus, a state not on this form is entangled.

A function that is identically zero for all product states but that is non-zero for at least some other state is an entanglement indicator. Here we consider functions that are indicators of entanglement between the spinorial degrees of freedom of Alice's particle and the spinorial degrees of freedom of Bob's particle.
Specifically, we consider such functions of the spinorial degree of freedom of Alice's particle at the spacetime coordinate $x_A$ and the spinorial degree of freedom of Bob's particle at the spacetime coordinate $x_B$. Additionally we require that the functions are Lorentz invariant in each of the two observers labs, i.e., we require that the properties these functions describe are independent of the choice of local reference frames.

We can utilize the properties of the bilinear forms and sesquilinear forms described in Section \ref{invariants} to construct locally Lorentz invariant polynomial indicators of spinorial entanglement.
The two bilinear forms $\psi^T(x)C\gamma^5\varphi(x)$ and $\psi^T(x)C\varphi(x)$ as well as the two sesquilinear forms $\psi^\dagger(x)\gamma^0\varphi(x)$ and $\psi^\dagger(x)\gamma^0\gamma^5\varphi(x)$ are each pointwise Lorentz invariant. Moreover, both $\psi^T(x)C\psi(x)$ and $\psi^T(x)C\gamma^5\psi(x)$ are identically zero since they are skew-symmetric.
Below we show how mixed locally Lorentz invariant polynomials can be constructed for two of Dirac particles in a way that is an extension of the method used to construct homogeneous polynomials for two Dirac particles in Ref. \cite{spinorent}.

We consider again the state in Eq. (\ref{bilbo}) and suppress the subscripts $A$ and $B$ in the description of the spinor basis elements and let $\psi_{jk}(x_A,x_B)\equiv \psi_{j_A,k_B}(x_A,x_B)$.
The state coefficients $\psi_{jk}(x_A,x_B)$ can be arranged as a matrix by letting $j$ be the row index and $k$ be the column index. As in Ref. \cite{spinorent} we denote this matrix by $\Psi_{AB}(x_A,x_B)$. Written out it is 
\begin{eqnarray}\label{bofur}
&&\Psi_{AB}(x_A,x_B)\nonumber\\
&&\equiv\sum_{jk}\psi_{jk}(x_A,x_B)\phi_{j}\otimes \phi_{k}^T\nonumber\\&&=
\begin{pmatrix}
\psi_{00}(x_A,x_B) & \psi_{01}(x_A,x_B) & \psi_{02}(x_A,x_B) & \psi_{03}(x_A,x_B)\\
\psi_{10}(x_A,x_B) & \psi_{11}(x_A,x_B) & \psi_{12}(x_A,x_B) & \psi_{13}(x_A,x_B)\\
\psi_{20}(x_A,x_B) & \psi_{21}(x_A,x_B) & \psi_{22}(x_A,x_B) & \psi_{23}(x_A,x_B) \\
\psi_{30}(x_A,x_B) & \psi_{31}(x_A,x_B) & \psi_{32}(x_A,x_B) & \psi_{33}(x_A,x_B)\\
\end{pmatrix}.\nonumber\\
\end{eqnarray}
The spinor representation of a proper orthochronous Lorentz transformation on Alice's particle $S_A$ acts on $\Psi_{AB}$ from the left and the spinor representation of a proper orthochronous Lorentz transformation on Bob's particle $S_B$ acts in transposed form $S_{B}^T$ from the right
\begin{eqnarray}
\Psi_{AB}\to S_A\Psi_{AB}S_{B}^T.
\end{eqnarray}
Therefore, we can see from the Lorentz invariance of the bilinear forms $\psi^T(x)C\gamma^5\varphi(x)$ and $\psi^T(x)C\varphi(x)$ that a proper orthochronous Lorentz transformation in Alice's Minkowski space preserves the matrices $\Psi_{AB}^TC\Psi_{AB}$ and $\Psi_{AB}^TC\gamma^5\Psi_{AB}$. Likewise we can see from the Lorentz invariance of the sesquilinear forms $\psi^\dagger(x)\gamma^0\varphi(x)$ and $\psi^\dagger(x)\gamma^0\gamma^5\varphi(x)$ that a proper orthochronous Lorentz transformation in Alice's Minkowski space preserves the matrices $\Psi_{AB}^\dagger \gamma^0\Psi_{AB}$ and $\Psi_{AB}^\dagger \gamma^0\gamma^5\Psi_{AB}$,
\begin{eqnarray}\label{uuvvcx}
\Psi_{AB}^TC\Psi_{AB}&&\to \Psi_{AB}^TS_{A}^TCS_{A}\Psi_{AB}=\Psi_{AB}^TC\Psi_{AB},\nonumber\\
\Psi_{AB}^TC\gamma^5\Psi_{AB}&&\to \Psi_{AB}^TS_{A}^TC\gamma^5 S_{A}\Psi_{AB}=\Psi_{AB}^TC\gamma^5 \Psi_{AB},\nonumber\\
\Psi_{AB}^\dagger \gamma^0\Psi_{AB}&&\to \Psi_{AB}^\dagger S_{A}^\dagger\gamma^0 S_{A}\Psi_{AB}=\Psi_{AB}^\dagger \gamma^0\Psi_{AB},\nonumber\\
\Psi_{AB}^\dagger \gamma^0\gamma^5\Psi_{AB}&&\to \Psi_{AB}^\dagger S_{A}^\dagger\gamma^0\gamma^5 S_{A}\Psi_{AB}=\Psi_{AB}^\dagger \gamma^0\gamma^5\Psi_{AB}.
\end{eqnarray}

Similarly, a proper orthochronous Lorentz transformation in Bob's Minkowski space preserves the matrices $\Psi_{AB}C\Psi_{AB}^T$ and $\Psi_{AB}C\gamma^5\Psi_{AB}^T$ as well as the matrices $\Psi_{AB}^* \gamma^0\Psi_{AB}^T$ and $\Psi_{AB}^* \gamma^0\gamma^5\Psi_{AB}^T$
\begin{eqnarray}\label{uuvvcx2}
\Psi_{AB}C\Psi_{AB}^T&&\to \Psi_{AB}S_{B}^TCS_{B}\Psi_{AB}^T=\Psi_{AB}C\Psi_{AB}^T,\nonumber\\
\Psi_{AB}C\gamma^5\Psi_{AB}^T&&\to \Psi_{AB}S_{B}^TC\gamma^5 S_{B}\Psi_{AB}^T=\Psi_{AB}C\gamma^5 \Psi_{AB}^T,\nonumber\\
\Psi_{AB}^* \gamma^0\Psi_{AB}^T&&\to \Psi_{AB}^* S_{B}^\dagger\gamma^0 S_{B}\Psi_{AB}^T=\Psi_{AB}^* \gamma^0\Psi_{AB}^T,\nonumber\\
\Psi_{AB}^* \gamma^0\gamma^5\Psi_{AB}^T&&\to \Psi_{AB}^* S_{B}^\dagger\gamma^0\gamma^5 S_{B}\Psi_{AB}^T=\Psi_{AB}^* \gamma^0\gamma^5\Psi_{AB}^T.
\end{eqnarray}

Moreover, if the state factorizes on the pair points $x_A,x_B$, i.e., if $\Psi_{AB}(x_A,x_B)=\psi(x_A)\otimes\varphi(x_B)^T$ we have that
\begin{eqnarray}\label{ttredf}
\Psi_{AB}C\Psi_{AB}^T&=&\psi(x_A)\otimes\varphi(x_B)^TC\varphi(x_B)\otimes\psi(x_A)^T=0,\nonumber\\
\Psi_{AB}^TC\Psi_{AB}&=&\varphi(x_B)\otimes\psi(x_A)^TC\psi(x_A)\otimes\varphi(x_B)^T=0,\nonumber\\
\Psi_{AB}C\gamma^5\Psi_{AB}^T&=&\psi(x_A)\otimes\varphi(x_B)^TC\gamma^5\varphi(x_B)\otimes\psi(x_A)^T=0,\nonumber\\
\Psi_{AB}^TC\gamma^5\Psi_{AB}&=&\varphi(x_B)\otimes\psi(x_A)^TC\gamma^5\psi(x_A)\otimes\varphi(x_B)^T=0,\nonumber\\
\end{eqnarray}
as well as
\begin{eqnarray}\label{sessi}
\Psi_{AB}\gamma^0\Psi_{AB}^\dagger &&=\psi(x_A)\otimes\varphi(x_B)^T\gamma^0\varphi(x_B)^*
\otimes\psi(x_A)^\dagger\nonumber\\&&=\psi(x_A)\otimes\psi(x_A)^\dagger \Tr[\varphi(x_B)^T\gamma^0\varphi(x_B)^*],\nonumber\\
\Psi_{AB}^\dagger\gamma^0\Psi_{AB}&&=\varphi(x_B)^*\otimes\psi(x_A)^\dagger\gamma^0\psi(x_A)\otimes\varphi(x_B)^T\nonumber\\&&=\varphi(x_B)^*\otimes\varphi(x_B)^T \Tr[\psi(x_A)^\dagger\gamma^0\psi(x_A)],\nonumber\\
\Psi_{AB}\gamma^0\gamma^5\Psi_{AB}^\dagger &&=\psi(x_A)\otimes\varphi(x_B)^T\gamma^0\gamma^5\varphi(x_B)^*\otimes\psi(x_A)^\dagger\nonumber\\&&=\psi(x_A)\otimes\psi(x_A)^\dagger \Tr[\varphi(x_B)^T\gamma^0\gamma^5\varphi(x_B)^*],\nonumber\\
\Psi_{AB}^\dagger\gamma^0\gamma^5\Psi_{AB}&&=\varphi(x_B)^*\otimes\psi(x_A)^\dagger\gamma^0\gamma^5\psi(x_A)\otimes\varphi(x_B)^T\nonumber\\&&=\varphi(x_B)^*\otimes\varphi(x_B)^T \Tr[\psi(x_A)^\dagger\gamma^0\gamma^5\psi(x_A)].\nonumber\\
\end{eqnarray}

The properties of the eight matrices $\Psi_{AB}^TC\Psi_{AB}$, $\Psi_{AB}C\Psi_{AB}^T$, $\Psi_{AB}^TC\gamma^5\Psi_{AB}$, $\Psi_{AB}C\gamma^5\Psi_{AB}^T$, $\Psi_{AB}^\dagger \gamma^0\Psi_{AB}$, $\Psi_{AB}\gamma^0\Psi_{AB}^\dagger $, $\Psi_{AB}^\dagger \gamma^0\gamma^5\Psi_{AB}$ and $\Psi_{AB} \gamma^0\gamma^5\Psi_{AB}^\dagger$ described in Eq. (\ref{uuvvcx}), Eq. (\ref{uuvvcx2}), Eq. (\ref{ttredf}), and Eq. (\ref{sessi}) makes it possible to construct polynomials that are Lorentz invariant in both Alice's and Bob's Minkowski space and take the value zero for all product states using matrix traces.
We now describe a method for constructing such polynomials.

Consider a four element string of matrices $\Psi_{AB}\phantom{0}X\phantom{0} \Psi_{AB}^T\phantom{0}X$ where $X$ is an unspecified matrix. Then consider the string that is made up of $n$ repetitions of this four element string.
Replace the first $2n-1$ $X$s in this string with some combination of the matrices $C$, $C\gamma^5$, $\gamma^0$, and $\gamma^0\gamma^5$. Next, complex conjugate the copies of $\Psi_{AB}$ and $\Psi_{AB}^T$ in a subset of the string chosen in such a way that each of the $C$ and the $C\gamma^5$ are now in between a $\Psi_{AB}$ and a
$\Psi_{AB}^T$ or alternatively a $\Psi_{AB}^*$ and a
$\Psi_{AB}^\dagger$, and each of the $\gamma^0$ and the $\gamma^0\gamma^5$ are in between a $\Psi_{AB}$ and a
$\Psi_{AB}^\dagger$ or alternatively a $\Psi_{AB}^*$ and a
$\Psi_{AB}^T$. Finally, if the string is now on the form $\Psi_{AB}\phantom{0}\dots\phantom{0} \Psi_{AB}^T\phantom{0}X$ or on the form $\Psi_{AB}^*\phantom{0}\dots\phantom{0} \Psi_{AB}^\dagger\phantom{0}X$ replace the final $X$ with either $C$ or $C\gamma^5$, and otherwise replace the final $X$ with either $\gamma^0$ or $\gamma^0\gamma^5$.
Then realize the string as a matrix product and take the trace over it.
This matrix trace is Lorentz invariant in both Alice's and Bob's lab. In particular it is a polynomial  in the state coefficients $\psi_{jk}$ and the complex conjugated state coefficients $\psi^*_{jk}$ of bidegree $(2n-m,m)$ where $m$ is the number of copies of $\Psi_{AB}^*$ and
$\Psi_{AB}^\dagger$ in the string.
This procedure can always be carried out for any combination of $C$, $C\gamma^5$, $\gamma^0$, and $\gamma^0\gamma^5$ replacing the first $2n-1$ $X$s. If only $C$ and $C\gamma^5$ are used in the procedure the result is either a homogeneous polynomial in the state coefficients $\psi_{jk}$ or a homogeneous polynomial in the complex conjugated state coefficients $\psi^*_{jk}$ and the procedure essentially coincides with that described in Ref. \cite{spinorent}.

If the string contains at least one $C$ or $C\gamma^5$ the corresponding matrix trace is identically zero for all product states as can be seen from Eq. (\ref{ttredf}). Thus such a matrix trace is a locally Lorentz invariant indicator of spinor entanglement. Note that if the bidegree $(n,m)$ of the polynomial is such that $m\neq n$ it follows that for each observer at least one $C$ or $C\gamma^5$ has been used.

If the string does not contain any $C$ or $C\gamma^5$ the corresponding matrix trace is not zero for all product states. However we may still be able to use this kind of strings to
create polynomials that are identically zero for all product states. 
Consider a polynomial constructed from a
string that does not contain any $C$ or $C\gamma^5$, or a product of such polynomials.
For any product state $\Psi_{AB}(x_A,x_B)=\psi(x_A)\otimes\varphi(x_B)^T$ a polynomial of this kind reduces, up to a sign, to some product of the Hermitian sesquilinear forms
$\varphi(x_B)^\dagger\gamma^0\varphi(x_B)$ and $\psi(x_A)^\dagger\gamma^0\psi(x_A)$, and the skew-Hermitian sesquilinear forms $\varphi(x_B)^\dagger\gamma^0\gamma^5\varphi(x_B)$ and $\psi(x_A)^\dagger\gamma^0\gamma^5\psi(x_A)$ as can be seen from Eq. (\ref{sessi}). 
Therefore if we have two linearly independent polynomials that reduce to the same product of sesquilinear forms on the product states we can create a polynomial that is identically zero for all product states by taking their difference. It follows that if we have a set of $n$ linearly independent polynomials that reduce to the same product of sesquilinear forms on the product states we can construct $n-1$ linearly independent polynomials that are identically zero for all product states.
Such linear combinations are locally Lorentz invariant indicators of spinor entanglement.

The locally Lorentz invariant matrix traces that have been previously constructed in Ref. \cite{spinorent} use only the bilinear forms in the construction and the resulting polynomials are all homogeneous. The use of also the sesquilinear forms as described above allows the construction of additional locally Lorentz invariant entanglement indicators as mixed polynomials that are sensitive to types of entanglement not indicated by the homogeneous polynomials.

Since the polynomials are constructed from bilinear and sesquilinear forms involving Dirac spinors they satisfy dynamical equations that can be derived from the dynamical equations of the individual bilinear and sesquilinear forms given in Section \ref{dynnn}. In particular we note that if for Alice's particle only bilinear forms are used to construct the polynomial it follows from Eq. (\ref{hugc2}) and Eq. (\ref{hugd2}) that under certain conditions the polynomial is invariant up to a U(1) phase under evolution with respect to Alice's time. These conditions, described in Section \ref{dynnn}, are that Alice's particle is in a momentum eigenstate, that the evolution preserves this momentum eigenspace and depending on the combination of bilinear forms used that either the mass $m$ of Alice's particle, the pseudoscalar coupling $g$, or both, are zero. 
However, as described in Section \ref{dynnn} the sesquilinear forms do not have the same kind of dynamical evolution as the bilinear forms. Therefore if for Alice's particle one or more sesquilinear forms are used to construct the polynomial it is in general not invariant up to a U(1) phase under evolution with respect to Alice's time under these conditions.
The analogous situation holds for Bob's particle. 
We note that if for some homogeneous polynomial constructed according to the above method the conditions described above hold for both Alice's and Bob's particles, the homogeneous polynomial is invariant up to a U(1) phase under evolution with respect to both Alice's and Bob's time as described in Ref. \cite{spinorent}.
However, the mixed polynomials do in general not have this property.

Finally, we note that by using $C$ and $C\gamma^5$ in the construction of the local Lorentz invariants we have in essence used the spinor representations of the T and CP transformation, respectively. Using such inversion transformations is similar to the idea of using "state inversion" transformations \cite{wootters2,uhlmann,rungta} in the construction of the Wootters concurrence \cite{wootters,wootters2} for the case of two non-relativistic spin-$\frac{1}{2}$ particles.
Moreover, by using $\gamma^0$ and $\gamma^0\gamma^5$ in the construction we have in essence used the spinor representations of the P and CT transformation, respectively.

\subsection{Polynomials of bidegree (2,0) and (0,2)}
In Ref. \cite{spinorent} four locally Lorentz invariant polynomials of bidegree (2,0) have been constructed based on the properties of the two bilinear forms $\psi^T(x)C\gamma^5\varphi(x)$ and $\psi^T(x)C\varphi(x)$. These polynomials are identically zero for any product state. They are thus locally Lorentz invariant spinor entanglement indicators. The four polynomials $I_1,I_2, I_{2A}$ and $I_{2B}$ are defined by
\begin{eqnarray}
I_1&&=\frac{1}{2}\Tr[\Psi_{AB}^TC\Psi_{AB} C],\nonumber\\
I_2&&=\frac{1}{2}\Tr[\Psi_{AB}^TC\gamma^5\Psi_{AB} C\gamma^5],\nonumber\\
I_{2A}&&=\frac{1}{2}\Tr[\Psi_{AB}^TC\Psi_{AB} C\gamma^5],\nonumber\\
I_{2B}&&=\frac{1}{2}\Tr[\Psi_{AB}^TC\gamma^5\Psi_{AB} C].
\end{eqnarray}
 
Explicitly written out the polynomials are
\begin{eqnarray}
I_1&&=\psi_{00} \psi_{11}-\psi_{01} \psi_{10} +\psi_{02}\psi_{13} - \psi_{03} \psi_{12} \nonumber\\&&+\psi_{20}      \psi_{ 31} -\psi_{21}\psi_{30}+\psi_{22 }\psi_{33}  - 
 \psi_{23}\psi_{32},\\
I_{2}&&=\psi_{13} \psi_{20}-\psi_{10}\psi_{23} +\psi_{11}\psi_{22}-\psi_{12} \psi_{21}\nonumber\\ &&+\psi_{02}\psi_{31} -\psi_{01}\psi_{32} +\psi_{00}\psi_{33} -\psi_{03}\psi_{30},\\
I_{2A}&&=\psi_{00}\psi_{13}-\psi_{03}\psi_{10} +\psi_{02}\psi_{11} -\psi_{01}\psi_{12}\nonumber\\  &&+\psi_{22}\psi_{31} - \psi_{21}\psi_{32} +\psi_{20}\psi_{33} -\psi_{23}\psi_{30} ,\\
I_{2B}&&=\psi_{11}\psi_{20} -\psi_{10}\psi_{21} +\psi_{13}\psi_{22} -\psi_{12}\psi_{23}\nonumber\\ &&+\psi_{00}\psi_{31}-\psi_{01}\psi_{30}  +\psi_{02}\psi_{33}-\psi_{03}\psi_{32}.
\end{eqnarray}
The complex conjugates of $I_1,I_2, I_{2A}$ and $I_{2B}$ are locally Lorentz invariant polynomial spinor entanglement indicators of bidegree (0,2).

Next we briefly comment on the time evolution of these polynomials with respect to Alice's or Bob's time. 
We can find the dynamical evolution of the Lorentz invariant polynomials from the dynamical evolution of the bilinear forms described in Section \ref{dynnn}. For example the derivative with respect to Alice's time of the polynomial $I_1$ is
\begin{eqnarray}\label{ewwwwwwo}
&&\partial^A_{0}(\Tr[\Psi_{AB}^T(x_A) C \Psi_{AB}(x_A) C])\nonumber\\
&=&-2iqA_{0}(x_A)\Tr[\Psi_{AB}^T(x_A) C\Psi_{AB}(x_A) C]\nonumber\\
&&-2im\Tr[\Psi_{AB}^T(x_A)\gamma^0 C \Psi_{AB}(x_A) C]\nonumber\\
&&-\sum_{\mu=1,2,3}\Tr[\Psi_{AB}^T(x_A) C \gamma^0\gamma^\mu(\partial^A_{\mu}\Psi_{AB}(x_A)) C]\nonumber\\
&&+\sum_{\mu=1,2,3}\Tr[(\partial^A_{\mu}\Psi_{AB}^T(x_A)) C \gamma^0\gamma^{\mu}\Psi_{AB}(x_A) C],
\end{eqnarray}
where $x_A$ is the coordinate in Alice's Minkowski space.
In particular we can see from  Eq. (\ref{ewwwwwwo}) that if $\Psi_{AB}(x_A)$ is a momentum eigenstate with respect to Alice's inertial reference frame so that $\partial^A_{\mu}\Psi_{AB}^T(x_A)\propto \Psi_{AB}^T(x_A)$ for all $\mu=1,2,3$ and the mass $m$ equals zero, we have the first order linear ordinary differential equation $\partial_{0}(\Tr[\Psi_{AB}^T(x_A) C\Psi_{AB}(x_A) C])=-2iq A_{0}(x_A)\Tr[\Psi_{AB}^T(x_A) C\Psi_{AB}(x_A) C]$. 
This follows since $\Tr[\Psi_{AB}^T(x_A) C \gamma^0\gamma^\mu\Psi_{AB}(x_A) C]=0$ for all $\mu=1,2,3$.
Solving the differential equation gives that $\Tr[\Psi_{AB}^T(t_A,\mathbf{x}_A) C\Psi_{AB}(t_A,\mathbf{x}_A) C]=K\exp(-2iq\int_{0}^{t_A} A_{0}(s,\mathbf{x}_A)ds)$ where $K$ is a constant.
Thus in this case $I_1$ as a function of Alice's time is invariant up to a U(1) phase as described in Ref. \cite{spinorent}.

Similarly the derivative with respect to Alice's time of the polynomial $I_2$ is
\begin{eqnarray}\label{ewwwwwwo2}
&&\partial^A_{0}(\Tr[\Psi_{AB}^T(x_A) C \Psi_{AB}(x_A) C])\nonumber\\
&=&-2iqA_{0}(x_A)\Tr[\Psi_{AB}^T(x_A) C\gamma^5\Psi_{AB}(x_A) C\gamma^5]\nonumber\\&&+g\phi(x_A)\Tr[\Psi_{AB}^T(x_A)\gamma^0 C\gamma^5 \Psi_{AB}(x_A) C\gamma^5]\nonumber\\
&&-\sum_{\mu=1,2,3}\Tr[\Psi_{AB}^T(x_A) C\gamma^5 \gamma^0\gamma^\mu(\partial^A_{\mu}\Psi_{AB}(x_A)) C\gamma^5]\nonumber\\
&&+\sum_{\mu=1,2,3}\Tr[(\partial^A_{\mu}\Psi_{AB}^T(x_A)) C\gamma^5 \gamma^0\gamma^{\mu}\Psi_{AB}(x_A) C\gamma^5],
\end{eqnarray}
where $x_A$ is the coordinate in Alice's Minkowski space.
From  Eq. (\ref{ewwwwwwo2}) we can see that if $\Psi_{AB}(x_A)$ is a momentum eigenstate with respect to Alice's inertial reference frame so that $\partial^A_{\mu}\Psi_{AB}^T(x_A)\propto \Psi_{AB}^T(x_A)$ for all $\mu=1,2,3$ and the pseudoscalar coupling $g$ equals zero, we have the first order linear ordinary differential equation $\partial_{0}(\Tr[\Psi_{AB}^T(x_A) C\gamma^5\Psi_{AB}(x_A) C\gamma^5])=-2iq A_{0}(x_A)\Tr[\Psi_{AB}^T(x_A) C\gamma^5\Psi_{AB}(x_A) C\gamma^5]$. 
This follows since $\Tr[\Psi_{AB}^T(x_A) C\gamma^5 \gamma^0\gamma^\mu\Psi_{AB}(x_A) C\gamma^5]=0$ for all $\mu=1,2,3$.
Solving the differential equation gives that $\Tr[\Psi_{AB}^T(t_A,\mathbf{x}_A) C\gamma^5\Psi_{AB}(t_A,\mathbf{x}_A) C\gamma^5]=K\exp(-2iq\int_{0}^{t_A} A_{0}(s,\mathbf{x}_A)ds)$ where $K$ is a constant.
Thus in this case $I_2$ as a function of Alice's time is invariant up to a U(1) phase as described in Ref. \cite{spinorent}.

Finding the derivatives with respect to Bob's time for $I_1$ and $I_2$ is completely analogous. Likewise
finding the derivatives of $I_{2A}$ and $I_{2B}$ with respect to Alice's or Bob's time can be done in the same way.

\subsection{Polynomials of bidegree (1,1)}

We can construct four linearly independent locally Lorentz invariant polynomials of bidegree (1,1) 
using the sesquilinear forms $\psi^\dagger(x)\gamma^0\varphi(x)$ and $\psi^\dagger(x)\gamma^0\gamma^5\varphi(x)$. In writing these polynomials we suppress the subscript $AB$ of $\Psi$ in the following.
The four polynomials are given by
\begin{eqnarray}\label{trew}
N_1&=&\Tr[\Psi^T\gamma^0\Psi^*\gamma^0],\nonumber\\
N_2&=&\Tr[\Psi^T\gamma^0\gamma^5\Psi^*\gamma^0],\nonumber\\
N_3&=&\Tr[\Psi^T\gamma^0\Psi^*\gamma^0\gamma^5],\nonumber\\
N_4&=&\Tr[\Psi^T\gamma^0\gamma^5\Psi^*\gamma^0\gamma^5].
\end{eqnarray}
The polynomials $N_1$ and $N_4$ are real valued while $N_2$ and $N_3$ are pure imaginary.
None of these polynomials is zero for all product states and thus they are not indicators of spinor entanglement. However products of these polynomials can be used in the construction of indicators of spinor entanglement of higher bidegrees as described in Section \ref{consrt}.

\subsection{Polynomials of bidegree (2,2)}
Here we construct locally Lorentz invariant polynomials of bidegree $(2,2)$ that are indicators of spinor entanglement.
We can construct such locally Lorentz invariant polynomials of bidegree (2,2) by using only the bilinear forms $\psi^T(x)C\gamma^5\varphi(x)$ and $\psi^T(x)C\varphi(x)$ or by using only the sesquilinear forms $\psi^\dagger(x)\gamma^0\varphi(x)$ and $\psi^\dagger(x)\gamma^0\gamma^5\varphi(x)$, or alternatively by using both the bilinear and the sesquilinear forms. In writing these polynomials we suppress the subscript $AB$ of $\Psi$ in the following.

Polynomials of bidegree (2,2) can be constructed as a products of the four bidegree (2,0) polynomials 
\begin{eqnarray}
I_1&=&\frac{1}{2}\Tr[\Psi^TC\Psi C],\nonumber\\
I_2&=&\frac{1}{2}\Tr[\Psi^TC\gamma^5\Psi C\gamma^5],\nonumber\\
I_{2A}&=&\frac{1}{2}\Tr[\Psi^TC\Psi C\gamma^5],\nonumber\\
I_{2B}&=&\frac{1}{2}\Tr[\Psi^TC\gamma^5\Psi C],
\end{eqnarray}
with their complex conjugates. 
In addition to the 16 polynomials obtained in this way we can construct eight more bidegree (2,2) polynomials by combining bilinear and sesquilinear forms. Three more bidegree (2,2) polynomials can be constructed from only sesquilinear forms.

We first consider the polynomials combining bilinear and sesquilinear forms.
A real valued polynomial of this kind is
\begin{eqnarray}
R_1&=&\frac{1}{2}\Tr[\Psi^TC\Psi\gamma^0 \Psi^\dagger C\Psi^*\gamma^0]\nonumber\\
&=&\frac{1}{2}\Tr[\Psi^TC\Psi\gamma^0\gamma^5 \Psi^\dagger C\Psi^*\gamma^0\gamma^5]-|I_{2A}|^2+|I_{1}|^2.
\end{eqnarray}
Written out it is
\begin{eqnarray}
R_1&=& |\psi_{03}\psi_{12}- \psi_{02}\psi_{13}+ \psi_{23}\psi_{32}- \psi_{22}\psi_{33}|^2\nonumber\\
  &&-|\psi_{12}\psi_{00}- \psi_{02}\psi_{10}+ \psi_{20}\psi_{32}- \psi_{22}\psi_{30}|^2 \nonumber\\&&-|\psi_{13}\psi_{00}- \psi_{03}\psi_{10}+ \psi_{20}\psi_{33}- 
    \psi_{23}\psi_{30}|^2 \nonumber\\&&+ |\psi_{10}\psi_{01}- 
    \psi_{11}\psi_{00}+ \psi_{21}\psi_{30}- \psi_{20}\psi_{31}|^2\nonumber\\&&-|\psi_{12}\psi_{01}- \psi_{02}\psi_{11}+ \psi_{21}\psi_{32}- \psi_{31}\psi_{22}|^2 \nonumber\\&&- |\psi_{13}\psi_{01}- \psi_{11}\psi_{03} + \psi_{21}\psi_{33}- 
    \psi_{23}\psi_{31}|^2.
\end{eqnarray}
Two complex valued polynomials can be constructed as
\begin{eqnarray}
R_2&=&\frac{1}{2}\Tr[\Psi^TC\gamma^5\Psi\gamma^0 \Psi^\dagger C\Psi^*\gamma^0]\nonumber\\
&=&\frac{1}{2}\Tr[\Psi^TC\gamma^5\Psi\gamma^0\gamma^5 \Psi^\dagger C\Psi^*\gamma^0\gamma^5]+I_{2B}I_1^*-I_{2}I_{2A}^*,
\end{eqnarray}
and its complex conjugate $R_2^*$. Written out $R_2$ is 
\begin{eqnarray}
 R_2&=&  (\psi_{12}\psi_{23}- \psi_{02}\psi_{33}+ \psi_{03}\psi_{32}- \psi_{13}\psi_{22})\nonumber\\&&\times (\psi^*_{03}\psi^*_{12}- \psi^*_{02}\psi^*_{13}+ 
    \psi^*_{23}\psi^*_{32}- \psi^*_{22}\psi^*_{33}) \nonumber\\&& + (\psi_{22}\psi_{10}- \psi_{12}\psi_{20}+ \psi_{02}\psi_{30}- 
    \psi_{00}\psi_{32})\nonumber\\&&\times ( \psi^*_{00}\psi^*_{12}-\psi^*_{02}\psi^*_{10}- \psi^*_{22}\psi^*_{30}+ \psi^*_{20}\psi^*_{32}) \nonumber\\&& + (\psi_{23}\psi_{10}- 
    \psi_{13}\psi_{20}+ \psi_{03}\psi_{30}- \psi_{33}\psi_{00})\nonumber\\&&\times ( \psi^*_{00}\psi^*_{13}-\psi^*_{03}\psi^*_{10}- \psi^*_{23}\psi^*_{30}+ 
    \psi^*_{20}\psi^*_{33}) \nonumber\\&& + (\psi_{11} \psi_{20}- \psi_{10}\psi_{21}+ \psi_{00}\psi_{31}- \psi_{01}\psi_{30}) \nonumber\\&&\times( 
    \psi^*_{00}\psi^*_{11}-\psi^*_{01}\psi^*_{10}- \psi^*_{21}\psi^*_{30}+ \psi^*_{20}\psi^*_{31}) \nonumber\\&& + (\psi_{11}\psi_{22}- \psi_{01}\psi_{32}+ \psi_{31}\psi_{02}- 
    \psi_{21}\psi_{12})\nonumber\\&&\times (\psi^*_{01}\psi^*_{12}-\psi^*_{02}\psi^*_{11}- \psi^*_{22}\psi^*_{31}+ \psi^*_{21}\psi^*_{32}) \nonumber\\&& + (\psi_{11}\psi_{23}- 
    \psi_{01}\psi_{33}+ \psi_{31}\psi_{03}- \psi_{21}\psi_{13})\nonumber\\&&\times (\psi^*_{01}\psi^*_{13}-\psi^*_{03}\psi^*_{11}- \psi^*_{23}\psi^*_{31}+ 
    \psi^*_{21}\psi^*_{33}).
\end{eqnarray}
A second real valued polynomial can be obtained as
\begin{eqnarray}
R_3&=&\frac{1}{2}\Tr[\Psi^TC\gamma^5\Psi\gamma^0 \Psi^\dagger C\gamma^5\Psi^*\gamma^0]\nonumber\\
&=&\frac{1}{2}\Tr[\Psi^TC\gamma^5\Psi\gamma^0\gamma^5 \Psi^\dagger C\gamma^5\Psi^*\gamma^0\gamma^5]+|I_{2B}|^2-|I_{2}|^2.
\end{eqnarray}
Written out it is
\begin{eqnarray}
R_3&=& |\psi_{20}\psi_{11}- \psi_{10}\psi_{21}+ \psi_{00}\psi_{31}- \psi_{01}\psi_{30}|^2 \nonumber\\&&-|\psi_{30}\psi_{03}- \psi_{33}\psi_{00}+ \psi_{10}\psi_{23}- \psi_{13}\psi_{20}|^2 \nonumber\\&&
-| \psi_{23}\psi_{11}- \psi_{33}\psi_{01}+ \psi_{31}\psi_{03}- 
    \psi_{21}\psi_{13}|^2\nonumber\\&& -| \psi_{30}\psi_{02}- 
    \psi_{32}\psi_{00}+ \psi_{22}\psi_{10}- \psi_{12}\psi_{20}|^2\nonumber\\&&  -| \psi_{31}\psi_{02}- \psi_{32}\psi_{01}- \psi_{21}\psi_{12} + \psi_{22}\psi_{11} |^2\nonumber\\&& +| \psi_{33}\psi_{02}- \psi_{32}\psi_{03}+ \psi_{22}\psi_{13}- 
    \psi_{23}\psi_{12}|^2. 
\end{eqnarray}

A third real valued polynomial can be obtained as
\begin{eqnarray}
R_4&=&\frac{1}{2}\Tr[\Psi C\Psi^T\gamma^0 \Psi^* C\Psi^\dagger\gamma^0]\nonumber\\
&=&\frac{1}{2}\Tr[\Psi C\Psi^T\gamma^0\gamma^5 \Psi^* C\Psi^\dagger\gamma^0\gamma^5]-|I_{2B}|^2+|I_{1}|^2.
\end{eqnarray}
This polynomial is related to $R_1$ by a permutation of Alice's and Bob's labs.
Written out it is
\begin{eqnarray}
R_4&=& |\psi_{30}\psi_{21}- \psi_{20}\psi_{31}+ \psi_{32}\psi_{23}- \psi_{22}\psi_{33}|^2\nonumber\\&&
  -|\psi_{21}\psi_{00}- \psi_{20}\psi_{01}+ \psi_{02}\psi_{23}- \psi_{22}\psi_{03}|^2 \nonumber\\&&-|\psi_{31}\psi_{00}- \psi_{30}\psi_{01}+ \psi_{02}\psi_{33}- 
    \psi_{32}\psi_{03}|^2 \nonumber\\&&+ |\psi_{10}\psi_{01}- 
    \psi_{11}\psi_{00}+ \psi_{12}\psi_{03}- \psi_{02}\psi_{13}|^2\nonumber\\&&-|\psi_{21}\psi_{10}- \psi_{20}\psi_{11}+ \psi_{12}\psi_{23}- \psi_{13}\psi_{22}|^2 \nonumber\\&&- |\psi_{31}\psi_{10}- \psi_{11}\psi_{30} + \psi_{12}\psi_{33}- 
    \psi_{32}\psi_{13}|^2.
\end{eqnarray}

Two more complex valued polynomials can be constructed as
\begin{eqnarray}
R_5&=&\frac{1}{2}\Tr[\Psi C\gamma^5\Psi^T\gamma^0 \Psi^* C\Psi^\dagger\gamma^0]\nonumber\\
&=&\frac{1}{2}\Tr[\Psi C\gamma^5\Psi^T\gamma^0\gamma^5 \Psi^* C\Psi^\dagger\gamma^0\gamma^5]+I_{2A}I_1^*-I_{2}I_{2B}^*,
\end{eqnarray}
and its complex conjugate $R_5^*$. The polynomial $R_5$ is related to $R_2$ by a permutation of Alice's and Bob's labs. Written out $R_5$ is 
\begin{eqnarray}
 R_5&=&  (\psi_{21}\psi_{32}- \psi_{20}\psi_{33}+ \psi_{30}\psi_{23}- \psi_{31}\psi_{22})\nonumber\\&&\times (\psi^*_{30}\psi^*_{21}- \psi^*_{20}\psi^*_{31}+ 
    \psi^*_{32}\psi^*_{23}- \psi^*_{22}\psi^*_{33}) \nonumber\\&& + (\psi_{22}\psi_{01}- \psi_{21}\psi_{02}+ \psi_{20}\psi_{03}- 
    \psi_{00}\psi_{23})\nonumber\\&&\times ( \psi^*_{00}\psi^*_{21}-\psi^*_{20}\psi^*_{01}- \psi^*_{22}\psi^*_{03}+ \psi^*_{02}\psi^*_{23}) \nonumber\\&& + (\psi_{32}\psi_{01}- 
    \psi_{31}\psi_{02}+ \psi_{03}\psi_{30}- \psi_{33}\psi_{00})\nonumber\\&&\times ( \psi^*_{00}\psi^*_{31}-\psi^*_{30}\psi^*_{01}- \psi^*_{32}\psi^*_{03}+ 
    \psi^*_{02}\psi^*_{33}) \nonumber\\&& + (\psi_{11} \psi_{02}- \psi_{01}\psi_{12}+ \psi_{00}\psi_{13}- \psi_{10}\psi_{03}) \nonumber\\&&\times( 
    \psi^*_{00}\psi^*_{11}-\psi^*_{01}\psi^*_{10}- \psi^*_{12}\psi^*_{03}+ \psi^*_{02}\psi^*_{13}) \nonumber\\ &&+ (\psi_{11}\psi_{22}- \psi_{10}\psi_{23}+ \psi_{13}\psi_{20}- 
    \psi_{12}\psi_{21})\nonumber\\&&\times (\psi^*_{10}\psi^*_{21}-\psi^*_{20}\psi^*_{11}- \psi^*_{22}\psi^*_{13}+ \psi^*_{12}\psi^*_{23}) \nonumber\\&& + (\psi_{11}\psi_{32}- 
    \psi_{10}\psi_{33}+ \psi_{13}\psi_{30}- \psi_{12}\psi_{31})\nonumber\\&&\times (\psi^*_{10}\psi^*_{31}-\psi^*_{30}\psi^*_{11}- \psi^*_{32}\psi^*_{13}+ 
    \psi^*_{12}\psi^*_{33}). 
\end{eqnarray}

Finally, a fourth real valued polynomial can be constructed as.
\begin{eqnarray}
R_6&=&\frac{1}{2}\Tr[\Psi C\gamma^5\Psi^T\gamma^0 \Psi^* C\gamma^5\Psi^\dagger\gamma^0]\nonumber\\
&=&\frac{1}{2}\Tr[\Psi C\gamma^5\Psi^T\gamma^0\gamma^5 \Psi^* C\gamma^5\Psi^\dagger\gamma^0\gamma^5]+|I_{2A}|^2-|I_{2}|^2.
\end{eqnarray}
This polynomial is related to $R_3$ by a permutation of Alice's and Bob's labs.
Written out it is
\begin{eqnarray}
R_6&=& |\psi_{02}\psi_{11}- \psi_{01}\psi_{12}+ \psi_{00}\psi_{13}- \psi_{10}\psi_{03}|^2 \nonumber\\&&-|\psi_{30}\psi_{03}- \psi_{33}\psi_{00}+ \psi_{01}\psi_{32}- \psi_{31}\psi_{02}|^2 \nonumber\\&&
-| \psi_{32}\psi_{11}- \psi_{33}\psi_{10}+ \psi_{13}\psi_{30}- 
    \psi_{12}\psi_{31}|^2\nonumber\\&& -| \psi_{03}\psi_{20}- 
    \psi_{23}\psi_{00}+ \psi_{22}\psi_{01}- \psi_{21}\psi_{02}|^2\nonumber\\&&  -| \psi_{13}\psi_{20}- \psi_{23}\psi_{10}- \psi_{12}\psi_{21} + \psi_{22}\psi_{11} |^2\nonumber\\&& +| \psi_{33}\psi_{20}- \psi_{23}\psi_{30}+ \psi_{22}\psi_{31}- 
    \psi_{32}\psi_{21}|^2. 
\end{eqnarray}

Next we construct three additional polynomials using only sesquilinear forms.
Each of these polynomials is constructed as the difference between two polynomials that reduce to the same polynomial on the set of product states.

One such real valued polynomial can be obtained as the difference
\begin{eqnarray}
T_1&=&\Tr[\Psi^T \gamma^0\Psi^*\gamma^0 \Psi^T \gamma^0\Psi^*\gamma^0]-N_1^2\nonumber\\
&=&-\Tr[\Psi^T \gamma^0\gamma^5\Psi^*\gamma^0 \Psi^T \gamma^0\gamma^5\Psi^*\gamma^0]+N_2^2+2R_3-2R_1
\nonumber\\
&=&-\Tr[\Psi^T \gamma^0\Psi^*\gamma^0\gamma^5 \Psi^T \gamma^0\Psi^*\gamma^0\gamma^5]+N_3^2+2R_6-2R_4\nonumber\\
&=&\Tr[\Psi^T \gamma^0\gamma^5\Psi^*\gamma^0\gamma^5 \Psi^T \gamma^0\gamma^5\Psi^*\gamma^0\gamma^5]-N_4^2+2R_3-2R_1\nonumber\\&&-2R_4+2R_6-|I_{2B}|^2+|I_{2}|^2+2|I_{1}|^2-2|I_{2A}|^2.
\end{eqnarray}
Written out the polynomial $T_1$ is 
\begin{eqnarray}
T_1&=& (|\psi_{00}|^2+ |\psi_{10}|^2- |\psi_{20}|^2- |\psi_{30}|^2)^2
\nonumber\\&&+(|\psi_{01}|^2+ |\psi_{11}|^2- |\psi_{21}|^2- |\psi_{31}|^2)^2
\nonumber\\&&+(|\psi_{02}|^2+ |\psi_{12}|^2- |\psi_{22}|^2- |\psi_{32}|^2)^2
\nonumber\\&&+(|\psi_{03}|^2+ |\psi_{13}|^2- |\psi_{23}|^2- |\psi_{33}|^2)^2
\nonumber\\&&+2|\psi_{01}\psi^*_{00}+\psi_{11}\psi^*_{10}- \psi_{21}\psi^*_{20}-  \psi_{31}\psi^*_{30}|^2 \nonumber\\&&-2|\psi_{02}\psi^*_{00}+\psi_{12}\psi^*_{10}- \psi_{22}\psi^*_{20}-  \psi_{32}\psi^*_{30}|^2
\nonumber\\&&-2|\psi_{02}\psi^*_{01}+\psi_{12}\psi^*_{11}- \psi_{22}\psi^*_{21}-  \psi_{32}\psi^*_{31}|^2
\nonumber\\&&-2|\psi_{03}\psi^*_{00}+\psi_{13}\psi^*_{10}- \psi_{23}\psi^*_{20}-  \psi_{33}\psi^*_{30}|^2
\nonumber\\&&-2|\psi_{03}\psi^*_{01}+\psi_{13}\psi^*_{11}- \psi_{23}\psi^*_{21}-  \psi_{33}\psi^*_{31}|^2
\nonumber\\&&+2|\psi_{03}\psi^*_{02}+\psi_{13}\psi^*_{12}- \psi_{23}\psi^*_{22}-  \psi_{33}\psi^*_{32}|^2
\nonumber\\&&-(|\psi_{00}|^2+ |\psi_{10}|^2- |\psi_{20}|^2- |\psi_{30}|^2\nonumber\\&&+|\psi_{01}|^2+ |\psi_{11}|^2- |\psi_{21}|^2- |\psi_{31}|^2
\nonumber\\&&-|\psi_{02}|^2- |\psi_{12}|^2+ |\psi_{22}|^2+ |\psi_{32}|^2    
\nonumber\\&&-|\psi_{03}|^2- |\psi_{13}|^2+ |\psi_{23}|^2+ |\psi_{33}|^2)^2.
\end{eqnarray}

A pure imaginary valued polynomial can be obtained as the difference
\begin{eqnarray}
T_2&=&\Tr[\Psi^T \gamma^0\gamma^5\Psi^*\gamma^0\gamma^5 \Psi^T \gamma^0\Psi^*\gamma^0]\nonumber\\&&-\Tr[\Psi^\dagger \gamma^0\gamma^5\Psi\gamma^0\gamma^5 \Psi^\dagger \gamma^0\Psi\gamma^0]\nonumber\\
&=&-\Tr[\Psi^T \gamma^0\gamma^5\Psi^*\gamma^0 \Psi^T \gamma^0\Psi^*\gamma^0\gamma^5]\nonumber\\&&+\Tr[\Psi^\dagger \gamma^0\gamma^5\Psi\gamma^0 \Psi^\dagger \gamma^0\Psi\gamma^0\gamma^5].
\end{eqnarray}
See Appendix \ref{cameron} for the written out form of $T_2$.

A sixth real valued polynomial can be obtained from products of the bidegree (1,1) polynomials $N_1$, $N_2$, $N_3$, and $N_4$ defined in Eq. (\ref{trew}), as
\begin{eqnarray}
N_1N_4-N_2N_3.
\end{eqnarray}
See Appendix \ref{cameron2} for the written out form of $N_1N_4-N_2N_3$.

The 16 polynomials $|I_1|^2,|I_2|^2,|I_{2A}|^2,|I_{2B}|^2,I_1I_2^*,I_1I_{2A}^*$, $I_1I_{2B}^*,I_2I_{1}^*,I_2I_{2A}^*,I_2I_{2B}^*,I_{2A}I_{1}^*,I_{2A}I_{2}^*,I_{2A}I_{2B}^*,I_{2B}I_{1}^*,I_{2B}I_{2}^*,I_{2B}I_{2A}^*$ together with the eight polynomials $R_1,R_2,R_2^*,R_3,R_4,R_5,R_5^*,R_6$  and the three polynomials $T_1,T_2$ and $N_1N_4-N_2N_3$ span a 27 dimensional polynomial space, i.e., the 27 polynomials are linearly independent.

\subsection{Polynomials of bidegree (3,1) and (1,3)}
Here we construct locally Lorentz invariant polynomials of bidegree $(3,1)$ that are indicators of spinor entanglement. Polynomial spinor entanglement indicators of bidegree $(1,3)$ can be obtained as the complex conjugates of these.
As described in section \ref{consrt} the bidegree $(3,1)$ polynomials are constructed such that for each observer one $C$ or $C\gamma^5$ is used. Thus all such polynomials are identically zero for the product states. In writing these polynomials we suppress the subscript $AB$ of $\Psi$ in the following.

We can construct polynomials of bidegree (3,1) as a products of one the four bidegree (2,0) polynomials  
\begin{eqnarray}
I_1&=&\frac{1}{2}\Tr[\Psi^TC\Psi C],\nonumber\\
I_2&=&\frac{1}{2}\Tr[\Psi^TC\gamma^5\Psi C\gamma^5],\nonumber\\
I_{2A}&=&\frac{1}{2}\Tr[\Psi^TC\Psi C\gamma^5],\nonumber\\
I_{2B}&=&\frac{1}{2}\Tr[\Psi^TC\gamma^5\Psi C],
\end{eqnarray}
multiplied by one of the four bidegree (1,1) polynomials  
\begin{eqnarray}
N_1&=&\Tr[\Psi^T\gamma^0\Psi^*\gamma^0],\nonumber\\
N_2&=&\Tr[\Psi^T\gamma^0\gamma^5\Psi^*\gamma^0],\nonumber\\
N_3&=&\Tr[\Psi^T\gamma^0\Psi^*\gamma^0\gamma^5],\nonumber\\
N_4&=&\Tr[\Psi^T\gamma^0\gamma^5\Psi^*\gamma^0\gamma^5].
\end{eqnarray}
The 16 possible such products all take the value zero for any product state. Beyond this we can construct four more polynomials of bidegree (3,1) that take the value zero for any product state.
All these constructions involve two instances of $C$ or $C\gamma^5$ and two instances of $\gamma^0$ or $\gamma^0\gamma^5$.

One complex valued polynomial of this kind is
\begin{eqnarray}
Q_1&=&\Tr[\Psi^TC\Psi C\Psi^T\gamma^0\Psi^*\gamma^0]-I_1N_1\nonumber\\
&=&-\Tr[\Psi^TC\gamma^5\Psi C\gamma^5\Psi^T\gamma^0\gamma^5\Psi^*\gamma^0\gamma^5]+I_2N_4
\nonumber\\
&=&-\Tr[\Psi^TC\Psi C\gamma^5\Psi^T\gamma^0\Psi^*\gamma^0\gamma^5]+I_{2A}N_3
\nonumber\\
&=&\Tr[\Psi^TC\gamma^5\Psi C\Psi^T\gamma^0\gamma^5\Psi^*\gamma^0]-I_{2B}N_2.
\end{eqnarray}
Written out it is given by
\begin{eqnarray}
Q_1&&=(\psi_{03}\psi_{00}^*+\psi_{01}\psi_{02}^*+\psi_{11}\psi_{12}^*+\psi_{13}\psi_{10}^*)(\psi_{22}\psi_{30}-\psi_{20}\psi_{32})\nonumber\\&&
+ (\psi_{02}\psi_{00}^*+\psi_{12}\psi_{10}^*- 
    \psi_{01}\psi_{03}^*-\psi_{11}\psi_{13}^*)(\psi_{20}\psi_{33}- \psi_{23}\psi_{30})\nonumber\\&&+ (\psi_{00}\psi_{00}^*+\psi_{10}\psi_{10}^*+\psi_{01}\psi_{01}^*+\psi_{11}\psi_{11}^*)(\psi_{23}\psi_{32}-
     \psi_{22}\psi_{33})
   \nonumber\\&& 
   + (\psi_{03}\psi_{01}^*-\psi_{00}\psi_{02}^*+\psi_{13}\psi_{11}^*-\psi_{10}\psi_{12}^*)(\psi_{22}\psi_{31}- \psi_{21}\psi_{32})\nonumber\\&&
   +(\psi_{02}\psi_{01}^*+\psi_{12}\psi_{11}^*+\psi_{00}\psi_{03}^*+\psi_{10}\psi_{13}^*)(\psi_{21}\psi_{33}-\psi_{23}\psi_{31} 
    ) \nonumber\\&& 
+(\psi_{02} \psi_{02}^*+\psi_{03} \psi_{03}^*+ \psi_{12}\psi_{12}^*+\psi_{13}\psi_{13}^*)( \psi_{20}\psi_{31}-\psi_{21}\psi_{30})\nonumber\\&&  
         + (\psi_{20}\psi_{20}^*+\psi_{21}\psi_{21}^*+\psi_{30}\psi_{30}^*+\psi_{31}\psi_{31}^*)(\psi_{02}\psi_{13}-\psi_{03}\psi_{12})\nonumber\\&&
         + (\psi_{22}\psi_{20}^*-\psi_{21}\psi_{23}^*+\psi_{32}\psi_{30}^*-\psi_{31}\psi_{33}^*)(\psi_{03}\psi_{10}- \psi_{00}\psi_{13})\nonumber\\&&+(\psi_{23}\psi_{20}^*+\psi_{33}\psi_{30}^*+\psi_{31}\psi_{32}^*+\psi_{21}\psi_{22}^*)( \psi_{00}\psi_{12}- 
    \psi_{02}\psi_{10}) 
   \nonumber\\&&  + (\psi_{22}\psi_{21}^*+\psi_{32}\psi_{31}^*+\psi_{20}\psi_{23}^*+\psi_{30}\psi_{33}^* ) ( \psi_{03}\psi_{11}- \psi_{01}\psi_{13})\nonumber\\&& + (\psi_{20}\psi_{22}^*-\psi_{33}\psi_{31}^*-\psi_{23}\psi_{21}^*+\psi_{30}\psi_{32}^*)(\psi_{02}\psi_{11}- \psi_{01}\psi_{12})\nonumber\\&& + 
    (\psi_{22}\psi_{22}^*+\psi_{23}\psi_{23}^*+\psi_{32}\psi_{32}^*+\psi_{33}\psi_{33}^*)(\psi_{01}\psi_{10}- \psi_{00}\psi_{11}). \nonumber\\  
    \end{eqnarray}
A second complex valued polynomial of this kind can be obtained as    
\begin{eqnarray}
Q_2&=&\Tr[\Psi^TC\gamma^5\Psi C\gamma^5\Psi^T\gamma^0\Psi^*\gamma^0]-I_2N_1\nonumber\\
&=&-\Tr[\Psi^TC\Psi C\Psi^T\gamma^0\gamma^5\Psi^*\gamma^0\gamma^5]+I_1N_4 \nonumber\\
&=&\Tr[\Psi^TC\Psi C\gamma^5\Psi^T\gamma^0\gamma^5\Psi^*\gamma^0]-I_{2A}N_2
\nonumber\\
&=&-\Tr[\Psi^TC\gamma^5\Psi C\Psi^T\gamma^0\Psi^*\gamma^0\gamma^5]+I_{2B}N_3.
\end{eqnarray}
Written out this polynomial is given by
\begin{eqnarray}
 Q_2&&=(\psi_{00}\psi^*_{00}- \psi_{03}\psi^*_{03}+ \psi_{33}\psi^*_{33}- \psi_{30}\psi^*_{30}) (\psi_{12}\psi_{21}- 
    \psi_{11}\psi_{22})\nonumber\\&& + (\psi_{03}\psi^*_{01}- \psi_{32}\psi^*_{30}- \psi_{33}\psi^*_{31}+ \psi_{02}\psi^*_{00}) (\psi_{11}\psi_{20}- 
    \psi_{10}\psi_{21})\nonumber\\&& + (\psi_{00}\psi^*_{01}- \psi_{32}\psi^*_{33}- \psi_{30}\psi^*_{31}+ \psi^*_{03}\psi_{02}) (\psi_{13}\psi_{21}- 
    \psi_{11}\psi_{23})\nonumber\\&& + (\psi_{03}\psi^*_{02}- \psi_{33}\psi^*_{32}- \psi_{31}\psi^*_{30}+ \psi^*_{00}\psi_{01}) ( \psi_{10}\psi_{22}- \psi_{12}\psi_{20} 
   )\nonumber\\&& + ( 
   \psi_{32}\psi^*_{32}- \psi_{02}\psi^*_{02}+ \psi^*_{01}\psi_{01}- \psi_{31}\psi^*_{31}) (\psi_{10}\psi_{23}-\psi_{13}\psi_{20} 
    )\nonumber\\&& + (\psi_{31}\psi^*_{33}- \psi_{00}\psi^*_{02}+ \psi_{30}\psi^*_{32}- \psi_{01}\psi^*_{03}) (\psi_{13}\psi_{22}- 
    \psi_{12}\psi_{23})\nonumber\\&& + (\psi_{10}\psi^*_{12}- \psi_{20}\psi^*_{22}- \psi_{21}\psi^*_{23}+ \psi_{11}\psi^*_{13}) ( 
   \psi_{03} \psi_{32}- \psi_{02}\psi_{33})\nonumber\\&& + (\psi_{11}\psi^*_{10}- \psi_{23}\psi^*_{22}+ \psi_{13}\psi^*_{12}- 
    \psi_{21} \psi^*_{20}) (\psi_{30}\psi_{02}-\psi_{32} \psi_{00} )\nonumber\\&& + (\psi_{12}\psi^*_{10}- \psi_{22} \psi^*_{20}- \psi_{23} \psi^*_{21}+
     \psi^*_{11}\psi_{13}) (\psi_{31} \psi_{00}- \psi_{01}\psi_{30})\nonumber\\&& + (\psi_{12}\psi^*_{13} - \psi_{22} \psi^*_{23}+ \psi_{10}\psi^*_{11}- 
    \psi_{20} \psi^*_{21}) (\psi_{33}\psi_{01}- \psi_{31} \psi_{03}) \nonumber\\&&+ (\psi_{11}\psi^*_{11}- \psi_{12}\psi^*_{12}- \psi_{21}\psi^*_{21}+ 
    \psi_{22}\psi^*_{22}) (\psi_{30}\psi_{03}- \psi_{00} \psi_{33})\nonumber\\&& + (\psi_{10}\psi^*_{10}- \psi_{13}\psi^*_{13}- \psi_{20}\psi^*_{20}+ 
    \psi_{23}\psi^*_{23}) (\psi_{01}\psi_{32}- \psi_{31}\psi_{02} ).\nonumber\\
\end{eqnarray}
A third complex valued polynomial can be constructed as 
\begin{eqnarray}
Q_3&=&\Tr[\Psi^TC\gamma^5\Psi C\Psi^T\gamma^0\Psi^*\gamma^0]-I_{2B}N_1\nonumber\\
&=&\Tr[\Psi^TC\Psi C\Psi^T\gamma^0\gamma^5\Psi^*\gamma^0]-I_{1}N_2\nonumber\\
&=&-\Tr[\Psi^T C\gamma^5\Psi C\gamma^5\Psi^T\gamma^0\Psi^*\gamma^0\gamma^5]+I_{2}N_3\nonumber\\
&=&-\Tr[\Psi^T C\Psi C\gamma^5\Psi^T\gamma^0\gamma^5\Psi^*\gamma^0\gamma^5]+I_{2A}N_4.
\end{eqnarray}
Written out it is given by
\begin{eqnarray}
Q_3&&= (\psi_{20}\psi^*_{00}+ \psi_{21}\psi^*_{01}+ \psi_{30}\psi^*_{10}+ 
    \psi_{31}\psi^*_{11})( \psi_{02}\psi_{13}-\psi_{03}\psi_{12})\nonumber\\&& +  (\psi_{23}\psi^*_{01}- \psi_{20}\psi^*_{02}+ \psi_{33}\psi^*_{11}- 
    \psi_{30}\psi^*_{12})(\psi_{01}\psi_{12}-\psi_{02}\psi_{11})\nonumber\\&& +  (\psi_{23}\psi^*_{00}+ \psi_{21}\psi^*_{02}+ \psi_{33}\psi^*_{10}+ 
    \psi_{31}\psi^*_{12})(\psi_{00}\psi_{12}-\psi_{02}\psi_{10}) \nonumber\\&&+  (\psi_{22}\psi^*_{01}+ \psi_{20}\psi^*_{03}+ \psi_{32}\psi^*_{11}+ 
    \psi_{30}\psi^*_{13}) (\psi_{03}\psi_{11}- \psi_{01}\psi_{13})\nonumber\\&&+  (\psi_{22}\psi^*_{00}- \psi_{21}\psi^*_{03}+ \psi_{32}\psi^*_{10}- 
    \psi_{31}\psi^*_{13})(\psi_{03}\psi_{10}- \psi_{00}\psi_{13})\nonumber\\&& +  (\psi_{22}\psi^*_{02}+ \psi_{23}\psi^*_{03}+ \psi_{32}\psi^*_{12}+ 
    \psi_{33}\psi^*_{13})(\psi_{01}\psi_{10}- \psi_{00}\psi_{11})\nonumber\\&& +  (\psi_{00}\psi^*_{20}+ \psi_{01}\psi^*_{21}+ \psi_{10}\psi^*_{30}+ 
    \psi_{11}\psi^*_{31})(\psi_{23}\psi_{32}- \psi_{22}\psi_{33}) \nonumber\\&&+  ( \psi_{00}\psi^*_{22}-\psi_{03}\psi^*_{21}- \psi_{13}\psi^*_{31}+ 
    \psi_{10}\psi^*_{32})( \psi_{21}\psi_{32}-\psi_{22}\psi_{31})\nonumber\\&& +  (\psi_{03}\psi^*_{20}+ \psi_{01}\psi^*_{22}+ \psi_{13}\psi^*_{30}+ 
    \psi_{11}\psi^*_{32})(\psi_{22}\psi_{30}- \psi_{20}\psi_{32})\nonumber\\&& +  (\psi_{02}\psi^*_{21}+ \psi_{00}\psi^*_{23}+ \psi_{12}\psi^*_{31}+ 
    \psi_{10}\psi^*_{33})( \psi_{21}\psi_{33}-\psi_{23}\psi_{31})\nonumber\\&& +  (\psi_{02}\psi^*_{20}- \psi_{01}\psi^*_{23}+ \psi_{12}\psi^*_{30}- 
    \psi_{11}\psi^*_{33})( \psi_{20}\psi_{33}-\psi_{23}\psi_{30})\nonumber\\&& + (\psi_{02}\psi^*_{22}+ \psi_{03}\psi^*_{23}+ \psi_{12}\psi^*_{32}+ 
    \psi_{13}\psi^*_{33}) ( \psi_{20}\psi_{31}-\psi_{21}\psi_{30}).\nonumber\\
\end{eqnarray}
Finally, a fourth complex valued polynomial of this kind is 
\begin{eqnarray}
Q_4&=&\Tr[\Psi^T C\Psi C\gamma^5\Psi^T\gamma^0\Psi^*\gamma^0]-I_{2A}N_1\nonumber\\
&=&-\Tr[\Psi^T C\Psi C\Psi^T\gamma^0\Psi^*\gamma^0\gamma^5]+I_{1}N_3\nonumber\\
&=&\Tr[\Psi^T C\gamma^5\Psi C\gamma^5\Psi^T\gamma^0\gamma^5\Psi^*\gamma^0]-I_{2}N_2\nonumber\\
&=&-\Tr[\Psi^T C\gamma^5\Psi C\Psi^T\gamma^0\gamma^5\Psi^*\gamma^0]+I_{2B}N_4.
\end{eqnarray}
The polynomial $Q_4$ is related to $Q_3$ by a permutation of Alice's and Bob's labs. Written out it is given by
\begin{eqnarray}
 Q_4&&=(\psi_{00}\psi^*_{00}- \psi_{03}\psi^*_{03}- \psi_{13} \psi^*_{13}+ \psi_{10} \psi^*_{10})( 
   \psi_{21}\psi_{32}- \psi_{22}\psi_{31} )\nonumber\\&& + (\psi_{03}\psi^*_{01}+ \psi_{12} \psi^*_{10}+ \psi_{13} \psi^*_{11}+ \psi_{02}\psi^*_{00})( 
   \psi_{20}\psi_{31}- \psi_{21}\psi_{30} )\nonumber\\&& + (\psi_{00}\psi^*_{01}+ \psi_{10} \psi^*_{11}+ \psi_{12} \psi^*_{13}+ \psi_{02}\psi^*_{03})( 
   \psi_{21}\psi_{33}- \psi_{23}\psi_{31})\nonumber\\&& + (\psi_{03}\psi^*_{02}+ \psi_{13} \psi^*_{12}+ \psi_{11} \psi^*_{10}+ \psi_{01}\psi^*_{00})( 
   \psi_{22}\psi_{30}- \psi_{20}\psi_{32})\nonumber\\&& + (\psi_{00} \psi^*_{02}+ \psi_{10} \psi^*_{12}+ \psi_{11} \psi^*_{13}+ 
    \psi_{01} \psi^*_{03})(\psi_{23}\psi_{32}- \psi_{22}\psi_{33})\nonumber\\&& + (\psi_{11} \psi^*_{11}- \psi_{12} \psi^*_{12}- \psi_{02}\psi^*_{02}+ 
    \psi_{01}\psi^*_{01})(\psi_{23}\psi_{30}- \psi_{20}\psi_{33})\nonumber\\&& + (\psi_{21}\psi^*_{20}+ \psi_{23}\psi^*_{22}+ \psi_{31}\psi^*_{30}+ 
    \psi_{33}\psi^*_{32})(\psi_{00}\psi_{12}- \psi_{02}\psi_{10})\nonumber\\&&+ (\psi_{22}\psi^*_{20}+ \psi_{23}\psi^*_{21}+ \psi_{32}\psi^*_{30}+ 
    \psi_{33}\psi^*_{31})(\psi_{10}\psi_{01}- \psi_{00}\psi_{11})\nonumber\\&& + (\psi_{20}\psi^*_{21}+ \psi_{22}\psi^*_{23}+ \psi_{30}\psi^*_{31}+ 
    \psi_{32}\psi^*_{33})(\psi_{03}\psi_{11}- \psi_{01}\psi_{13})\nonumber\\&&+ (\psi_{21}\psi^*_{21}- \psi_{22}\psi^*_{22}+ \psi_{31}\psi^*_{31}- 
    \psi_{32}\psi^*_{32})(\psi_{00}\psi_{13}- \psi_{03}\psi_{10})\nonumber\\&& + (\psi_{20}\psi^*_{22}+ \psi_{21}\psi^*_{23}+ \psi_{30}\psi^*_{32}+ 
    \psi_{31}\psi^*_{33})(\psi_{02}\psi_{13}- \psi_{03}\psi_{12})\nonumber\\&& + (\psi_{20} \psi^*_{20}- \psi_{23}\psi^*_{23}+ \psi_{30}\psi^*_{30}- 
    \psi_{33}\psi^*_{33})(\psi_{02}\psi_{11}- \psi_{01}\psi_{12}).\nonumber\\
\end{eqnarray}
The four polynomials $Q_1,Q_2,Q_3,Q_4$ and the 16 products $I_1N_1,I_1N_2,I_1N_3,I_1N_4,I_2N_1,I_2N_2,I_2N_3,I_2N_4,I_{2A}N_1,I_{2A}N_2,$ $I_{2A}N_3,I_{2A}N_4,I_{2B}N_1,I_{2B}N_2,I_{2B}N_3,$ and $I_{2B}N_4$ are linearly independent and together span a 20 dimensional polynomial space. Polynomials of bidegree (1,3) can be straightforwardly obtained as the complex conjugates of the above bidegree (3,1) polynomials.

\subsection{Positive and negative energy subspaces in the case of zero momenta and zero four-potentials}

We can consider the case of zero particle momentum and zero electromagnetic four-potential, often identified as the non-relativistic limit of a free particle. If Alice's and Bob's particles are both in this limit and also in either the local positive or negative energy subspace the shared state is invariant under some combination of projections $P_+^A$ or $P_-^A$ by Alice and $P_+^B$ or $P_-^B$ by Bob. In this case only a $2\times 2$ submatrix of $\Psi_{AB}$ is nonzero.
As a consequence the polynomial $I_{1}$ and its complex conjugate $I_{1}^*$ reduce, up to a sign and a relabelling of the indices, to the polynomial $\mathcal{C}$ defined by
\begin{eqnarray}\label{woot}
\mathcal{C}=\psi_{00} \psi_{11}-\psi_{01} \psi_{10},
\end{eqnarray}
and its complex conjugate $\mathcal{C}^*$, respectively. The polynomial $\mathcal{C}$ is the Wootters concurrence \cite{wootters,wootters2} used to describe the spin entanglement between two non-relativistic spin-$\frac{1}{2}$ particles.

All other polynomials of bidegree (2,0) and (0,2) described above are zero for this case. In particular any polynomial that involves one or more bilinear forms containing $C\gamma^5$ is zero in this case since $P_+C\gamma^5P_+=P_-C\gamma^5P_-=0$.

The  bidegree (2,2) polynomials $R_1$ and $R_4$ both reduce, up to a sign and a relabelling of the indices, to $|\mathcal{C}|^2$ and $T_1$ reduces to $2|\mathcal{C}|^2$ while the other bidegree (2,2) polynomials constructed above $R_2$, $R_3$, $R_5,R_6,T_2$ and $N_1N_4-N_2N_3$ are zero in this case since $P_+C\gamma^5P_+=P_-C\gamma^5P_-=0$ and $P_+\gamma^0\gamma^5P_+=P_-\gamma^0\gamma^5P_-=0$. The bidegree (3,1) polynomial $I_1N_1$ reduces, up to a sign and a relabelling of the indices, to $\mathcal{C}(|\psi_{00}|^2+|\psi_{11}|^2+|\psi_{01}|^2+|\psi_{10}|^2)$ while all the other bidegree (3,1) polynomials constructed above are zero since $P_+C\gamma^5P_+=P_-C\gamma^5P_-=0$ and $P_+\gamma^0\gamma^5P_+=P_-\gamma^0\gamma^5P_-=0$.

\subsection{Weyl particles}
We can consider the case where Alice's and Bob's particles are both Weyl particles, i.e., both have definite chirality. Then the shared state of the two particles is invariant under some combination of projections $P_L^A$ or $P_R^A$ by Alice and $P_L^B$ or $P_R^B$ by Bob. In this case the state coefficients of the shared state satisfy $\psi_{jk}=(-1)^{|LA|}\psi_{(j-2) k}=(-1)^{|LB|}\psi_{j (k-2)}$, where $|LA|=1$ if the state is invariant under $P_L^A$ and otherwise zero, $|LB|=1$ if the state is invariant under $P_L^B$ and otherwise zero, and $j$,$k$ are defined modulo 4.

The bidegree (2,0) polynomials $I_{1}$, $I_{2}$, $I_{2A}$ and $I_{2B}$, reduce, up to a sign, to $4\mathcal{C}$ where $\mathcal{C}$ is the Wootters concurrence in Eq. (\ref{woot}) whenever the state is invariant under some combination of these projectors as described in Ref. \cite{spinorent}.

The bidegree (2,2) polynomials $R_1,R_2,R_3,R_4,R_5,R_6,T_1,T_2$ and $N_1N_4-N_2N_3$ all reduce to zero when Alice's and Bob's particles are Weyl particles since $P_L\gamma^0P_L=P_R\gamma^0P_R=0$ and $P_L\gamma^0\gamma^5P_L=P_R\gamma^0\gamma^5P_R=0$. Likewise the bidegree (3,1) polynomials $Q_1,Q_2,Q_3$ and $Q_4$ all reduce to zero in this case for the same reason.

\subsection{Examples of spinor entangled two-particle states}\label{exxx}
Here we consider a few examples of spinor entangled two-particle states to illustrate how the different polynomials indicate different types of entanglement. In particular we consider both states indicated by the homogeneous polynomials and states not indicated by the homogeneous polynomials.

The generation of entanglement between the spinorial degrees of freedom of two Dirac particles has been studied in Reference \cite{pachos}. In particular the so called spinor "EPR-state"
\begin{eqnarray}
\frac{1}{\sqrt{2}}({\phi_0^A}\otimes{\phi_1^B}-i{\phi_1^A}\otimes{\phi_0^B}),
\end{eqnarray}
was considered.
This state or equivalent states have been considered also in References \cite{alsing,mano,moradi,geng}. As described in Ref. \cite{spinorent} we can see that for this state the bidegree (2,0) polynomial $I_1$ is non-zero and has the absolute value $1/2$ which is its maximal absolute value for normalized states. The other bidegree (2,0) polynomials $I_2,I_{2A}$, and $I_{2B}$ are identically zero. The bidegree (2,2) polynomials $R_1$ and $R_4$ both attain the absolute value $1/4$ and $T_1$ attains the absolute value $1/2$ while $R_2,R_{3},R_{5},R_{6},T_2$ and $N_1N_4-N_2N_3$ are all identically zero. The bidegree (3,1) polynomials $Q_1,Q_2,Q_{3},Q_{4}$ are all identically zero for this state.

In a similar way we can construct a state
\begin{eqnarray}
\frac{1}{\sqrt{2}}({\phi_1^A}\otimes{\phi_3^B}-{\phi_2^A}\otimes{\phi_0^B}),
\end{eqnarray} 
for which the bidegree (2,0) polynomial $I_2$ is non-zero and attains the absolute value $1/2$, its maximal absolute value for normalized states. The other bidegree (2,0) polynomials $I_1,I_{2A}$, and $I_{2B}$ are identically zero. The bidegree (2,2) polynomials $R_3$ and $R_6$ both attain the absolute value $1/4$ and $T_1$ attains the absolute value $1/2$ while $R_1,R_{2},R_{4},R_{5},T_2$ and $N_1N_4-N_2N_3$ are all identically zero. The bidegree (3,1) polynomials $Q_1,Q_2,Q_{3},Q_{4}$ are all identically zero for this state.

The state 
\begin{eqnarray}
\frac{1}{\sqrt{2}}({\phi_0^A}\otimes{\phi_0^B}-{\phi_1^A}\otimes{\phi_3^B}),
\end{eqnarray}
is such that the bidegree (2,0) polynomial $I_{2A}$ is non-zero and attains the absolute value $1/2$, its maximal absolute value for normalized states. The other bidegree (2,0) polynomials $I_1,I_{2}$, and $I_{2B}$ are identically zero. The bidegree (2,2) polynomials $R_1$ and $R_6$ both attain the absolute value $1/4$ and $T_1$ attains the absolute value $1/2$ while $R_{2},R_3,R_{4},R_{5},T_2$ and $N_1N_4-N_2N_3$ are all identically zero. The bidegree (3,1) polynomials $Q_1,Q_2,Q_{3},Q_{4}$ are all identically zero for this state.

The state 
\begin{eqnarray}
\frac{1}{\sqrt{2}}({\phi_1^A}\otimes{\phi_1^B}-{\phi_2^A}\otimes{\phi_0^B}),
\end{eqnarray} 
is such that the bidegree (2,0) polynomial $I_{2B}$ is non-zero and attains the absolute value $1/2$, its maximal absolute value for normalized states. 
The other bidegree (2,0) polynomials $I_1,I_{2}$, and $I_{2A}$ are identically zero. The bidegree (2,2) polynomials $R_3$ and $R_4$ both attain the absolute value $1/4$ and $T_1$ attains the absolute value $1/2$ while $R_{1},R_2,R_{5},R_{6},T_2$ and $N_1N_4-N_2N_3$ are all identically zero. The bidegree (3,1) polynomials $Q_1,Q_2,Q_{3},Q_{4}$ are all identically zero for this state.  
 
The bidegree (2,0) polynomials $I_1,I_{2},I_{2A}$, and $I_{2B}$ previously described in Ref. \cite{spinorent}
do not indicate all forms of spinor entanglement. We therefore consider states for which $I_1=I_{2}=I_{2A}=I_{2B}=0$ but the entanglement is indicated by the bidegree (2,2) polynomials $R_1,R_{3},R_{4},R_{6}$, and $T_{1}$.

One such state is
\begin{eqnarray}\label{xccx}
\frac{1}{\sqrt{2}}({\phi_0^A}\otimes{\phi_1^B}+{\phi_1^A}\otimes{\phi_3^B}).
\end{eqnarray}
For this state the bidegree (2,2) polynomial $R_1$ is nonzero with absolute value $1/4$ and
 $T_1$ is nonzero with absolute value $1/2$ while $R_2,R_{3},R_{4},R_{5},R_{6},T_2$ and $N_1N_4-N_2N_3$ are all identically zero. The bidegree (2,0) polynomials $I_1,I_2,I_{2A},I_{2B}$ and the bidegree (3,1) polynomials $Q_1,Q_2,Q_{3},Q_{4}$ are identically zero. Similarly we can construct a state

\begin{eqnarray}\label{xccx2}
\frac{1}{\sqrt{2}}({\phi_0^A}\otimes{\phi_2^B}+{\phi_3^A}\otimes{\phi_0^B}).
\end{eqnarray}
For this state the bidegree (2,2) polynomial $R_3$ is nonzero with absolute value $1/4$ and
 $T_1$ is nonzero with absolute value $1/2$ while $R_1,R_{2},R_{4},R_{5},R_{6},T_2$ and $N_1N_4-N_2N_3$ are all identically zero. The bidegree (2,0) polynomials $I_1,I_2,I_{2A},I_{2B}$ and the bidegree (3,1) polynomials $Q_1,Q_2,Q_{3},Q_{4}$ are identically zero.

The state
\begin{eqnarray}\label{xccx3}
\frac{1}{\sqrt{2}}({\phi_0^A}\otimes{\phi_0^B}+{\phi_2^A}\otimes{\phi_1^B}),
\end{eqnarray}
is such that the bidegree (2,2) polynomial $R_4$ is nonzero with absolute value $1/4$ and
 $T_1$ is nonzero with absolute value $1/2$ while $R_1,R_{2},R_{3},R_{5},R_{6},T_2$ and $N_1N_4-N_2N_3$ are all identically zero. The bidegree (2,0) polynomials $I_1,I_2,I_{2A},I_{2B}$ and the bidegree (3,1) polynomials $Q_1,Q_2,Q_{3},Q_{4}$ are identically zero.

Finally, we can consider the state
\begin{eqnarray}\label{xccx4}
\frac{1}{\sqrt{2}}({\phi_0^A}\otimes{\phi_3^B}+{\phi_2^A}\otimes{\phi_0^B}).
\end{eqnarray}
For this state the bidegree (2,2) polynomial $R_6$ is nonzero with absolute value $1/4$ and
 $T_1$ is nonzero with absolute value $1/2$ while $R_1,R_{2},R_{3},R_{4},R_{5},T_2$ and $N_1N_4-N_2N_3$ are all identically zero. The bidegree (2,0) polynomials $I_1,I_2,I_{2A},I_{2B}$ and the bidegree (3,1) polynomials $Q_1,Q_2,Q_{3},Q_{4}$ are identically zero.

Next we consider states for which $I_1=I_{2}=I_{2A}=I_{2B}=0$ but the entanglement is indicated by the bidegree (2,2) polynomials $T_1,T_2$ and $N_1N_4-N_2N_3$. 
One such state is
\begin{eqnarray}\label{xccx5}
\frac{1}{\sqrt{2}}({\phi_0^A}\otimes{\phi_2^B}+{\phi_2^A}\otimes{\phi_0^B}).
\end{eqnarray}
For this state the bidegree (2,2) polynomial $T_1$ is nonzero with absolute value $1/2$ 
and $N_1N_4-N_2N_3$ is nonzero with absolute value $1$.
while $R_1,R_{2},R_{3},R_{4},R_{5},R_{6},T_2$ are all identically zero. The bidegree (2,0) polynomials $I_1,I_2,I_{2A},I_{2B}$ and the bidegree (3,1) polynomials $Q_1,Q_2,Q_{3},Q_{4}$ are identically zero.

Another such state is 
\begin{eqnarray}\label{xccx6}
\frac{1}{\sqrt{3}}({\phi_0^A}\otimes{\phi_2^B}+(1+i){\phi_2^A}\otimes{\phi_0^B}).
\end{eqnarray}
For this state the bidegree (2,2) polynomial $T_1$ is nonzero with absolute value $4/3$, the bidegree (2,2) polynomial $T_2$ is nonzero with absolute value $4/3$ and $N_1N_4-N_2N_3$ is nonzero with absolute value $6/9$,
while $R_1,R_{2},R_{3},R_{4},R_{5},R_{6}$ are all identically zero. The bidegree (2,0) polynomials $I_1,I_2,I_{2A},I_{2B}$ and the bidegree (3,1) polynomials $Q_1,Q_2,Q_{3},Q_{4}$ are identically zero.

The bidegree (2,2) polynomials $R_2$ and $R_5$ and the bidegree (3,1) polynomials $Q_1,Q_2,Q_{3},Q_{4}$ can only indicate entangled states with three or more nonzero state coefficients. We therefore consider a few such examples.

The state
\begin{eqnarray}\label{utoy}
\frac{1}{\sqrt{3}}({\phi_0^A}\otimes{\phi_2^B}+{\phi_1^A}\otimes{\phi_0^B}+{\phi_2^A}\otimes{\phi_2^B}),
\end{eqnarray}
is such that the bidegree (2,2) polynomials $R_1$ $R_2$ and $R_3$ are nonzero with absolute value $1/9$ while $R_4,R_{5},R_{6},T_1,T_2$ and $N_1N_4-N_2N_3$ are all identically zero. The bidegree (2,0) polynomials $I_1,I_2,I_{2A},I_{2B}$ and the bidegree (3,1) polynomials $Q_1,Q_2,Q_{3},Q_{4}$ are identically zero.

Similarly, the state
\begin{eqnarray}\label{utoya}
\frac{1}{\sqrt{3}}({\phi_2^A}\otimes{\phi_0^B}+{\phi_0^A}\otimes{\phi_1^B}+{\phi_2^A}\otimes{\phi_2^B}),
\end{eqnarray}
is such that the bidegree (2,2) polynomials $R_4$ $R_5$ and $R_6$ are nonzero with absolute value $1/9$ while $R_1,R_{2},R_{3},T_1,T_2$ and $N_1N_4-N_2N_3$ are all identically zero. The bidegree (2,0) polynomials $I_1,I_2,I_{2A},I_{2B}$ and the bidegree (3,1) polynomials $Q_1,Q_2,Q_{3},Q_{4}$ are identically zero.

The state
\begin{eqnarray}\label{toi}
\frac{1}{\sqrt{5}}({\phi_0^A}\otimes{\phi_2^B}+{\phi_1^A}\otimes{\phi_1^B}-{\phi_2^A}\otimes{\phi_2^B}+{\phi_3^A}\otimes{\phi_0^B}+{\phi_3^A}\otimes{\phi_1^B}),\nonumber\\
\end{eqnarray}
is such that the bidegree (2,2) polynomials $R_1$, $R_2$,  $R_3$ and $R_4$ are nonzero with absolute value $1/25$ and $T_1$ is nonzero with absolute value $2/25$ while $R_{5},R_{6},T_2$ and $N_1N_4-N_2N_3$ are all identically zero. The bidegree (3,1) polynomials $Q_2$ and $Q_{4}$ are nonzero with absolute value $1/25$ while $Q_1,Q_{3}$ are identically zero.
All the bidegree (2,0) polynomials $I_1,I_2,I_{2A},I_{2B}$ are identically zero.
 
The state
\begin{eqnarray}
\frac{1}{\sqrt{5}}({\phi_2^A}\otimes{\phi_0^B}+{\phi_1^A}\otimes{\phi_1^B}-{\phi_2^A}\otimes{\phi_2^B}+{\phi_0^A}\otimes{\phi_3^B}+{\phi_1^A}\otimes{\phi_3^B}),\nonumber\\
\end{eqnarray}
is such that the bidegree (2,2) polynomials $R_1$, $R_4$,  $R_5$ and $R_6$ are nonzero with absolute value $1/25$ and $T_1$ is nonzero with absolute value $2/25$ while $R_{2},R_{3},T_2$ and $N_1N_4-N_2N_3$ are all identically zero. The bidegree (3,1) polynomials $Q_2$ and $Q_{3}$ are nonzero with absolute value $1/25$ while $Q_1,Q_{4}$ are identically zero.
All the bidegree (2,0) polynomials $I_1,I_2,I_{2A},I_{2B}$ are identically zero.

\section{Constructing polynomial spinor entanglement indicators for multiple Dirac particles}\label{ent}

Here we generalize the method given in Section \ref{consrt} for constructing locally Lorentz invariant spinor entanglement indicators to the case of multiple spacelike separated Dirac particles. Just like for the case of two particles the method for multiple particles is an extension of the method described in References \cite{spinorent,multispinor}.

To construct locally Lorentz invariant polynomials we consider a number of spacelike separated observers each with their own laboratory holding a Dirac particle. For convenience we give the first three observers the names Alice, Bob, and Charlie, respectively. We then let the particles be in a joint state and assume that any operations on one observers particle can be made jointly with any operations on the other observers particles, i.e., we assume that operations made by the different observers commute. Further we make the assumption that a tensor product structure can be used to describe the shared system and that the tensor products of local basis spinors $\phi_{j_A}\otimes \phi_{k_B}\otimes\phi_{l_C}\otimes\dots$ can be used as a basis. 
Then the state can be expanded in this basis as
\begin{eqnarray}
&&\psi_{ABC\dots}(x_A,x_B,x_C,\dots)\nonumber\\
&=&\sum_{j_A,k_B,l_C,\dots }\psi_{j_A,k_B,l_C,\dots}(x_A,x_B,x_C,\dots)\phi_{j_A}\otimes \phi_{k_B}\otimes \phi_{l_C}\otimes\dots,\nonumber\\
\end{eqnarray}
where the $\psi_{j_A,k_B,l_C,\dots}(x_A,x_B,x_C,\dots)$ are complex valued functions of the tuple of points $x_A,x_B,x_C,\dots$. 

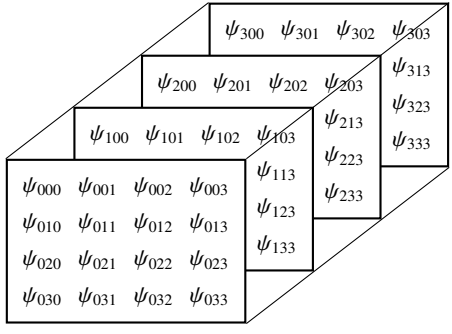
\begin{figure}
\centering
\begin{tikzpicture}[every node/.style={anchor=north east,thick,fill=white,minimum width=4mm,minimum height=4mm}]
\matrix (mA) [draw,matrix of math nodes]
{
\psi_{300} & \psi_{301} & \psi_{302} & \psi_{303} \\
\psi_{310} & \psi_{311} & \psi_{312} & \psi_{313} \\
\psi_{320} & \psi_{321} & \psi_{322} & \psi_{323} \\
\psi_{330} & \psi_{331} & \psi_{332} & \psi_{333} \\
};

\matrix (mB) [draw,matrix of math nodes] at ($(mA.south west)+(2.3,1.5)$)
{
\psi_{200} & \psi_{201} & \psi_{202} & \psi_{203} \\
\psi_{210} & \psi_{211} & \psi_{212} & \psi_{213} \\
\psi_{220} & \psi_{221} & \psi_{222} & \psi_{223} \\
\psi_{230} & \psi_{231} & \psi_{232} & \psi_{233} \\
};

\matrix (mC) [draw,matrix of math nodes] at ($(mB.south west)+(2.3,1.5)$)
{
\psi_{100} & \psi_{101} & \psi_{102} & \psi_{103} \\
\psi_{110} & \psi_{111} & \psi_{112} & \psi_{113} \\
\psi_{120} & \psi_{121} & \psi_{122} & \psi_{123} \\
\psi_{130} & \psi_{131} & \psi_{132} & \psi_{133} \\
};

\matrix (mD) [draw,matrix of math nodes] at ($(mC.south west)+(2.3,1.5)$)
{
\psi_{000} & \psi_{001} & \psi_{002} & \psi_{003} \\
\psi_{010} & \psi_{011} & \psi_{012} & \psi_{013} \\
\psi_{020} & \psi_{021} & \psi_{022} & \psi_{023} \\
\psi_{030} & \psi_{031} & \psi_{032} & \psi_{033} \\
};
\draw (mA.north east)--(mD.north east);
\draw (mA.north west)--(mD.north west);
\draw (mA.south east)--(mD.south east);
\end{tikzpicture}
\caption{Visual representation of the $4\times 4\times 4$ tensor $\Psi^{ABC}$ describing the state of three Dirac particles. \label{tretens}}
\end{figure}

We suppress the dependence on $x_A,x_B,x_C,\dots$ in the description of the state and let $\psi_{jkl\dots}\equiv \psi_{j_A,k_B,l_C,\dots}(x_A,x_B,x_C,\dots)$.
Then we arrange these coefficients $\psi_{jkl\dots}$ as a tensor, i.e., a multi-dimensional array, by letting the spinor basis indices $jkl\dots$ be the tensor component indices. For the case of a two particles the state coefficients $\psi_{jk}$ form a two-dimensional tensor, i.e., a matrix, where $j$ and $k$ are the column and row indices respectively as described in Section \ref{consrt} (See also Ref. \cite{spinorent}). For three particles we can arrange the state coefficients $\psi_{jkl}$ as a three-dimensional tensor where where $j$, $k$ and $l$ are likewise the indices of the three different dimensions, respectively. See Fig. \ref{tretens} for a visual representation. In the same way an $n$-particle state corresponds to an $n$-dimensional tensor. As in Ref. \cite{multispinor} we denote this tensor by $\Psi^{ABC\dots}$, and its components by $\Psi^{ABC\dots}_{jkl\dots}\equiv\psi_{jkl\dots}$.

Transformations $S^A$ on Alice's particle act on the first index of  $\Psi^{ABC\dots}$.  Transformations $S^B$ on Bob's particle act on the second index, and transformations $S^C$ on Charlie's particle act on the third index. Following this pattern transformations on the $n$th observer's particle act on the $n$th index

\begin{eqnarray}
\Psi^{ABC\dots}_{jkl\dots}\to\sum_{mno\dots} S^A_{jm}S^{B}_{kn}S^{C}_{lo}\dots\Psi^{ABC\dots}_{mno\dots} .
\end{eqnarray}

If we take two copies of $\Psi^{ABC\dots}$ together with the matrix $C$ we can construct products of tensor components as $\Psi^{ABC\dots}_{jkl\dots}C_{jm}\Psi^{ABC\dots}_{mno\dots}$ and then sum over the indices $j$ and $m$. This sum is now an invariant under the spinor representation of the proper orthochronous Lorentz group acting on Alice's particle. This follows directly from the properties of the bilinear form $\psi^T C \varphi$ described in section \ref{invariants} since $\sum_{jm}\Psi^{ABC\dots}_{jkl\dots}C_{jm}\Psi^{ABC\dots}_{mno\dots}$ is such a bilinear form for every fixed set of indices $kl\dots$ and $no\dots$.
In particular, for any spinor representation of a proper orthochronous Lorentz transformation $S(\Lambda)$ in Alice's lab we have

\begin{eqnarray}
\sum_{jm}\Psi^{ABC\dots}_{jkl\dots}C_{jm}\Psi^{ABC\dots}_{mno\dots}\to &&\sum_{jmpq}\Psi^{ABC\dots}_{jkl\dots}S^T(\Lambda)_{jp}C_{pq}S(\Lambda)_{qm}\Psi^{ABC\dots}_{mno\dots}\nonumber\\
=&&\sum_{jm}\Psi^{ABC\dots}_{jkl\dots}C_{jm}\Psi^{ABC\dots}_{mno\dots}.
\end{eqnarray}
Following Ref. \cite{multispinor} we call the construction $\sum_{jm}\Psi^{ABC\dots}_{jkl\dots}C_{jm}\Psi^{ABC\dots}_{mno\dots}$ a {\it tensor sandwich contraction} where $C$ is being sandwiched between the two copies of $\Psi^{ABC\dots}$.
In the same way we can make such sandwich contractions over a pair of indices corresponding to any of the other observers particle with the matrix $C$ being sandwiched.
Furthermore, we can make sandwich contractions also with $C\gamma^5$ sandwiched between two copies of $\Psi^{ABC\dots}$.

If we consider one copy of $\Psi^{ABC\dots}$ and one copy of $\Psi^{* ABC\dots}$ together with the matrix $\gamma^0$ we can construct products of tensor components as $\Psi^{* ABC\dots}_{jkl\dots}\gamma^0_{jm}\Psi^{ABC\dots}_{mno\dots}$ and then take the sum over $j$ and $m$. This sum is invariant under the spinor representation of the proper orthochronous Lorentz group acting on Alice's particle. This follows from the properties of the sesquilinear form $\psi^\dagger \gamma^0 \varphi$ described in section \ref{invariants} since $\sum_{jm}\Psi^{* ABC\dots}_{jkl\dots}\gamma^0_{jm}\Psi^{ ABC\dots}_{mno\dots}$ is such a sesquilinear form for every fixed set of indices $kl\dots$ and $no\dots$.
In particular, for a spinor representation of a proper orthochronous Lorentz transformation $S(\Lambda)$ in Alice's lab we have

\begin{eqnarray}
\sum_{jm}\Psi^{* ABC\dots}_{jkl\dots}\gamma^0_{jm}\Psi^{ ABC\dots}_{mno\dots}\to &&\sum_{jmpq}\Psi^{* ABC\dots}_{jkl\dots}S^\dagger(\Lambda)_{jp}\gamma^0_{pq}S(\Lambda)_{qm}\Psi^{ ABC\dots}_{mno\dots}\nonumber\\
=&&\sum_{jm}\Psi^{* ABC\dots}_{jkl\dots}\gamma^0_{jm}\Psi^{ABC\dots}_{mno\dots}.
\end{eqnarray}
Similarly we can make this kind of sandwich contraction over a pair of indices corresponding to any of the other observers particle with the matrix $\gamma^0$ being sandwiched.
Furthermore, we can make sandwich contractions also with $\gamma^0\gamma^5$ sandwiched between the copy of $\Psi^{ABC\dots}$ and the copy of $\Psi^{* ABC\dots}$.

Now consider a number $n$ of copies of $\Psi^{ABC\dots}$ and a number $m$ of copies of $\Psi^{* ABC\dots}$ such that $n+m$ is even. Next, for a given observer split the set of indices corresponding to that observer's particle into $(n+m)/2$ pairs. Then consider the pairs of indices that belong to two copies of $\Psi^{ABC\dots}$ or alternatively to two copies of $\Psi^{* ABC\dots}$. For these pairs sandwich contract every pair of indices with either $C$ or $C\gamma^5$ sandwiched in between the two copies of $\Psi^{ABC\dots}$ or the two copies of $\Psi^{* ABC\dots}$, as described above. Then consider the pairs of indices that belong to one copy of $\Psi^{ABC\dots}$ and one copy of $\Psi^{* ABC\dots}$. For these pairs sandwich contract every pair of indices with either $\gamma^0$ or $\gamma^0\gamma^5$ sandwiched in between the copies of $\Psi^{ABC\dots}$ and $\Psi^{* ABC\dots}$, as described above.
Then repeat this procedure for every other observer. The result is a function that is invariant under the local spinor representation of the proper orthochronous Lorentz group in every lab. In particular it is a polynomial of bidegree $(n,m)$ in the state coefficients $\psi_{jkl\dots}$ and the complex conjugated state coefficients $\psi^*_{jkl\dots}$. Note that this procedure is similar to Cayley's $\Omega$ process for constructing polynomial invariants under the special linear group \cite{cayley,omega} and the algorithm given in Ref. \cite{toumazet} to construct invariants under unitary and special unitary groups. If no copies of $\Psi^{* ABC\dots}$ are used in the procedure the result is a homogeneous polynomial in the state coefficients $\psi_{jkl\dots}$ and the procedure coincides with that described in Ref. \cite{multispinor}.

The computational difficulty in constructing locally Lorentz invariant polynomials through tensor sandwich contractions rises sharply with the number of particles and the number of copies of $\Psi^{ABC\dots}$ and $\Psi^{* ABC\dots}$ involved in the contractions, i.e., the bidegree of the polynomials. 
However, for low polynomial bidegrees such invariants can be obtained with relatively modest effort, in particular for three particles.

Testing the linear independence of the constructed polynomials can also be a challenging problem.
However, one way to make this task less difficult is to subdivide the polynomials based on their behaviour under parity inversion P in the different labs.
To see why we can consider a given polynomial constructed through sandwich contractions. For the indices corresponding to a given observer's particle let $r$ be the sum of the number of the matrices $C\gamma^5$ sandwiched and the number of the matrices $\gamma^0\gamma^5$ sandwiched. If $r$ is an even number, the polynomial is invariant under the parity inversion P in the lab of the given observer. If $r$ is odd the polynomial changes sign under P. Thus we can see that a polynomial cannot be linearly dependent on any polynomial with a different behaviour under the parity inversion P in any of the labs.

\subsection{Indicators of spinor entanglement involving all the particles}
\label{entaq}

Consider a polynomial constructed through tensor sandwich contractions.
If for each observer at least one pair of indices have been contracted with either $C$ or $C\gamma^5$ the polynomial is an indicator of spinor entanglement if it is nonzero. 
Due to the antisymmetry of $C$ and $C\gamma^5$ it follows that such a polynomial is identically zero in a tuple of points $x_A,x_B,x_C,\dots$ if $\Psi^{ABC\dots}(x_A,x_B,x_C,\dots)$ can be factored as $\Psi^{A}(x_A)\otimes\Psi^{BC\dots}(x_B,x_C,\dots)$, i.e., if the state is a product state over the partition $A|BC\dots$. For the same reason such a polynomial is identically zero if any other observer's particle is in a product state with the rest of the particles.
However, a polynomial with this property is not necessarily zero if the state is a product state over a partition of the particles into sets where each set has two or more particles. Instead for this case the polynomial factorizes into a product of polynomials defined on the different sets of particles in the partitioning.
Thus all nonzero polynomials of this kind are indicators of some type of spinor entanglement that involves all the particles. Either multipartite spinor entanglement where every particle is entangled with all the other particles or at least entanglement where every particle is spinor entangled with a subset of the other particles.

If the bidegree $(n,m)$ of a polynomial constructed as a tensor sandwich contraction is such that $m\neq n$ it follows that for each observer at least one pair of indices is sandwich contracted with either $C$ or $C\gamma^5$ sandwiched inbetween two copies of $\Psi^{ABC\dots}$ or two copies of $\Psi^{* ABC\dots}$.
All nonzero polynomials of this kind are thus indicators of either multipartite spinor entanglement involving all the particles or entanglement where every particle is spinor entangled with a subset of the other particles.

\subsection{Indicators of spinor entanglement not involving all the particles}
\label{utb}
Consider a mixed polynomial constructed through tensor sandwich contractions.
If for Alice no pair of indices have been contracted with $C$ or $C\gamma^5$ the polynomial is 
not necessarily zero in a tuple of points $x_A,x_B,x_C,\dots$ when $\Psi^{ABC\dots}(x_A,x_B,x_C,\dots)$ can be factored as $\Psi^{A}(x_A)\otimes\Psi^{BC\dots}(x_B,x_C,\dots)$, i.e., if the state is a product state over the partition $A|BC\dots$. In this case the polynomial factorizes into a polynomial that is a tensor sandwich contraction of the tensor $\Psi^{BC\dots}(x_B,x_C,\dots)$ and some product of the sesquilinear forms $\psi(x_A)^\dagger\gamma^0\psi(x_A)$ and $\psi(x_A)^\dagger\gamma^0\gamma^5\psi(x_A)$. Therefore such a polynomial can be nonzero even if Alice's particle is not entangled with any other particle. Thus it is not an indicator of spinor entanglement that involves Alice's particle.
The analogous property holds for any other observer if for this observer no pair of indices has been contracted with $C$ or $C\gamma^5$.

A polynomial of this kind is only an indicator of entanglement of the particles for which at least one pair of the respective observers indices have been contracted with either $C$ or $C\gamma^5$.
Therefore in this way we can construct polynomials that are indicators of spinor entanglement for any given subset of the particles by choosing which observers have at least one pair of their indices contracted with either $C$ or $C\gamma^5$.

If for at least one of the observers at least one pair of indices have been contracted with either $C$ or $C\gamma^5$ a polynomial of this kind is still zero if the state is a product state over every bipartition. 
All polynomials of this kind where for at least one of the observers at least one pair of indices have been contracted with either $C$ or $C\gamma^5$ are thus indicators of spinor entanglement involving some subset of the particles.

Ultimately, we may consider also tensor sandwich contractions where all the sandwiched matrices are either $\gamma^0$ or $\gamma^0\gamma^5$. Such polynomials are not zero for all the states where no particles are spinor entangled. However analogously to the construction is Section \ref{consrt} for the bipartite case we may be able to construct differences between such polynomials that are identically zero when none of the particles are spinor entangled.

Consider a polynomial that is a tensor sandwich contraction where all the sandwiched matrices are either $\gamma^0$ or $\gamma^0\gamma^5$, or a product of such polynomials.
For any product state $\psi(x_A)\otimes\varphi(x_B)\otimes\zeta(x_C)\otimes\dots$ 
a polynomial of this kind reduces, up to a sign, to a product of the Hermitian sesquilinear forms
$\psi(x_A)^\dagger\gamma^0\psi(x_A),\varphi(x_B)^\dagger\gamma^0\varphi(x_B),\zeta(x_C)^\dagger\gamma^0\zeta(x_C),\dots$, and the skew-Hermitian sesquilinear forms $\psi(x_A)^\dagger\gamma^0\gamma^5\psi(x_A),\varphi(x_B)^\dagger\gamma^0\gamma^5\varphi(x_B),\zeta(x_C)^\dagger\gamma^0\gamma^5\zeta(x_C),\dots$. Therefore if we have two linearly independent polynomials that reduce to the same product of sesquilinear forms on the product states we can create a polynomial that is identically zero for all product states by taking their difference. It follows that if we have a set of $n$ linearly independent polynomials that reduce to the same product of sesquilinear forms on the product states we can construct $n-1$ linearly independent polynomials that are identically zero for all product states. Such linear combinations are locally Lorentz invariant indicators of spinor entanglement.

\subsection{Dynamical evolution of the polynomials}

As described above the polynomials constructed through tensor sandwich contractions are made from bilinear and sesquilinear forms involving Dirac spinors. Therefore they satisfy dynamical equations that can be derived from the dynamical equations of the individual bilinear and sesquilinear forms given in Section \ref{dynnn}. In particular we note that if for Alice all pairs of indices have been contracted with either $C$ or $C\gamma^5$ it follows from Eq. (\ref{hugc2}) and Eq. (\ref{hugd2}) that under certain conditions the polynomial is invariant up to a U(1) phase under evolution with respect to Alice's time. These conditions, described in Section \ref{dynnn}, are that Alice's particle is in a momentum eigenstate, that the evolution preserves this momentum eigenspace and, depending on the combination of the matrices $C$ and $C\gamma^5$ used, that either the mass $m$ of Alice's particle, the pseudoscalar coupling $g$, or both are zero. 
However, as described in Section \ref{dynnn} the sesquilinear forms do not have the same kind of dynamical evolution as the bilinear forms. Therefore if for Alice one or more pairs of indices have been contracted with either $\gamma^0$ or $\gamma^0\gamma^5$ the polynomial is in general not invariant up to a U(1) phase under evolution with respect to Alice's time under these conditions.
The analogous situation holds for any other observer. 
We note that if for some homogeneous polynomial constructed through tensor sandwich contractions the conditions described above hold for all the observers, the homogeneous polynomial is invariant up to a U(1) phase under evolution with respect to any observers time as described in Refs. \cite{spinorent,multispinor}.
However, the mixed polynomials do in general not have this property.

\section{Spinor entanglement indicators for three Dirac particles}\label{three}

Here we consider the case of three spacelike separated Dirac particles and construct locally Lorentz invariant polynomials that are indicators of spinor entanglement. In particular we describe how to construct such polynomials of bidegree (2,2), (3,1) and (3,3) and give some examples.

For three Dirac particles we arrange the state coefficients as a $4\times 4\times 4$ tensor $\Psi^{ABC}$.
Local transformations $S^A,S^B,S^C$ that act on Alice's, Bob's and Charlie's particle, respectively, are described as
\begin{eqnarray}
\Psi^{ABC}_{ijk}\to\sum_{lmn} S^A_{il}S^B_{jm}S^C_{kn}\Psi^{ABC}_{lmn}.
\end{eqnarray}

Local Lorentz invariants of bidegree (2,0) can be constructed as tensor sandwich contractions of the form $\sum_{ijklmn}X_{li}X_{mj}X_{nk}\Psi^{ABC}_{ijk}\Psi^{ABC}_{lmn}$ involving two copies of $\Psi^{ABC}$ where either $X=C$ or $X=C\gamma^5$ for each instance. However, as described in Ref. \cite{multispinor} the 8 bidegree (2,0) polynomials that can be constructed in this way are all identically zero due to the antisymmetry of $C$ and $C\gamma^5$. It follows that all local Lorentz invariants of bidegree (0,2) that can be constructed as tensor sandwich contractions are also identically zero. 

Local Lorentz invariants of bidegree (1,1) can be constructed as tensor sandwich contractions of the form $\sum_{ijklmn}X_{li}X_{mj}X_{nk}\Psi^{* ABC}_{ijk}\Psi^{ABC}_{lmn}$ involving one copy of $\Psi^{ABC}$ and  one copy of $\Psi^{* ABC}$ with either $X=\gamma^0$ or $X=\gamma^0\gamma^5$ for each instance. However, the Lorentz invariants constructed in this way do not take the value zero for all product states and are thus not indicators of spinor entanglement. 

Local Lorentz invariants of bidegree (4,0) that are indicators of threepartite spinor entanglement have been constructed in
Ref. \cite{multispinor}. Their complex conjugates are likewise indicators of threepartite spinor entanglement of bidegree (0,4). 

Locally Lorentz invariant polynomials of bidegree (2,2) that are indicators of spinor entanglement can be constructed.
However, no locally Lorentz invariant indicators of spinor entanglement of bidegree (2,2) can be constructed as tensor sandwich contractions in such a way that for each observer at least one pair of indices have been contracted with either $C$ or $C\gamma^5$ sandwiched. 
This follows since in this case for each observer there are two pairs of indices to be contracted and these tensor sandwich contractions are either both bilinear forms with some combination of $C$ and $C\gamma^5$ being sandwiched or both sesquilinear forms with some combination of $\gamma^0$ and $\gamma^0\gamma^5$ being sandwiched.  If for all observers both pairs of indices are contracted as bilinear forms the resulting Lorentz invariant factorizes into a product of a polynomial of bidegree (2,0) and a polynomial of bidegree (0,2) but all such polynomials constructed as tensor sandwich contractions are identically zero as described in Ref. \cite{multispinor}. 
If for any observer both pairs of indices are contracted as sesquilinear forms the resulting Lorentz invariant is nonzero for some state that is a product state over the partition between this observers particle and the rest of the particles. Such contractions of bidegree (2,2) may therefore be used to indicate spinor entanglement that involve only two of the particles as described in Section \ref{utb}.

The polynomials of bidegree (3,1) and their complex conjugates of bidegree (1,3) are identically zero for any state that is a product state over a bipartition between one observer and the other observers as described in Section \ref{entaq}. Thus these polynomials only indicate spinor entanglement that involves all the particles.

For bidegree (3,3) more flexibility exists in choosing the tensor contractions and one can construct both polynomials that indicate spinor entanglement that involve only two of the particles as well as polynomials that only indicate spinor entanglement that involves all the particles.

\subsection{Polynomials of bidegree (1,1)}\label{bideg11}

We can construct eight linearly independent locally Lorentz invariant polynomials of bidegree (1,1) through tensor sandwich contractions.
In writing these sandwich contractions we suppress the superscript $ABC$ of $\Psi$ and $\Psi^*$ in the following.
We also leave out the summation sign, with the understanding that repeated indices are summed over.  
Furthermore, we use the abbreviated notation $\gamma^0\gamma^5\equiv \gamma^{05}$ in giving the formal expressions for the polynomials.
\begin{eqnarray}
V_1&=&\gamma^0_{li}\gamma^0_{mj}\gamma^0_{nk}\Psi^{*}_{ijk}\Psi_{lmn},\nonumber\\
V_2&=&\gamma^{05}_{li}\gamma^0_{mj}\gamma^0_{nk}\Psi^{*}_{ijk}\Psi_{lmn},\nonumber\\
V_3&=&\gamma^0_{li}\gamma^{05}_{mj}\gamma^0_{nk}\Psi^{*}_{ijk}\Psi_{lmn},\nonumber\\
V_4&=&\gamma^0_{li}\gamma^0_{mj}\gamma^{05}_{nk}\Psi^{*}_{ijk}\Psi_{lmn},\nonumber\\
V_5&=&\gamma^{05}_{li}\gamma^{05}_{mj}\gamma^0_{nk}\Psi^{*}_{ijk}\Psi_{lmn},\nonumber\\
V_6&=&\gamma^{05}_{li}\gamma^0_{mj}\gamma^{05}_{nk}\Psi^{*}_{ijk}\Psi_{lmn},\nonumber\\
V_7&=&\gamma^0_{li}\gamma^{05}_{mj}\gamma^{05}_{nk}\Psi^{*}_{ijk}\Psi_{lmn},\nonumber\\
V_8&=&\gamma^{05}_{li}\gamma^{05}_{mj}\gamma^{05}_{nk}\Psi^{*}_{ijk}\Psi_{lmn}.
\end{eqnarray}

None of these polynomials is zero for all product states and thus they are not indicators of spinor entanglement. However products of these polynomials can be used in the construction of indicators of spinor entanglement of higher bidegrees as described in Section \ref{utb}.

\subsection{Polynomials of bidegree (2,2)}\label{bideg22}

Here we describe how to construct locally Lorentz invariant polynomials of bidegree (2,2) that are indicators of spinor entanglement.
Since the tensor sandwich contractions of bidegree (2,0) are all zero there are no nonzero tensor sandwich contractions of bidegree (2,2) that factorize as a product of a bidegree (2,0) polynomial and a  bidegree (0,2) polynomial. There are however nonzero tensor sandwich contractions that factorize as a product of two bidegree (1,1) polynomials. But we have already described the tensor sandwich contractions of bidegree (1,1).
Thus it remains to consider the tensor sandwich contractions involving two copies of $\Psi^{ABC}$ and two copies of $\Psi^{* ABC}$ that do not factorize into a product of two polynomials. 
We subdivide these contractions based on how many sandwiched matrices that are either $C$ or $C\gamma^5$.

There are three different ways to pair up the tensor indices such that four of the sandwiched matrices are either $C$ or $C\gamma^5$. In writing these sandwich contractions we suppress the superscript $ABC$ of $\Psi$ and $\Psi^*$ in the following.
We also leave out the summation sign, with the understanding that repeated indices are summed over.   The three different ways to contract the indices are $B_{2,1}$, $D_{2,1}$, and $Z_{2,1}$ given by
\begin{eqnarray}\label{tens2}
B_{2,1}&=& X_{ij}X_{mk}X_{nl}\Psi_{jkl}\Psi_{qmn}X_{qr}X_{ps}X_{ut}\Psi^*_{rst}\Psi^*_{ipu},\nonumber\\
D_{2,1}&=& X_{ij}X_{mk}X_{nl}\Psi_{jkl}\Psi_{ipn}X_{qr}X_{ps}X_{ut}\Psi^*_{rst}\Psi^*_{qmu},\nonumber\\
Z_{2,1}&=& X_{ij}X_{mk}X_{nl}\Psi_{jkl}\Psi_{imu}X_{qr}X_{ps}X_{ut}\Psi^*_{rst}\Psi^*_{qpn}.
\end{eqnarray}
The first way to contract the indices $B_{2,1}$ is invariant with respect to a permutation of laboratories B and C. Similarly, the second way to contract the indices $D_{2,1}$ is invariant with respect to a permutation of laboratories A and C and the third way to contract the indices $Z_{2,1}$ is invariant with respect to a permutation of laboratories A and B.

\begin{figure}[htb]
\setcounter{totalnumber}{4}
\subfloat[\label{subfig:B3}]{
  \includegraphics[scale=0.75]{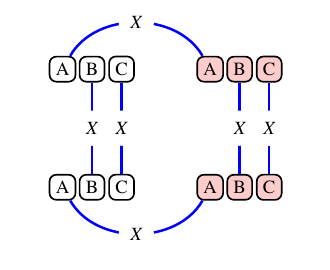}}
  \hfill
\subfloat[\label{subfig:D3}]{
  \includegraphics[scale=0.75]{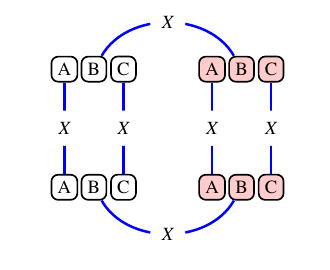}}
  \hfill  
\subfloat[\label{subfig:Z3}]{
  \includegraphics[scale=0.75]{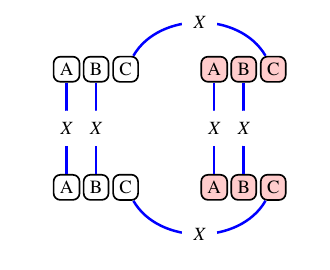}}

  \caption{Graph representations of the three tensor sandwich contractions $B_{2,1}$ (a), $D_{2,1}$ (b), $Z_{2,1}$ (c). Each of the two copies of $\Psi^{ABC}$ is represented by three unfilled boxes corresponding to the three tensor indices and each of the two copies of $\Psi^{*ABC}$ is represented by three filled boxes. The tensor sandwich contractions are represented by blue lines connecting the contracted indices interrupted by $X$ representing the sandwiched matrix.}  
\label{risz3}    
\end{figure}

The three different tensor contractions in Eq. (\ref{tens2}) can be represented as graphs which provides an additional way to understand them.
See Fig. \ref{risz3} for the graph representation of $B_{2,1}$, $D_{2,1}$, and $Z_{2,1}$.

There are six different ways to pair up the tensor indices such that two of the sandwiched matrices are either $C$ or $C\gamma^5$. In writing these sandwich contractions we suppress the superscript $ABC$ of $\Psi$ and $\Psi^*$ in the following.
We also leave out the summation sign, with the understanding that repeated indices are summed over.   The six different ways to contract the indices are given by
\begin{eqnarray}\label{tens3}
B_{1,2}&=& X_{ij}X_{mk}X_{nl}\Psi^*_{jkl}\Psi_{qmn}X_{qr}X_{ps}X_{ut}\Psi_{rst}\Psi^*_{ipu},\nonumber\\
D_{1,2}&=& X_{ij}X_{mk}X_{nl}\Psi^*_{jkl}\Psi_{ipn}X_{qr}X_{ps}X_{ut}\Psi_{rst}\Psi^*_{qmu},\nonumber\\
Z_{1,2}&=& X_{ij}X_{mk}X_{nl}\Psi^*_{jkl}\Psi_{imu}X_{qr}X_{ps}X_{ut}\Psi_{rst}\Psi^*_{qpn},\nonumber\\
X_{1,2A}&=&X_{ij}X_{mk}X_{nl}\Psi^*_{jkl}\Psi_{ipt}X_{qr}X_{ps}X_{ut}\Psi_{rsn}\Psi^*_{qmu},\nonumber\\
X_{1,2B}&=&X_{ij}X_{mk}X_{nl}\Psi_{jkl}\Psi_{ipt}X_{qr}X_{ps}X_{ut}\Psi_{rsn}^*\Psi^*_{qmu},\nonumber\\
X_{1,2C}&=&X_{ij}X_{mk}X_{nl}\Psi_{jkl}\Psi^*_{ipt}X_{qr}X_{ps}X_{ut}\Psi_{rsn}\Psi^*_{qmu}.
\end{eqnarray}
The first way to contract the indices $B_{1,2}$ is invariant with respect to a permutation of laboratories B and C. Similarly, the second way to contract the indices $D_{1,2}$ is invariant with respect to a permutation of laboratories A and C and the third way to contract the indices $Z_{1,2}$ is invariant with respect to a permutation of laboratories A and B. The three ways to contract the indices $X_{1,2A}$,$X_{1,2B}$ and $X_{1,2C}$ are not invariant under any permutation of laboratories but are instead related to each other by such permutations.

\begin{figure}[htb]
\setcounter{totalnumber}{4}
\subfloat[\label{subfig:B4}]{
  \includegraphics[scale=0.75]{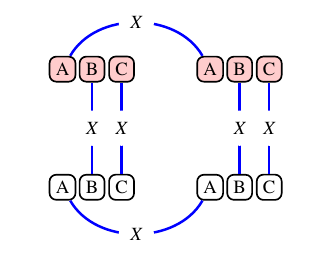}}
  \hfill
\subfloat[\label{subfig:D4}]{
  \includegraphics[scale=0.75]{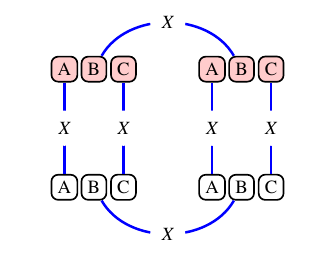}}
  \hfill  
\subfloat[\label{subfig:Z4}]{
  \includegraphics[scale=0.75]{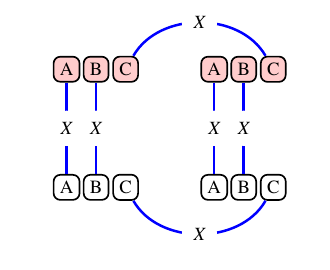}} 
  \hfill
\subfloat[\label{subfig:X4}]{
  \includegraphics[scale=0.75]{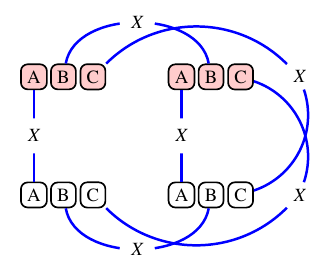}} 
\hfill
\subfloat[\label{subfig:X4b}]{
  \includegraphics[scale=0.75]{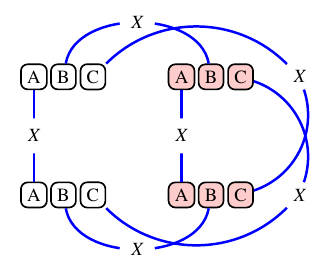}}   
\hfill
\subfloat[\label{subfig:X4c}]{
  \includegraphics[scale=0.75]{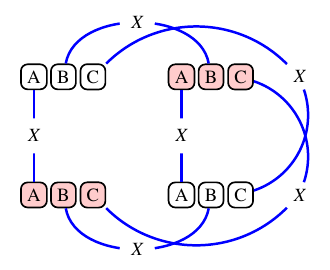}} 
  \caption{Graph representations of the six tensor sandwich contractions $B_{1,2}$ (a), $D_{1,2}$ (b), $Z_{1,2}$ (c), $X_{1,2A}$ (d), $X_{1,2B}$ (e), and $X_{1,2C}$ (f). Each of the two copies of $\Psi^{ABC}$ is represented by three unfilled boxes corresponding to the three tensor indices and each of the two copies of $\Psi^{*ABC}$ is represented by three filled boxes. The tensor sandwich contractions are represented by blue lines connecting the contracted indices interrupted by $X$ representing the sandwiched matrix.}  
\label{risz4}    
\end{figure}

The six different tensor contractions in Eq. (\ref{tens3}) can be represented as graphs.
See Fig. \ref{risz4} for the graph representation of $B_{1,2}$, $D_{1,2}$, $Z_{1,2}$, $X_{1,2A}$, $X_{1,2B}$ and $X_{1,2C}$ .

Next we consider the polynomials constructed from tensor sandwich contractions where none of the sandwiched matrices are $C$ or $C\gamma^5$. Polynomials that are identically zero for all product states can be built from linear combinations of such polynomials.
First we consider the polynomials of this kind that can be built from
linear combinations of products of the bidegree (1,1) polynomials. We can construct 9 linearly independent such polynomials that are identically zero for all product states
\begin{eqnarray}\label{tognglist}
&&V_1V_8-V_3V_6,\nonumber\\
 &&V_2V_7-V_3V_6,\nonumber\\
   &&V_2V_7-V_4V_5,\nonumber\\
 &&V_1V_7-V_3V_4,\nonumber\\
 &&V_1V_5-V_2V_3,\nonumber\\
 &&V_1V_6-V_2V_4,\nonumber\\
 &&V_2V_8-V_5V_6,\nonumber\\
 &&V_4V_8-V_6V_7,\nonumber\\
  &&V_3V_8-V_5V_7.
\end{eqnarray}
Note that the three first polynomials in Eq. (\ref{tognglist}), $V_1V_8-V_3V_6$, $V_2V_7-V_3V_6$ and $V_2V_7-V_4V_5$ have been selected
as linear combinations that are zero for all product states from a 4 dimensional subspace of polynomials that reduce to the same product of sesquilinear forms on the product states. The other six polynomials have each been chosen as a linear combination from a 2 dimensional subspace of polynomials that reduce to the same product of sesquilinear forms on the product states.

\begin{figure}[htb]
\setcounter{totalnumber}{4}
\subfloat[\label{subfig:B2}]{
  \includegraphics[scale=0.75]{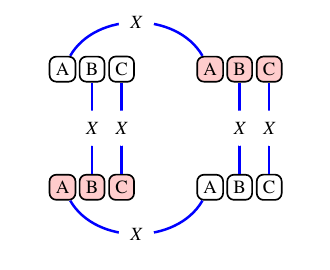}}
  \hfill
\subfloat[\label{subfig:D2}]{
  \includegraphics[scale=0.75]{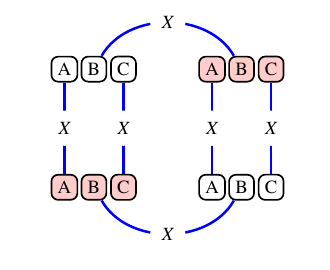}}
  \hfill  
\subfloat[\label{subfig:Z2}]{
  \includegraphics[scale=0.75]{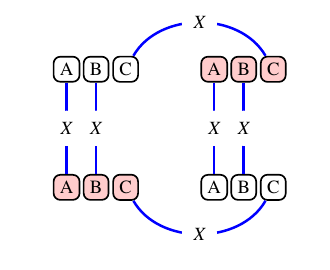}}

  \caption{Graph representations of the three tensor sandwich contractions $B_{0,3}$ (a), $D_{0,3}$ (b), $Z_{0,3}$ (c). Each of the two copies of $\Psi^{ABC}$ is represented by three unfilled boxes corresponding to the three tensor indices and each of the two copies of $\Psi^{*ABC}$ is represented by three filled boxes. The tensor sandwich contractions are represented by blue lines connecting the contracted indices interrupted by $X$ representing the sandwiched matrix.}  
\label{risz2}    
\end{figure}

We then consider the tensor sandwich contractions that do not factorize into two bidegree (1,1) polynomials.
There are three different ways to pair up the tensor indices such that none of the sandwiched matrices are $C$ or $C\gamma^5$ and the polynomial does not factorize.
In writing these sandwich contractions we suppress the superscript $ABC$ of $\Psi$ and $\Psi^*$ in the following.
We also leave out the summation sign, with the understanding that repeated indices are summed over.   The three different ways to contract the indices are
\begin{eqnarray}\label{tens4}
B_{0,3}&=& X_{ij}X_{mk}X_{nl}\Psi_{jkl}\Psi^*_{qmn}X_{qr}X_{ps}X_{ut}\Psi_{rst}\Psi^*_{ipu},\nonumber\\
D_{0,3}&=& X_{ij}X_{mk}X_{nl}\Psi_{jkl}\Psi^*_{ipn}X_{qr}X_{ps}X_{ut}\Psi_{rst}\Psi^*_{qmu},\nonumber\\
Z_{0,3}&=& X_{ij}X_{mk}X_{nl}\Psi_{jkl}\Psi^*_{imu}X_{qr}X_{ps}X_{ut}\Psi_{rst}\Psi^*_{qpn}.
\end{eqnarray}
The first way to contract the indices $B_{0,3}$ is invariant with respect to a permutation of laboratories B and C. Similarly, the second way to contract the indices $D_{0,3}$ is invariant with respect to a permutation of laboratories A and C and the third way to contract the indices $Z_{0,3}$ is invariant with respect to a permutation of laboratories A and B.
The three different tensor contractions in Eq. (\ref{tens4}) can be represented as graphs.
See Fig. \ref{risz2} for the graph representation of $B_{0,3}$, $D_{0,3}$, and $Z_{0,3}$.

For each way to contract indices $B_{0,3}$, $D_{0,3}$, and $Z_{0,3}$ we can consider the different ways to choose the $X$s as $\gamma^0$ or $\gamma^0\gamma^5$ and construct these polynomials. From the set of these polynomials we can then select a subset where all the polynomials reduce to the same product of sesquilinear forms for the product states. To that set we can add all the products of two bidegree (1,1) polynomials that also reduce to the same product of sesquilinear forms for the product states. If the dimension of the polynomial space spanned by these polynomials is $n$ we can construct $n-1$ linearly independent polynomials of bidegree (2,2) that are zero if no particles are spinor entangled. Thus these $n-1$ polynomials are locally Lorentz invariant indicators of spinor entanglement. Then we can repeat this procedure for all the different products of sesquilinear forms.

We have now outlined how to construct polynomials of bidegree (2,2) that are indicators of spinor entanglement for three Dirac particles.
For each way to pair up the tensor indices there is $2^6=64$ ways to choose the $X$s sandwiched between two copies of $\Psi^{ABC}$ or between two copies of $\Psi^{* ABC}$ as either $C$ or $ C\gamma^5$ and the $X$s sandwiched between one copy of $\Psi^{ABC}$ and one copy of $\Psi^{*ABC}$ as either $\gamma^0$ or $ \gamma^0\gamma^5$.
Thus there is a total of $64\times 12=768$ ways to construct polynomials by this method. 
Note that this does not necessarily mean that one can construct 768 linearly independent polynomials.
Here we do not give a complete list of such polynomials. Instead we consider a limited selection of polynomials.
In particular we consider the 12 polynomials that can be constructed by using only $C$ and $\gamma^0$. These are given by
\begin{eqnarray}\label{tonglist}
B_{0,3}^1&=& \gamma^0_{ij}\gamma^0_{mk}\gamma^0_{nl}\Psi_{jkl}\Psi^*_{qmn}\gamma^0_{qr}\gamma^0_{ps}\gamma^0_{ut}\Psi_{rst}\Psi^*_{ipu}-V_1^2,\nonumber\\
D_{0,3}^1&=& \gamma^0_{ij}\gamma^0_{mk}\gamma^0_{nl}\Psi_{jkl}\Psi^*_{ipn}\gamma^0_{qr}\gamma^0_{ps}\gamma^0_{ut}\Psi_{rst}\Psi^*_{qmu}-V_1^2,\nonumber\\
Z_{0,3}^1&=& \gamma^0_{ij}\gamma^0_{mk}\gamma^0_{nl}\Psi_{jkl}\Psi^*_{imu}\gamma^0_{qr}\gamma^0_{ps}\gamma^0_{ut}\Psi_{rst}\Psi^*_{qpn}-V_1^2,\nonumber\\
B_{2,1}^1&=& \gamma^0_{ij}C_{mk}C_{nl}\Psi_{jkl}\Psi_{qmn}\gamma^0_{qr}C_{ps}C_{ut}\Psi^*_{rst}\Psi^*_{ipu},\nonumber\\
D_{2,1}^1&=& C_{ij}\gamma^0_{mk}C_{nl}\Psi_{jkl}\Psi_{ipn}C_{qr}\gamma^0_{ps}C_{ut}\Psi^*_{rst}\Psi^*_{qmu},\nonumber\\
Z_{2,1}^1&=& C_{ij}C_{mk}\gamma^0_{nl}\Psi_{jkl}\Psi_{imu}C_{qr}C_{ps}\gamma^0_{ut}\Psi^*_{rst}\Psi^*_{qpn},\nonumber\\
B_{1,2}^1&=& C_{ij}\gamma^0_{mk}\gamma^0_{nl}\Psi^*_{jkl}\Psi_{qmn}C_{qr}\gamma^0_{ps}\gamma^0_{ut}\Psi_{rst}\Psi^*_{ipu},\nonumber\\
D_{1,2}^1&=& \gamma^0_{ij}C_{mk}\gamma^0_{nl}\Psi^*_{jkl}\Psi_{ipn}\gamma^0_{qr}C_{ps}\gamma^0_{ut}\Psi_{rst}\Psi^*_{qmu},\nonumber\\
Z_{1,2}^1&=& \gamma^0_{ij}\gamma^0_{mk}C_{nl}\Psi^*_{jkl}\Psi_{imu}\gamma^0_{qr}\gamma^0_{ps}C_{ut}\Psi_{rst}\Psi^*_{qpn},\nonumber\\
X_{1,2A}^1&=&\gamma^0_{ij}C_{mk}\gamma^0_{nl}\Psi^*_{jkl}\Psi_{ipt}\gamma^0_{qr}C_{ps}\gamma^0_{ut}\Psi_{rsn}\Psi^*_{qmu},\nonumber\\
X_{1,2B}^1&=&C_{ij}\gamma^0_{mk}\gamma^0_{nl}\Psi_{jkl}\Psi_{ipt}C_{qr}\gamma^0_{ps}\gamma^0_{ut}\Psi_{rsn}^*\Psi^*_{qmu},\nonumber\\
X_{1,2C}^1&=&\gamma^0_{ij}\gamma^0_{mk}C_{nl}\Psi_{jkl}\Psi^*_{ipt}\gamma^0_{qr}\gamma^0_{ps}C_{ut}\Psi_{rsn}\Psi^*_{qmu}.
\end{eqnarray}

The 12 polynomials $B_{0,3}^1$, $D_{0,3}^1$, $Z_{0,3}^1$, $B_{2,1}^1$, $D_{2,1}^1$, $Z_{2,1}^1$, $B_{1,2}^1$, $D_{2,1}^1$, $Z_{2,1}^1$, $X_{1,2A}^1$, $X_{1,2B}^1$, $X_{1,2C}^1$ together with the 9 polynomials $V_1V_8-V_3V_6 $, $V_2V_7-V_3V_6 $, $ V_2V_7-V_4V_5 $, $ V_1V_7-V_3V_4 $, $ V_1V_5-V_2V_3 $, $ V_1V_6-V_2V_4 $, $ V_2V_8-V_5V_6 $, $ V_4V_8-V_6V_7 $, $  V_3V_8-V_5V_7$ span a 21 dimensional polynomial space. Thus the 21 polynomials are linearly independent.

Note that the three polynomials $B_{0,3}^1$, $D_{0,3}^1$, $Z_{0,3}^1$ have been chosen as linear combinations that are zero for all product states from a 4 dimensional subspace of polynomials that reduce to the same product of Hermitian sesquilinear forms
$\psi(x_A)^\dagger\gamma^0\psi(x_A),\varphi(x_B)^\dagger\gamma^0\varphi(x_B),\zeta(x_C)^\dagger\gamma^0\zeta(x_C)$ on the product states $\psi(x_A)\otimes\varphi(x_B)\otimes\zeta(x_C)$.

The polynomials  $B_{2,1}^1$, $D_{2,1}^1$, and $Z_{2,1}^1$ each have the property of being identically zero if either out of a specific pair of particles is in a product state with the other particles but may be nonzero if the third particle is in a product state with the other particles.
The polynomial $B_{2,1}^1$ is identically zero if the particle in Bob's lab or the particle in Charlie's lab is in a product state with the other particles,
but may still be nonzero if Alice's particle is in a product state with the other particles.
The polynomial $D_{2,1}^1$ is identically zero if the particle in Alice's lab or the particle in Charlie's lab is in a product state with the other particles,
but may still be nonzero if Bob's particle is in a product state with the other particles.
The polynomial $Z_{2,1}^1$ is identically zero if the particle in Alice's lab or the particle in Bob's lab is in a product state with the other particles,
but may still be nonzero if Charlie's particle is in a product state with the other particles.

The polynomials  $B_{1,2}^1$, $D_{1,2}^1$, $Z_{1,2}^1$, $X_{1,2A}^1$, $X_{1,2B}^1$, $X_{1,2C}^1$ as well as $B_{0,3}^1$, $D_{0,3}^1$, and $Z_{0,3}^1$ each have the property of being identically zero if one specific particle is in a product state with the other particles but may be nonzero if either of the other two particles is in a product state with the other particles.
The polynomials $B_{1,2}^1$, $X_{1,2B}^1$ and $B_{0,3}^1$ are identically zero if the particle in Alice's lab is in a product state with the other particles, but may still be nonzero if Bob's or Charlie's particle is in a product state with the other particles.
The polynomials $D_{1,2}^1$, $X_{1,2A}^1$ and $D_{0,3}^1$ are identically zero if the particle in Bob's lab is in a product state with the other particles, but may still be nonzero if Alice's or Charlie's particle is in a product state with the other particles.
The polynomials $Z_{1,2}^1$, $X_{1,2C}^1$ and $Z_{0,3}^1$ are identically zero if the particle in Charlie's lab is in a product state with the other particles, but may still be nonzero if Bob's or Alice's particle is in a product state with the other particles.

We can consider the case where one of the particles is in a product state with the other particles and one of the polynomials $B_{2,1}^1$, $D_{2,1}^1$, or $Z_{2,1}^1$ is nonzero. For each such case the nonzero polynomial reduces to a product of two sesquilinear forms times a polynomial that up to a constant factor and a relabelling of the labs is the two-particle polynomial $|I_1|^2$.

Similarly, we can consider the case where one of the particles is in a product state with the other particles and one of the polynomials $B_{1,2}^1$, $D_{1,2}^1$, $Z_{1,2}^1$, $X_{1,2A}^1$, $X_{1,2B}^1$, or $X_{1,2C}^1$ is nonzero. For each such case the nonzero polynomial reduces to a product of two sesquilinear forms times a polynomial that up to a constant factor and a relabelling of the labs is the two-particle polynomial $R_1$ (or alternatively $R_4$).

Finally, we can consider the case where one of the particles is in a product state with the other particles and one of the polynomials $B_{0,3}^1$, $D_{0,3}^1$, or $Z_{0,3}^1$ is nonzero. For each such case the nonzero polynomial reduces to a product of two sesquilinear forms times a polynomial that up to a constant factor and a relabelling of the labs is the two-particle polynomial $T_1$.

\subsubsection{Free particles at rest in energy subspaces and Weyl particles}

We can consider the case of free particles at rest and further restrict to only states in the local positive or negative energy subspaces. 
Then the shared state is invariant under some combination of projections $P_+^A$ or $P_-^A$ by Alice, $P_+^B$ or $P_-^B$ by Bob and $P_+^C$ or $P_-^C$ by Charlie. Then only a $2\times 2 \times 2$ subtensor of $\Psi^{ABC}$ is nonzero.

In this case the polynomial $D^1_{0,3}$ reduces, up to a sign and a relabelling of the indices, to the polynomial $J_3-J_1^2$
\begin{eqnarray}
J_3-J_1^2&=& 
 2 (\psi_{010}\psi^*_{000}+ \psi_{011}\psi^*_{001}+ \psi_{110}\psi^*_{100}+ \psi_{111}\psi^*_{101})\nonumber\\&&\times (\psi_{000}\psi^*_{010}+ 
    \psi_{001}\psi^*_{011}+ \psi_{100}\psi^*_{110}+ \psi_{101}\psi^*_{111})\nonumber\\     
    &&-2 (|\psi_{010}|^2+ |\psi_{011}|^2+ |\psi_{110}|^2+ |\psi_{111}|^2)\nonumber\\&&\times (|\psi_{000}|^2+ |\psi_{001}|^2+ |\psi_{100}|^2+ |\psi_{101}|^2). 
\end{eqnarray}
Here the polynomial $J_1=|\psi_{000}|^2+ |\psi_{001}|^2+ |\psi_{010}|^2+ |\psi_{011}|^2+ |\psi_{100}|^2+ |\psi_{101}|^2+ |\psi_{110}|^2+ |\psi_{111}|^2$ is the state norm squared. The polynomial $J_3$ has been described in References \cite{kempe,sud,toni,tarrach,toumazet}.

Similarly, the polynomial $B^1_{0,3}$ reduces, up to a sign and a relabelling of the indices, to the polynomial $J_2-J_1^2$
\begin{eqnarray}
J_2-J_1^2&=&2 ( \psi_{100}\psi^*_{000}+ \psi_{101}\psi^*_{001}+ \psi_{110}\psi^*_{010}+ \psi_{111}\psi^*_{011})\nonumber\\&&\times ( 
   \psi_{000}\psi^*_{100}+ \psi_{001}\psi^*_{101}+ \psi_{010}\psi^*_{110}+ \psi_{011}\psi^*_{111})\nonumber\\&&  
 -2 ( |\psi_{000}|^2+ |\psi_{001}|^2+ |\psi_{010}|^2+ |\psi_{011}|^2)\nonumber\\&&\times ( 
   |\psi_{100}|^2+ |\psi_{101}|^2+ |\psi_{110}|^2+ |\psi_{111}|^2).   
\end{eqnarray}
The polynomial $J_2$ has been described in References \cite{kempe,sud,toni,tarrach,toumazet}.

The polynomial $Z^1_{0,3}$ reduces, up to a sign and a relabelling of the indices, to the polynomial $J_4-J_1^2$
\begin{eqnarray}
J_4-J_1^2&=&   
   2 (\psi_{001} \psi^*_{000} + \psi_{011} \psi^*_{010} + \psi_{101} \psi^*_{100} +  \psi_{111} \psi^*_{110})\nonumber\\&&\times ( \psi_{000} \psi^*_{001} + 
    \psi_{010} \psi^*_{011} + \psi_{100} \psi^*_{101} +  \psi_{110} \psi^*_{111})\nonumber\\&& - 
 2 (|\psi_{ 000} |^2 + |\psi_{ 010} |^2 + |\psi_{100}|^2 + |\psi_{ 110} |^2) \nonumber\\&&\times( |\psi_{001} |^2 + 
    |\psi_{011} |^2 + |\psi_{ 101} |^2 + |\psi_{ 111}|^2).
\end{eqnarray}
The polynomial $J_4$ has been described in References \cite{kempe,sud,toni,tarrach,toumazet}.
The three polynomials $J_2$, $J_3$, and $J_4$ have been used in References \cite{kempe,sud,toni,tarrach} to characterize the spin entanglement of three non-relativistic spin-$\frac{1}{2}$ particles.

The remaining nine polynomials also reduce to linear combinations of  $J_3$, $J_2$, $J_4$, and $J_1^2$.
The polynomial $D^1_{2,1}$ reduces to $J_3-J_2-J_4+J_1^2$,
the polynomial $Z^1_{2,1}$ reduces to $J_4-J_3-J_2+J_1^2$, and $B^1_{2,1}$ reduces to $J_2-J_3-J_4+J_1^2$.
The polynomial $B^1_{1,2}$ reduces to $-J_2+J_1^2$, $Z^1_{1,2}$ reduces to $-J_4+J_1^2$, and $D^1_{1,2}$ reduces to $-J_3+J_1^2$.
The polynomial $X^1_{1,2A}$ reduces to $J_4-J_3$, $X^1_{1,2B}$ reduces to $J_2-J_3$, and $X^1_{1,2C}$ reduces to $J_2-J_4$.

The 9 polynomials in Eq. (\ref{tognglist}) all reduce to zero since $P_+\gamma^0\gamma^5P_+=P_-\gamma^0\gamma^5P_-=0$.

If we consider the case of Weyl particles we see that all the polynomials in Eq. (\ref{tonglist}) and all the polynomials in Eq. (\ref{tognglist}) reduce to zero since $P_L\gamma^0P_L=P_R\gamma^0P_R=0$ and $P_L\gamma^0\gamma^5P_L=P_R\gamma^0\gamma^5P_R=0$.

\subsection{Polynomials of bidegree (3,1) and (1,3)}\label{bideg3}
Here we describe how to construct locally Lorentz invariant polynomials of bidegree (3,1) that are indicators of spinor entanglement. The complex conjugates of these polynomials are locally Lorentz invariant indicators of spinor entanglement with bidegree (1,3).

Since the tensor sandwich contractions of bidegree (2,0) are all zero there are no nonzero tensor sandwich contractions of bidegree (3,1) that factorize as a product of a bidegree (2,0) polynomial and a  bidegree (1,1) polynomial.
Thus it remains to consider the tensor sandwich contractions involving three copies of $\Psi^{ABC}$ and one copy of $\Psi^{* ABC}$ that do not factorize into a product of two polynomials. There are four different such ways to pair up the tensor indices. In writing these sandwich contractions we suppress the superscript $ABC$ of $\Psi$ and $\Psi^*$.
We also leave out the summation sign, with the understanding that repeated indices are summed over.   The four different ways to contract the indices are

\begin{eqnarray}\label{tens}
B&=& X_{ij}X_{mk}X_{nl}\Psi_{jkl}\Psi_{qmn}X_{qr}X_{ps}X_{ut}\Psi_{rst}\Psi^*_{ipu},\nonumber\\
D&=& X_{ij}X_{mk}X_{nl}\Psi_{jkl}\Psi_{ipn}X_{qr}X_{ps}X_{ut}\Psi_{rst}\Psi^*_{qmu},\nonumber\\
Z&=& X_{ij}X_{mk}X_{nl}\Psi_{jkl}\Psi_{imu}X_{qr}X_{ps}X_{ut}\Psi_{rst}\Psi^*_{qpn},\nonumber\\
X&=&X_{ij}X_{mk}X_{nl}\Psi_{jkl}\Psi_{ipt}X_{qr}X_{ps}X_{ut}\Psi_{rsn}\Psi^*_{qmu}.
\end{eqnarray}

\begin{figure}[htb]
\setcounter{totalnumber}{4}
\subfloat[\label{subfig:B}]{
  \includegraphics[scale=0.75]{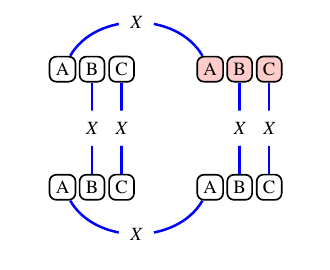}}
  \hfill
\subfloat[\label{subfig:D}]{
  \includegraphics[scale=0.75]{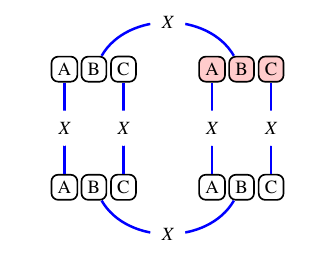}}
  \hfill  
\subfloat[\label{subfig:Z}]{
  \includegraphics[scale=0.75]{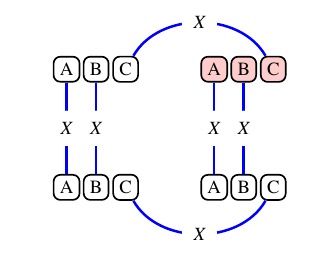}} 
  \hfill 
\subfloat[\label{subfig:X}]{
  \includegraphics[scale=0.75]{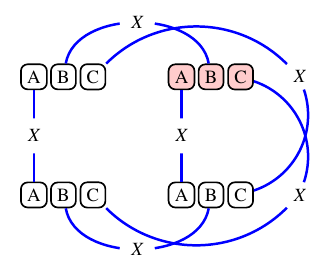}} 
  
  \caption{Graph representations of the four tensor sandwich contractions $B$ (a), $D$ (b), $Z$ (c), and  $X$ (d). Each of the three copies of $\Psi^{ABC}$ is represented by three unfilled boxes corresponding to the three tensor indices and the single copy of $\Psi^{*ABC}$ is represented by three filled boxes. The tensor sandwich contractions are represented by blue lines connecting the contracted indices interrupted by $X$ representing the sandwiched matrix.}  
\label{risz}    
\end{figure}

The first way to contract the indices $B$ is invariant with respect to a permutation of laboratories B and C. Similarly, the second way to contract the indices $D$ is invariant with respect to a permutation of laboratories A and C and the third way to contract the indices $Z$ is invariant with respect to a permutation of laboratories A and B. The final way to contract the indices $X$ is invariant with respect to any permutation of the laboratories. 
The four different tensor contractions in Eq. (\ref{tens}) can be represented as graphs which provides an additional way to understand them.
See Fig. \ref{risz} for the graph representation of $B$, $D$, $Z$, and $X$.
As described in section \ref{entaq} all the bidegree $(3,1)$ polynomials constructed in this way are such that for each observer one $C$ or $C\gamma^5$ is used. Thus all such polynomials are identically zero for the product states, and indicate only spinor entanglement that involves all the particles.

For each way to pair up the tensor indices there is $2^6=64$ ways to choose the $X$s sandwiched between two copies of $\Psi^{ABC}$ as either $C$ or $ C\gamma^5$ and the $X$s sandwiched between one copy of $\Psi^{ABC}$ and one copy of $\Psi^{*ABC}$ as either $\gamma^0$ or $ \gamma^0\gamma^5$.
Thus there is a total of $64\times 4=256$ ways to construct polynomials by this method. Here we do not give a complete list of such polynomials. Instead we consider a more limited selection.
If we consider only the polynomials that are invariant under the parity transformation P in all labs there is only
$2^3=8$ ways to construct such polynomials for each way to pair up the tensor indices. Thus we can consider these $8\times 4=32$ polynomials.

One can construct a set of 20 linearly independent 
polynomials that are invariant under P in all labs. These can be described by the 32 polynomials together with 12 linear dependencies. In the following we use the abbreviated notation $C^5\equiv C\gamma^5$ and $\gamma^0\gamma^5\equiv \gamma^{05}$ in giving the formal expressions for the polynomials. The polynomials are 

\begin{eqnarray}\label{longlist}
X_1&=&C_{il}C_{jp}C_{nq}\gamma^0_{kt}\gamma^0_{ms}\gamma^0_{or}\Psi_{ijk}\Psi_{lmn}\Psi_{opq}\Psi^*_{rst},\nonumber\\
X_{2}&=&C_{il}C_{jp}C^5_{nq}\gamma^{0}_{or}\gamma^{0}_{ms}\gamma^{05}_{kt}\Psi_{ijk}\Psi_{lmn}\Psi_{opq}\Psi^*_{rst},\nonumber\\
X_{3}&=&C_{il}^5C_{jp}C_{nq}\gamma^{05}_{or}\gamma^{0}_{ms}\gamma^0_{kt}\Psi_{ijk}\Psi_{lmn}\Psi_{opq}\Psi^*_{rst},\nonumber\\
X_{4}&=&C_{il}C^5_{jp}C_{nq}\gamma^0_{kt}\gamma^{05}_{ms}\gamma^0_{or}\Psi_{ijk}\Psi_{lmn}\Psi_{opq}\Psi^*_{rst},\nonumber\\
X_{5}&=&C^5_{il}C^5_{jp}C_{nq}\gamma^{05}_{or}\gamma^{05}_{ms}\gamma^{0}_{kt}\Psi_{ijk}\Psi_{lmn}\Psi_{opq}\Psi^*_{rst},\nonumber\\
X_{6}&=&C_{il}C^5_{jp}C^5_{nq}\gamma^{0}_{or}\gamma^{05}_{ms}\gamma^{05}_{kt}\Psi_{ijk}\Psi_{lmn}\Psi_{opq}\Psi^*_{rst},\nonumber\\
X_{7}&=&C^5_{il}C_{jp}C^5_{nq}\gamma^{05}_{or}\gamma^{0}_{ms}\gamma^{05}_{kt}\Psi_{ijk}\Psi_{lmn}\Psi_{opq}\Psi^*_{rst},\nonumber\\
X_{8}&=&C^5_{il}C^5_{jp}C^5_{nq}\gamma^{05}_{or}\gamma^{05}_{ms}\gamma^{05}_{kt}\Psi_{ijk}\Psi_{lmn}\Psi_{opq}\Psi^*_{rst},\nonumber\\
Z_1&=&C_{il}C_{jm}C_{nq}\gamma^0_{kt}\gamma^0_{ps}\gamma^0_{or}\Psi_{ijk}\Psi_{lmn}\Psi_{opq}\Psi^*_{rst},\nonumber\\
Z_2&=&C_{il}C_{jm}C^5_{nq}\gamma^0_{or}\gamma^0_{ps}\gamma^{05}_{kt}\Psi_{ijk}\Psi_{lmn}\Psi_{opq}\Psi^*_{rst},\nonumber\\
Z_3&=&C^5_{il}C_{jm}C_{nq}\gamma^{05}_{or}\gamma^0_{ps}\gamma^{0}_{kt}\Psi_{ijk}\Psi_{lmn}\Psi_{opq}\Psi^*_{rst},\nonumber\\
Z_4&=&C_{il}C^5_{jm}C_{nq}\gamma^{0}_{or}\gamma^{05}_{ps}\gamma^{0}_{kt}\Psi_{ijk}\Psi_{lmn}\Psi_{opq}\Psi^*_{rst},\nonumber\\
Z_5&=&C^5_{il}C^5_{jm}C_{nq}\gamma^{05}_{or}\gamma^{05}_{ps}\gamma^{0}_{kt}\Psi_{ijk}\Psi_{lmn}\Psi_{opq}\Psi^*_{rst},\nonumber\\
Z_6&=&C_{il}C^5_{jm}C^5_{nq}\gamma^{0}_{or}\gamma^{05}_{ps}\gamma^{05}_{kt}\Psi_{ijk}\Psi_{lmn}\Psi_{opq}\Psi^*_{rst},\nonumber\\
Z_7&=&C^5_{il}C^5_{jm}C_{nq}\gamma^{05}_{or}\gamma^{0}_{ps}\gamma^{05}_{kt}\Psi_{ijk}\Psi_{lmn}\Psi_{opq}\Psi^*_{rst},\nonumber\\
Z_8&=&C^5_{il}C^5_{jm}C^5_{nq}\gamma^{05}_{or}\gamma^{05}_{ps}\gamma^{05}_{kt}\Psi_{ijk}\Psi_{lmn}\Psi_{opq}\Psi^*_{rst},\nonumber\\
B_1&=&C_{io}C_{jm}C_{kn}\gamma^0_{lr}\gamma^0_{ps}\gamma^0_{qt}\Psi_{ijk}\Psi_{lmn}\Psi_{opq}\Psi^*_{rst},\nonumber\\
B_2&=&C_{io}C_{jm}C^5_{kn}\gamma^{0}_{lr}\gamma^0_{ps}\gamma^{05}_{qt}\Psi_{ijk}\Psi_{lmn}\Psi_{opq}\Psi^*_{rst},\nonumber\\
B_3&=&C^5_{io}C_{jm}C_{kn}\gamma^{05}_{lr}\gamma^0_{ps}\gamma^0_{qt}\Psi_{ijk}\Psi_{lmn}\Psi_{opq}\Psi^*_{rst},\nonumber\\
B_4&=&C_{io}C^5_{jm}C_{kn}\gamma^{0}_{lr}\gamma^{05}_{ps}\gamma^0_{qt}\Psi_{ijk}\Psi_{lmn}\Psi_{opq}\Psi^*_{rst},\nonumber\\
B_5&=&C^5_{io}C^5_{jm}C_{kn}\gamma^{05}_{lr}\gamma^{05}_{ps}\gamma^0_{qt}\Psi_{ijk}\Psi_{lmn}\Psi_{opq}\Psi^*_{rst},\nonumber\\
B_6&=&C_{io}C^5_{jm}C^5_{kn}\gamma^{0}_{lr}\gamma^{05}_{ps}\gamma^{05}_{qt}\Psi_{ijk}\Psi_{lmn}\Psi_{opq}\Psi^*_{rst},\nonumber\\
B_7&=&C^5_{io}C_{jm}C^5_{kn}\gamma^{05}_{lr}\gamma^{0}_{ps}\gamma^{05}_{qt}\Psi_{ijk}\Psi_{lmn}\Psi_{opq}\Psi^*_{rst},\nonumber\\
B_8&=&C^5_{io}C^5_{jm}C^5_{kn}\gamma^{05}_{lr}\gamma^{05}_{ps}\gamma^{05}_{qt}\Psi_{ijk}\Psi_{lmn}\Psi_{opq}\Psi^*_{rst},\nonumber\\
D_1&=&C_{il}C_{jp}C_{kn}\gamma^{0}_{or}\gamma^{0}_{ms}\gamma^{0}_{qt}\Psi_{ijk}\Psi_{lmn}\Psi_{opq}\Psi^*_{rst},\nonumber\\
D_2&=&C_{il}C_{jp}C^5_{kn}\gamma^{0}_{or}\gamma^{0}_{ms}\gamma^{05}_{qt}\Psi_{ijk}\Psi_{lmn}\Psi_{opq}\Psi^*_{rst},\nonumber\\
D_3&=&C^5_{il}C_{jp}C_{kn}\gamma^{05}_{or}\gamma^{0}_{ms}\gamma^{0}_{qt}\Psi_{ijk}\Psi_{lmn}\Psi_{opq}\Psi^*_{rst},\nonumber\\
D_4&=&C_{il}C^5_{jp}C_{kn}\gamma^{0}_{or}\gamma^{05}_{ms}\gamma^{0}_{qt}\Psi_{ijk}\Psi_{lmn}\Psi_{opq}\Psi^*_{rst},\nonumber\\
D_5&=&C^5_{il}C_{jp}C^5_{kn}\gamma^{05}_{or}\gamma^{0}_{ms}\gamma^{05}_{qt}\Psi_{ijk}\Psi_{lmn}\Psi_{opq}\Psi^*_{rst},\nonumber\\
D_6&=&C_{il}C^5_{jp}C^5_{kn}\gamma^{0}_{or}\gamma^{05}_{ms}\gamma^{05}_{qt}\Psi_{ijk}\Psi_{lmn}\Psi_{opq}\Psi^*_{rst},\nonumber\\
D_7&=&C^5_{il}C^5_{jp}C_{kn}\gamma^{05}_{or}\gamma^{05}_{ms}\gamma^{0}_{qt}\Psi_{ijk}\Psi_{lmn}\Psi_{opq}\Psi^*_{rst},\nonumber\\
D_8&=&C^5_{il}C^5_{jp}C^5_{kn}\gamma^{05}_{or}\gamma^{05}_{ms}\gamma^{05}_{qt}\Psi_{ijk}\Psi_{lmn}\Psi_{opq}\Psi^*_{rst}.
\end{eqnarray}
and the linear dependencies are described by the 12 independent equations
\begin{eqnarray}
X_7 + X_3 - B_7 - B_3 + Z_7 + Z_3&=&0,\nonumber\\
X_6 + X_4- B_4 - B_6 + Z_6 + Z_4 &=&0,\nonumber\\
 X_8 + X_5 - B_8 - B_5 + Z_8 + Z_5&=&0,\nonumber\\
X_1 + X_2-B_1-B_2+Z_1+Z_2 &=&0,\nonumber\\
X_5 + X_4-Z_5 - Z_4  + D_5 + D_4&=&0,\nonumber\\
X_8 + X_6 - Z_8 - Z_6+ D_8 + D_6 &=&0,\nonumber\\
X_7 + X_2-Z_7 - Z_2  + D_7 + D_2&=&0,\nonumber\\
X_1 + X_3-Z_1 - Z_3  + D_1 + D_3&=&0,\nonumber\\
 X_1 + X_4 - D_1 - D_4 + B_1 + B_4&=&0,\nonumber\\
  X_7 + X_8-D_8 - D_7 + B_7 + B_8 &=&0,\nonumber\\
X_6 + X_2 - D_6 - D_2 + B_6 + B_2&=&0,\nonumber\\
X_5 + X_3 - D_5 - D_3 + B_5 + B_3&=&0.
\end{eqnarray}
Polynomials of bidegree (1,3) can be straightforwardly obtained as the complex conjugates of the bidegree (3,1) polynomials in Eq. (\ref{longlist}).

\subsubsection{Free particles at rest in energy subspaces and Weyl particles}

We can consider the case of free particles at rest and further restrict to only states in the local positive or negative energy subspaces. 
Then the shared state is invariant under some combination of projections $P_+^A$ or $P_-^A$ by Alice, $P_+^B$ or $P_-^B$ by Bob and $P_+^C$ or $P_-^C$ by Charlie. In this case only a $2\times 2 \times 2$ subtensor of $\Psi^{ABC}$ is nonzero.
Then the polynomials $Z_1$, $D_1$ and $B_1$ reduce, up to a sign and a relabelling of the indices, to the polynomial $is_2$ where 
\begin{eqnarray}
s_2&=& (\psi_{000}\psi_{111}-\psi_{011}\psi_{100}+ \psi_{010}\psi_{101}- \psi_{001}\psi_{110})\nonumber\\&&\times(|\psi_{000}|^2+ 
     |\psi_{001}|^2+  |\psi_{100}|^2+  |\psi_{101}|^2)\nonumber\\&&+ 2( \psi_{001}\psi_{100}- 
    \psi_{000}\psi_{101})\nonumber\\&&\times( \psi_{010}\psi^*_{000}+  \psi_{011}\psi^*_{001}+ \psi_{110}\psi^*_{100}+ 
     \psi_{111}\psi^*_{101})\nonumber\\&&+ 2( \psi_{010}\psi_{111}-\psi_{011}\psi_{110})\nonumber\\&&\times( \psi_{000}\psi^*_{010}+ 
     \psi_{001}\psi^*_{011}+  \psi_{100}\psi^*_{110}+  \psi_{101}\psi^*_{111})\nonumber\\&&+ (\psi_{011}\psi_{100}- 
    \psi_{010}\psi_{101}+ \psi_{001}\psi_{110}- \psi_{000}\psi_{111})\nonumber\\&&\times( |\psi_{010}|^2+  |\psi_{011}|^2+ 
     |\psi_{110}|^2+  |\psi_{111}|^2).  
\end{eqnarray}
The polynomial $s_2$ has been described by Luque, Thibon, and Toumazet in Ref. \cite{toumazet}. The polynomial $X_1$ reduces to zero in this case and all other polynomials in Eq. (\ref{longlist}) are zero since $P_+C\gamma^5P_+=P_-C\gamma^5P_-=0$ and $P_+\gamma^0\gamma^5P_+=P_-\gamma^0\gamma^5P_-=0$.

If we consider the case of Weyl particles we see that all the polynomials in Eq. (\ref{longlist}) reduce to zero since $P_L\gamma^0P_L=P_R\gamma^0P_R=0$ and $P_L\gamma^0\gamma^5P_L=P_R\gamma^0\gamma^5P_R=0$.

\subsection{Polynomials of bidegree (3,3)}\label{bideg33}

\begin{figure}[htb]
\setcounter{totalnumber}{4}
\subfloat[\label{subfig:B33}]{
  \includegraphics[scale=0.75]{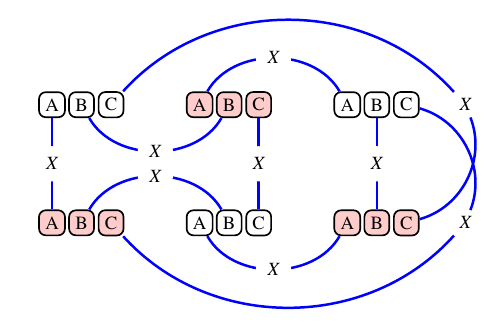}}
  \hfill
\subfloat[\label{subfig:B33c}]{
  \includegraphics[scale=0.75]{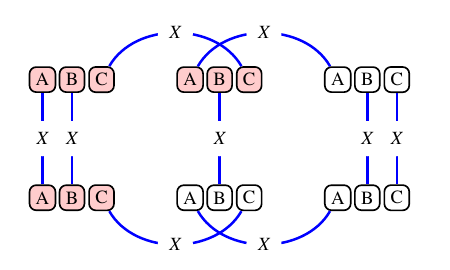}}  
  
  \caption{Graph representations of the tensor sandwich contractions $K$ (a) and $W$ (b). Each of the three copies of $\Psi^{ABC}$ is represented by three unfilled boxes corresponding to the three tensor indices and each of the three copies of $\Psi^{*ABC}$ is represented by three filled boxes. The tensor sandwich contractions are represented by blue lines connecting the contracted indices interrupted by $X$ representing the sandwiched matrix.}  
\label{risz6}    
\end{figure}

There is a large number of ways to construct polynomials of bidegree (3,3) as tensor sandwich contractions. Therefore we do not consider all such ways but instead consider two examples $K$ and $W$ to illustrate qualitatively different ways to tensor sandwich contract three copies of $\Psi^{ABC}$ and three copies of $\Psi^{* ABC}$. 
In writing these sandwich contractions we suppress the superscript $ABC$ of $\Psi$ and $\Psi^*$.
We also leave out the summation sign, with the understanding that repeated indices are summed over.  
The two ways to contract the indices are  

\begin{eqnarray}\label{kemp}
K&=&X_{il}X_{js}X_{kz}X_{mp}X_{nw}X_{ox}X_{qt}X_{ru}X_{vy}\Psi_{ijk}\Psi_{lmn}^*\Psi_{opq}\Psi_{rst}^*\Psi_{uvw}\Psi^*_{xyz},\nonumber\\
W&=&X_{il}X_{jm}X_{kq}X_{nt}X_{ou}X_{ps}X_{rx}X_{wz}X_{vy}\Psi^*_{ijk}\Psi_{lmn}^*\Psi_{opq}^*\Psi_{rst}\Psi_{uvw}\Psi_{xyz}.\nonumber\\
\end{eqnarray}
The two different tensor contractions in Eq. (\ref{kemp}) can be represented as graphs which provides an additional way to understand them.
See Fig. \ref{risz6} for the graph representation of $K$ and  $W$.

For each way $K$ and $W$ to pair up the tensor indices there is $2^9=512$ ways to choose the $X$s sandwiched between two copies of $\Psi^{ABC}$ or between two copies of $\Psi^{* ABC}$ as either $C$ or $ C\gamma^5$ and the $X$s sandwiched between one copy of $\Psi^{ABC}$ and one copy of $\Psi^{*ABC}$ as either $\gamma^0$ or $ \gamma^0\gamma^5$.
Thus there is a total of $512\times 2=1024$ ways to construct polynomials from $K$ and $W$ by this method. Here we do not give a complete list of these polynomials. Instead we consider a more limited selection of only two polynomials.

For the two ways $K$ and $W$ to contract the indices we can choose all the $X$s as $\gamma^0$ or $C$ and construct the two polynomials
\begin{eqnarray}\label{kemplist}
K_1&=&\gamma^0_{il}\gamma^0_{js}\gamma^0_{kz}\gamma^0_{mp}\gamma^0_{nw}\gamma^0_{ox}\gamma^0_{qt}\gamma^0_{ru}\gamma^0_{vy}\Psi_{ijk}\Psi_{lmn}^*\Psi_{opq}\Psi_{rst}^*\Psi_{uvw}\Psi^*_{xyz}\nonumber\\
&&-(V_1)^3,\nonumber\\
W_1&=&C_{il}C_{jm}C_{kq}\gamma^0_{nt}\gamma^0_{ou}\gamma^0_{ps}C_{rx}C_{wz}C_{vy}\Psi^*_{ijk}\Psi_{lmn}^*\Psi_{opq}^*\Psi_{rst}\Psi_{uvw}\Psi_{xyz}.\nonumber\\
\end{eqnarray}
Here we have defined $K_1$ as a difference of two polynomials that reduce to the same polynomial on the product states.

The polynomial $K_1$ is identically zero if all particles are in a product state with the other particles but is nonzero for states where only two particles are entangled.
The polynomial $W_1$ on the other hand is identically zero if any particle is in a product state with the other particles since for each observer two pairs of indices are contracted with $C$ sandwiched.
Thus $W_1$ is an example of a polynomial with bidegree on the form (k,k) that by construction is sensitive only to threepartite entanglement. 
The lowest bidegree on the form (k,k) for which polynomials where for each observer at least one pair of indices are contracted with $C$ or $C\gamma^5$ can be constructed is (3,3).

\subsubsection{Free particles at rest in energy subspaces and Weyl particles}

We can consider the case of free particles at rest and further restrict to only states in the local positive or negative energy subspaces. 
Then the shared state is invariant under some combination of projections $P_+^A$ or $P_-^A$ by Alice, $P_+^B$ or $P_-^B$ by Bob and $P_+^C$ or $P_-^C$ by Charlie. In this case only a $2\times 2 \times 2$ subtensor of $\Psi^{ABC}$ is nonzero. Then the polynomial $K_1$ reduces, up to a sign and a relabelling of the indices, to the polynomial $J_5-J_1^3$ where 
\begin{eqnarray}
J_5-J_1^3&=&\sum_{ijkmnoqrv=1,0} \psi_{ijk}\psi_{imn}^*\psi_{omq}\psi_{rjq}^*\psi_{rvn}\psi^*_{ovk}\nonumber\\&&-\left(\sum_{lps=1,0}| \psi_{lps}|^2\right)^3.
\end{eqnarray}
The polynomial $J_5$ is the Kempe invariant that has been described in Reference \cite{kempe} (See also References \cite{sud,toni,tarrach,toumazet}). 
The polynomial $W_1$ reduces, up to a sign and a relabelling of the indices, to the polynomial $-2/3J_5+1/2(J_2+J_3+J_4)J_1-5/6J_1^3$. 
The Kempe invariant $J_5$ together with the polynomials $J_2$, $J_3$, and $J_4$ have been used in References \cite{kempe,sud,toni,tarrach} to characterize the spin entanglement of three non-relativistic spin-$\frac{1}{2}$ particles.

If we consider the case of Weyl particles we see that both $K_1$ and $W_1$ reduce to zero since $P_L\gamma^0P_L=P_R\gamma^0P_R=0$.

\subsection{Examples of spinor entangled three-particle states}\label{examb}

Here we consider a few examples of tripartite spinor entangled states that are not indicated by the bidegree (4,0) polynomials in Ref. \cite{multispinor} but are indicated by some mixed polynomials. We only consider the mixed polynomials that by construction are sensitive only to entanglement that involves all particles, i.e., the polynomials of bidegree (3,1) in Eq. (\ref{longlist}) and the polynomial of bidegree (3,3) $W_1$ described in Eq. (\ref{kemplist}).

One such state is
\begin{eqnarray}\label{req1}
\frac{1}{{2}}&&({\phi_0^A}\otimes{\phi_0^B}\otimes{\phi_1^C}+{\phi_0^A}\otimes{\phi_1^B}\otimes{\phi_0^C}\nonumber\\&&+{\phi_1^A}\otimes{\phi_0^B}\otimes{\phi_0^C}+{\phi_0^A}\otimes{\phi_0^B}\otimes{\phi_0^C}).
\end{eqnarray}
For this state $|B_1|=|Z_1|=|D_1|=1/8$, but all the other bidegree (3,1) polynomials in Eq. (\ref{longlist}) are identically zero. The bidegree (3,3) invariant $W_1$ takes the absolute value $|W_1|=1/16$. Similarly we can construct a state
\begin{eqnarray}\label{req2}
\frac{1}{{2}}&&({\phi_0^A}\otimes{\phi_0^B}\otimes{\phi_3^C}+{\phi_0^A}\otimes{\phi_3^B}\otimes{\phi_0^C}\nonumber\\&&+{\phi_3^A}\otimes{\phi_0^B}\otimes{\phi_0^C}+{\phi_2^A}\otimes{\phi_2^B}\otimes{\phi_2^C}).
\end{eqnarray}
For this state $|B_8|=|Z_8|=|D_8|=1/8$, but all the other bidegree (3,1) polynomials in Eq. (\ref{longlist}) and the bidegree (3,3) polynomial $W_1$ are identically zero. 
A third state of this kind is
\begin{eqnarray}\label{req3}
\frac{1}{{2}}&&({\phi_0^A}\otimes{\phi_2^B}\otimes{\phi_0^C}+{\phi_0^A}\otimes{\phi_0^B}\otimes{\phi_1^C}\nonumber\\&&+{\phi_0^A}\otimes{\phi_3^B}\otimes{\phi_2^C}+{\phi_1^A}\otimes{\phi_0^B}\otimes{\phi_2^C}).
\end{eqnarray}
For this state $|B_6|=|Z_6|=|D_6|=1/8$, but all the other bidegree (3,1) polynomials in Eq. (\ref{longlist}) and the bidegree (3,3) polynomial $W_1$ are identically zero. 

Next we can consider a state that is not indicated by any of the  bidegree (4,0) polynomials in Ref. \cite{multispinor} or any of the bidegree (3,1) polynomials in Eq. (\ref{longlist}), the so called $W$-state \cite{dur}
\begin{eqnarray}\label{req1w}
\frac{1}{{\sqrt{3}}}({\phi_0^A}\otimes{\phi_0^B}\otimes{\phi_1^C}+{\phi_0^A}\otimes{\phi_1^B}\otimes{\phi_0^C}+{\phi_1^A}\otimes{\phi_0^B}\otimes{\phi_0^C}).\nonumber\\
\end{eqnarray}
For this state the polynomial $W_1$ is nonzero with absolute value $4/27$.%

\section{Types of entanglement only indicated by mixed polynomials}\label{tpo}

As described in Section \ref{ent} and Section \ref{three} some mixed polynomials constructed as tensor sandwich contractions are indicators of spinor entanglement that does not involve all particles. In contrast the homogeneous polynomials constructed as tensor sandwich contractions are only indicators of spinor entanglement that involves all particles. However mixed polynomials can indicate also some forms of entanglement that involve all particles but is not indicated by any homogeneous locally Lorentz invariant polynomial.

We have seen in Section \ref{exxx} and in Section \ref{examb} that there exist spinor entangled states for both two and three particles that are indicated by mixed polynomials but not by the homogeneous polynomials constructed in Reference \cite{spinorent} and Reference \cite{multispinor}, respectively. 
For many of these states one can show that no locally Lorentz invariant homogeneous polynomial that indicates their spinor entanglement exists (See Appendix \ref{semi} for a discussion). Thus there exist spinor entangled states that are only indicated by mixed polynomials. Moreover one can show that some of these states are only indicated by mixed polynomials with bidegree $(k,k)$ for some $k$ and not by any mixed polynomial with bidegree $(k,l)$ where $k\neq l$ (See Appendix \ref{semi} for a discussion). This situation is analogous to that of the spin entanglement of non-relativistic spin-$\frac{1}{2}$ particles where some types of entanglement are only indicated by mixed polynomials \cite{kempe,toni,sud,tarrach,dur,toumazet}. Similarly to the  case of non-relativistic spin-$\frac{1}{2}$ particles \cite{carsud} some spinor entangled states have continuous local unitary symmetries that are incompatible with any mixed locally Lorentz invariant polynomial with bidegree $(k,l)$ where $k\neq l$ taking a nonzero value (See Appendix \ref{semi} for a discussion). Such states exist for any number of Dirac particles. 
For non-relativistic spin-$\frac{1}{2}$ particles however the homogeneous polynomials are sufficient for indicating all entangled states of two particles \cite{wootters,wootters2}.

\section{Discussion and Conclusions}\label{diss}
In this work we have considered the problem of constructing locally Lorentz invariant indicators of spinor entanglement for a system of Dirac particles held by spacelike separated observers. The approach followed builds upon the ideas developed in Refs. \cite{spinorent,multispinor} for constructing homogeneous locally Lorentz invariant polynomials. We reviewed some properties of the Dirac equation, the Dirac gamma matrices, as well as the spinor representation of the Lorentz group and the charge conjugation transformation. We then described the properties of Lorentz invariant skew-symmetric bilinear forms and Lorentz invariant sesquilinear forms.
The physical assumption was made that the local curvature of spacetime can be neglected in each observers laboratory and that each particle can be described as being in a Minkowski space.  
Moreover, we assumed that the tensor products of the individual particle states can be used as a basis for the states of two or more particles.

Given the physical assumptions we used the properties of the skew-symmetric bilinear forms and the sesquilinear forms to describe a method for constructing polynomials in the state coefficients and their complex conjugates for a system of spacelike separated Dirac particles.
We refer to the polynomials with nonzero degree in both the state coefficients and their complex conjugates as {\it mixed polynomials} following Ref. \cite{oka}.
The method described here for constructing mixed or homogeneous polynomials is an extension of the method described in Refs. \cite{spinorent,multispinor} and the two methods coincide when constructing homogeneous polynomials. 

The polynomials constructed by the method given in this work, both homogeneous and mixed, are invariant under the spinor representations of the local proper orthochronous Lorentz groups. Moreover, each such locally Lorentz invariant polynomial is identically zero for any state where each of the particles is in a product state with the other particles. 
Therefore, the polynomials are considered indicators of the entanglement of the spinor degrees of freedom.

Polynomials can be constructed to be identically zero if any of the spinors is in a product state with the other spinors. Such polynomials are indicators only of spinor entanglement that involve all the particles.
Alternatively, for any proper subset of the particles mixed polynomials can be constructed to indicate spinor entanglement that involves this given subset. Only mixed polynomials can be constructed to have this property since no homogeneous polynomial can indicate spinor entanglement that involve only a proper subset of the particles.
Moreover, the mixed polynomials can be constructed to indicate types of spinor entanglement that involve all the particles but is not indicated by any homogeneous polynomial such as the entanglement of the W-state \cite{dur}.

For the case of two spacelike separated Dirac particles polynomials of bidegree (2,2) and (3,1) were constructed in addition to the polynomials of bidegree (2,0) previously described in Ref. \cite{spinorent} and their linear independence was tested. A set of eleven linearly independent polynomials of bidegree (2,2) and a set of four linearly independent polynomials of bidegree (3,1) were described. Examples of spinor entangled two-particle states that are not indicated by any homogeneous polynomials but indicated by either a polynomial of bidegree (2,2) or a polynomial of (3,1) were given.

For the case of three spacelike separated Dirac particles it was described how polynomials of bidegree (2,2), (3,1), and (3,3) can be constructed in addition to the
to the polynomials of bidegree (4,0) previously described in Ref. \cite{multispinor}. 
For any proper subset of the particles there is a polynomial of bidegree (2,2) that indicates spinor entanglement involving this subset. The polynomials of bidegree (3,1) on the other hand indicate only spinor entanglement that involves all the particles. Polynomials of bidegree (3,3) can be constructed to indicate spinor entanglement that involves only a proper subset of the particles or alternatively to indicate only spinor entanglement that involves all particles.
A select set with 21 linearly independent polynomials of bidegree (2,2), 20 linearly independent polynomials of bidegree (3,1), and 2 polynomials of bidegree (3,3) was given. 
Examples of states of three Dirac particles that are spinor entangled in a way that involves all the particles but that are not indicated by any homogeneous polynomial were given. These states are still indicated by mixed polynomials of bidegree (3,1) and (3,3) constructed to indicate only spinor entanglement involving all the particles. One such example given was the W-state \cite{dur} that is not indicated by any polynomial of bidegree (3,1) but is indicated by a polynomial of bidegree (3,3).

For both two and three Dirac particles we considered the case of zero particle momenta and zero electromagnetic four-potentials. Here we further restricted our consideration to particles in the local positive and negative energy subspaces corresponding to the Dirac equation. Particles in these subspaces are often identified with non-relativistic free spin-$\frac{1}{2}$ particles and antiparticles, respectively. 
When the particles belong to such subspaces the polynomials constructed in this work all reduce to linear combinations of the previously described polynomials that have been constructed for two and three non-relativistic spin-$\frac{1}{2}$ particles, or alternatively reduce to zero. 
For two Dirac particles the polynomials of bidegree (2,0) reduce to a multiple of the Wootters concurrence \cite{wootters,wootters2} or alternatively reduce to zero. Similarly, the polynomials of bidegree (2,2) reduce to a multiple of the Wootters concurrence times its complex conjugate or alternatively reduce to zero.
For three Dirac particles the polynomials of bidegree (3,1) either reduce to a multiple of the polynomial $s_2$ described in Reference \cite{toumazet} or alternatively reduce to zero. 
The polynomials of bidegree (2,2) reduce to linear combinations of the polynomials of bidegree (2,2) constructed for three non-relativistic spin-$\frac{1}{2}$ particles previously described in References \cite{kempe,sud,toni,tarrach,toumazet}.
The polynomials of bidegree (3,3) reduce to linear combinations of products of the lower bidegree polynomials constructed for non-relativistic spin-$\frac{1}{2}$ particles and the bidegree (3,3) Kempe invariant described in Reference \cite{kempe}.
Dirac particles with definite momenta can always be described in their respective rest frames. Therefore the previously described polynomials constructed for a system of non-relativistic spin-$\frac{1}{2}$ particles can always be used for the case of free particles with definite momenta as long as the particles belong to either the local positive or negative energy subspaces in their respective rest frames.

We considered also the case of Weyl particles, i.e., particles with definite chirality.
For the case of two Weyl particles the Lorentz invariant polynomials of bidegree (2,0) reduce to either a multiple of the Wootters concurrence \cite{wootters,wootters2} or alternatively are identically zero as described in Ref. \cite{spinorent}. However, all the polynomials of bidegree (2,2) that do not factorize as a polynomial of bidegree (2,0) and a polynomial of bidegree (0,2), and all the polynomials of bidegree (3,1) reduce to zero for Weyl particles. Likewise, for the case of three Weyl particles the polynomials of bidegree (2,2),
the polynomials of bidegree (3,1), and the polynomials of bidegree (3,3) all reduce to zero.

Since the spinor entanglement indicators are polynomials in the state coefficients and their complex conjugates they evolve dynamically. We therefore described how equations describing their evolution can be derived from the Dirac equation. In particular we described how to find their evolution with respect to the different observers time.

In Ref. \cite{spinorent} it was descried how convex roof extensions \cite{lima,wakker,uhlmannn} can be used to extend the absolute value of a locally Lorentz invariant polynomial for two Dirac spinors to a locally Lorentz invariant function on the set of states that are incoherent mixtures, i.e., the mixed states. Convex roof extensions can be made also for the case of the absolute values of the polynomials constructed in this work. A convex roof extension of this kind is by definition identically zero for any incoherent mixture of product states, i.e., for any separable state.
Therefore such convex roof extensions can provide indicators for different types of spinor entanglement of incoherent mixtures.

Finally we note that it is an open question if all locally Lorentz invariant polynomials that are indicators of spinor entanglement can be constructed by the method presented in this work. 

\appendix

\section{Dimension of operationally motivated Hilbert spaces}\label{opp}
In quantum mechanics subspaces of the Hilbert space correspond to experimental propositions about the system being described. Mutually exclusive propositions are described by orthogonal subspaces (See e.g. Reference \cite{birkhoff} or Reference \cite{neumann} for a discussion).

Therefore, from an operational point of view one needs at most as many orthogonal basis vectors for the Hilbert space as there is experimental propositions that can be made about the system.
In any experiment at most a finite number of preparations and measurements are made.
Therefore an operationally constructed Hilbert space can always be chosen as finite dimensional.

\section{The invariance groups of the bilinear and sesquilinear forms}\label{lie}

Each of the bilinear forms $\psi^TC\varphi$ and $\psi^TC\gamma^5\varphi$ as well as the sesquilinear forms  $\psi^\dagger \gamma^0\varphi$ and $\psi^\dagger \gamma^0\gamma^5\varphi$ are invariant under larger groups than the spinor representation of the proper orthochronous Lorentz group.
To see this we can consider all possible products of different numbers of distinct gamma matrices. For such products we have for $C$ that
\begin{eqnarray}\label{c}
(\gamma^\mu)^TC&=&C\gamma^\mu,\nonumber\\
(\gamma^\mu\gamma^\nu)^TC&=&-C\gamma^\mu\gamma^\nu,\nonumber\\
(\gamma^\mu\gamma^\nu\gamma^\rho)^TC&=&-C\gamma^\mu\gamma^\nu\gamma^\rho,\nonumber\\
(\gamma^\mu\gamma^\nu\gamma^\rho\gamma^\sigma)^TC&=&C\gamma^\mu\gamma^\nu\gamma^\rho\gamma^\sigma,
\end{eqnarray}
and for $C\gamma^5$ that
\begin{eqnarray}\label{c5}
(\gamma^\mu)^TC\gamma^5&=&-C\gamma^5\gamma^\mu,\nonumber\\
(\gamma^\mu\gamma^\nu)^TC\gamma^5&=&-C\gamma^5\gamma^\mu\gamma^\nu,\nonumber\\
(\gamma^\mu\gamma^\nu\gamma^\rho)^TC\gamma^5&=&C\gamma^5\gamma^\mu\gamma^\nu\gamma^\rho,\nonumber\\
(\gamma^\mu\gamma^\nu\gamma^\rho\gamma^\sigma)^TC\gamma^5&=&C\gamma^5\gamma^\mu\gamma^\nu\gamma^\rho\gamma^\sigma.
\end{eqnarray}

Similarly we have for $\gamma^0$ that
\begin{eqnarray}\label{g0}
(\gamma^\mu)^\dagger \gamma^0&=&\gamma^0\gamma^\mu,\nonumber\\
(\gamma^\mu\gamma^\nu)^\dagger \gamma^0&=&-\gamma^0\gamma^\mu\gamma^\nu,\nonumber\\
(\gamma^\mu\gamma^\nu\gamma^\rho)^\dagger \gamma^0&=&-\gamma^0\gamma^\mu\gamma^\nu\gamma^\rho,\nonumber\\
(\gamma^\mu\gamma^\nu\gamma^\rho\gamma^\sigma)^\dagger \gamma^0&=&\gamma^0\gamma^\mu\gamma^\nu\gamma^\rho\gamma^\sigma,
\end{eqnarray}
and for $\gamma^0\gamma^5$ that
\begin{eqnarray}\label{g05}
(\gamma^\mu)^\dagger \gamma^0\gamma^5&=&-\gamma^0\gamma^5\gamma^\mu,\nonumber\\
(\gamma^\mu\gamma^\nu)^\dagger \gamma^0\gamma^5&=&-\gamma^0\gamma^5\gamma^\mu\gamma^\nu,\nonumber\\
(\gamma^\mu\gamma^\nu\gamma^\rho)^\dagger \gamma^0\gamma^5&=&\gamma^0\gamma^5\gamma^\mu\gamma^\nu\gamma^\rho,\nonumber\\
(\gamma^\mu\gamma^\nu\gamma^\rho\gamma^\sigma)^\dagger \gamma^0\gamma^5&=&\gamma^0\gamma^5\gamma^\mu\gamma^\nu\gamma^\rho\gamma^\sigma.
\end{eqnarray}

From the algebraic relations in Eq. (\ref{c}) we can conclude that the bilinear form $\psi^TC\varphi$ is invariant under a connected matrix Lie group of real dimension 20. This group is generated by the exponentials of the real Lie algebra spanned by the 10 skew-Hermitian matrices $\gamma^5\gamma^0$, $i\gamma^5\gamma^1$,$i\gamma^5\gamma^2$,$i\gamma^5\gamma^3$, $i\gamma^0 \gamma^1$, $i\gamma^0 \gamma^2$, $i\gamma^0 \gamma^3$, $\gamma^1 \gamma^2$, $\gamma^1 \gamma^3$, $\gamma^2 \gamma^3$ and the 10 Hermitian matrices $i\gamma^5\gamma^0$, $\gamma^5\gamma^1$, $\gamma^5\gamma^2$, $\gamma^5\gamma^3$, $\gamma^0 \gamma^1$, $\gamma^0 \gamma^2$, $\gamma^0 \gamma^3$, $i\gamma^1 \gamma^2$, $i\gamma^1 \gamma^3$, $i\gamma^2 \gamma^3$. This Lie group is isomorphic to the symplectic group of $4\times 4$ matrices $\mathrm{Sp}(4,\mathbb{C})$ (See e.g. Ref. \cite{hall} Ch. 1.2.4.).

Similarly, from the algebraic relations in Eq. (\ref{c5}) we can conclude that the bilinear form $\psi^TC\gamma^5\varphi$ is invariant under a connected matrix Lie group of real dimension 20. This group is generated by the exponentials of the real Lie algebra spanned by the 10 skew-Hermitian matrices $i\gamma^0$, $\gamma^1$, $\gamma^2$, $\gamma^3$, $i\gamma^0 \gamma^1$, $i\gamma^0 \gamma^2$, $i\gamma^0 \gamma^3$, $\gamma^1 \gamma^2$, $\gamma^1 \gamma^3$, $\gamma^2 \gamma^3$ and the 10 Hermitian matrices $\gamma^0$, $i\gamma^1$, $i\gamma^2$, $i\gamma^3$, $\gamma^0 \gamma^1$, $\gamma^0 \gamma^2$, $\gamma^0 \gamma^3$, $i\gamma^1 \gamma^2$, $i\gamma^1 \gamma^3$, $i\gamma^2 \gamma^3$. This Lie group is isomorphic to the symplectic group of $4\times 4$ matrices $\mathrm{Sp}(4,\mathbb{C})$ (See e.g. Ref. \cite{hall} Ch. 1.2.4.).
 
From the algebraic relations in Eq. (\ref{g0}) we can conclude that the sesquilinear form $\psi^\dagger\gamma^0\varphi$ is invariant under a connected matrix Lie group of real dimension 16. This group is generated by the exponentials of the real Lie algebra spanned by the 8 skew-Hermitian matrices $i\gamma^0$, $i\gamma^5\gamma^1$, $i\gamma^5\gamma^2$, $i\gamma^5\gamma^3$, $\gamma^1 \gamma^2$, $\gamma^1 \gamma^3$, $\gamma^2 \gamma^3$, $iI$ and the 8 Hermitian matrices $i\gamma^1$, $i\gamma^2$, $i\gamma^3$, $i\gamma^5\gamma^0$, $\gamma^0 \gamma^1$, $\gamma^0 \gamma^2$, $\gamma^0 \gamma^3$, $\gamma^5$ . This Lie group is $\mathrm{U}(2,2)$ the generalized unitary group of signature 2,2 (See e.g. Ref. \cite{wall} Ch. 1.1.3).

Similarly, from the algebraic relations in Eq. (\ref{g05}) we can conclude that the sesquilinear form $\psi^\dagger\gamma^0\gamma^5\varphi$ is invariant under a connected matrix Lie group of real dimension 16. This group is generated by the exponentials of the real Lie algebra spanned by the 8 skew-Hermitian matrices  $\gamma^1$, $\gamma^2$, $\gamma^3$, $\gamma^1 \gamma^2$, $\gamma^1 \gamma^3$, $\gamma^2 \gamma^3$, $\gamma^5\gamma^0$, $iI$ and the 8 Hermitian matrices  $\gamma^0$,  $\gamma^0 \gamma^1$, $\gamma^0 \gamma^2$, $\gamma^0 \gamma^3$, $\gamma^5\gamma^1$, $\gamma^5\gamma^2$, $\gamma^5\gamma^3$, $\gamma^5$. This Lie group is isomorphic to $\mathrm{U}(2,2)$ the generalized unitary group of signature 2,2 (See e.g. Ref. \cite{wall} Ch. 1.1.3).

We can see that the largest connected Lie subgroup that is shared between the four different invariance groups of the bilinear and sesquilinear forms is
a connected matrix Lie group of real dimension 6. This group is generated by the exponentials of the real Lie algebra spanned by the 3 skew-Hermitian matrices $\gamma^1 \gamma^2$, $\gamma^1 \gamma^3$, $\gamma^2 \gamma^3$ and the 3 Hermitian matrices $\gamma^0 \gamma^1$, $\gamma^0 \gamma^2$, $\gamma^0 \gamma^3$. This is the spinor representation of the proper orthochronous Lorentz group.
Thus the largest connected matrix Lie group that preserve both the bilinear forms $\psi^TC\varphi$ and $\psi^TC\gamma^5\varphi$ as well as both the sesquilinear forms $\psi^\dagger\gamma^0\varphi$ and $\psi^\dagger\gamma^0\gamma^5\varphi$ is the spinor representation of the proper orthochronous Lorentz group.

\section{Semistability, balancedness and polynomial invariants}\label{semi}
Here we describe how mixed locally Lorentz invariant polynomials can indicate types of entanglement not indicated by any homogeneous locally Lorentz invariant polynomial.

Consider the connected Lie group $G$ generated by the exponentials of the real Lie algebra spanned by the 3 skew-Hermitian matrices $\gamma^1 \gamma^2$, $\gamma^1 \gamma^3$, $\gamma^2 \gamma^3$ and the 3 Hermitian matrices $i\gamma^1 \gamma^2$, $i\gamma^1 \gamma^3$, $i\gamma^2 \gamma^3$. The group $G$ is the complexification of the spinor representation of the group of spatial rotations.

A homogeneous polynomial that is invariant under the spinor representations of the local proper orthochronous Lorentz groups can take a nonzero value for a given spinor entangled $n$-particle state only if the state
satisfies a condition called {\it semistability} \cite{mumford} with respect to the action of the group $G^{\otimes n}\equiv G\otimes G\otimes G\otimes\dots$. This condition of semistability is that there exists a homogeneous polynomial invariant under $G^{\otimes n}$ that is nonzero for the state.

A polynomial is invariant under the spinor representations of the local proper orthochronous Lorentz groups only if it is invariant under the spinor representations of the local groups of spatial rotations. Moreover, the spinor representation of the local groups of spatial rotations has the same algebra of homogeneous polynomial invariants as its complexification $G^{\otimes n}$ (See e.g. Ref. \cite{multispinor} Theorem 1).
Thus, semistability of the state with respect to $G^{\otimes n}$ is necessary for the existence of a homogeneous polynomial invariant under the spinor representations of the local proper orthochronous Lorentz groups that is nonzero for the state.
We therefore describe a method to evaluate if a given state is semistable with respect to $G^{\otimes n}$.

Consider the expansion of an $n$-particle state $\psi_{ABC\dots}$ in the basis of tensor products $\phi_{j_A}\otimes \phi_{k_B}\otimes \phi_{l_C}\otimes\dots$ of the basis spinors defined in Eq. (\ref{basis})
\begin{eqnarray}
&&\psi_{ABC\dots}(x_A,x_B,x_C,\dots)\nonumber\\
&=&\sum_{j_A,k_B,l_C,\dots }\psi_{j_A,k_B,l_C\dots}(x_A,x_B,x_C,\dots)\phi_{j_A}\otimes \phi_{k_B}\otimes \phi_{l_C}\otimes\dots.\nonumber\\
\end{eqnarray}
Then consider the Abelian $n$-parameter subgroup $G_z^{\otimes n}$ of the spinor representations of the local proper orthochronous Lorentz transformations
that is generated locally by the exponentials of $S^{12}=1/2 \gamma^1\gamma^2$ for each particle. 
The group $G_z^{\otimes n}$ is the spinor representation of a group of local spatial rotations around a given axis for each particle and is a subgroup also of $G^{\otimes n}$.
The elements $g$ of $G_z^{\otimes n}$ are given by $g=\exp(\alpha \gamma^1\gamma^2)\otimes \exp(\beta \gamma^1\gamma^2)\otimes\dots$ for $\alpha,\beta,\dots\in \mathbb{R}$. All elements $g$ in $G_z^{\otimes n}$ are diagonal matrices in the given basis.
Thus the group $G_z^{\otimes n}$ is such that each basis vector $\phi_{j_A}\otimes \phi_{k_B}\otimes \phi_{l_C}\otimes\dots$ is an eigenvector with nonzero eigenvalue for all elements in $G_z^{\otimes n}$. 

Next define $S_{\psi}$ to be the set of basis vectors $\phi_{j_A}\otimes \phi_{k_B}\otimes \phi_{l_C}\otimes\dots$ such that $\psi_{j_A,k_B,l_C,\dots}(x_A,x_B,x_C,\dots)\neq 0$ in the expansion of the state $\psi_{ABC\dots}$.
Then consider the $n$ generators $g^h$ for $h=1,2,\dots,n$ of the Lie algebra of $G_z^{\otimes n}$ defined by $g^h=I\otimes I \otimes I\otimes\dots \otimes I\otimes \gamma^1\gamma^2\otimes I \otimes\dots \otimes I$ where the $h$th entry in the tensor product is $\gamma^1\gamma^2$ and all other entries are the identity matrix.
For a basis vector in the set $S_{\psi}$ we can now consider the action of $g^h$ on it. Each vector in $S_{\psi}$ is an eigenvector of $g^h$ with eigenvalue either $i$ or $-i$.
Next, for each vector in $S_{\psi}$ we act with each of the $n$ generators $g^h$ in the order $g^1$ to $g^n$ and collect the corresponding ordered eigenvalues in an $n$ component vector. This vector is called a {\it weight vector}.
We can repeat this procedure for each vector in $S_{\psi}$ and obtain a weight vector corresponding to each element of $S_{\psi}$.
The collection of these weight vectors span a polytope, the so called {\it weight polytope}.

The condition of semistability of the state with respect to $G^{\otimes n}$ can now be formulated in terms of the weight polytope.
The state $\psi_{ABC\dots}$ is semistable if and only if for every combination of local spatial rotations of the state the resulting state $\psi_{ABC\dots}'$ is such that corresponding set
$S_{\psi'}$ of basis vectors has the property that the weight polytope contains the zero vector (See e.g. Ref. \cite{dolgachev}). 
Evaluating this condition for semistability is not necessarily easy in general. However, if one has a basis expansion of a state for which the weight polytope does not contain the zero vector it follows immediately that the state is not semistable.

Assume for a given state $\psi_{ABC\dots}$ that there is a choice of local spatial rotations such that for the locally rotated state $\psi_{ABC\dots}'$ the weight polytope corresponding to the set of basis vectors $S_{\psi'}$ does not contain the zero vector. Then there exist no
homogeneous polynomial that is invariant under the spinor representations of the local proper orthochronous Lorentz groups and takes a nonzero value for $\psi_{ABC\dots}$.

The condition that the weight polytope contains the zero vector is equivalent to the existence of a convex combination of the weight vectors that equals zero. This condition has been noted in the context of entanglement theory for non-relativistic spin-$\frac{1}{2}$ particles in Ref. \cite{coffman} and termed {\it balancedness} in Ref. \cite{osterloh}. 

For a state that is not semistable with respect to $G^{\otimes n}$, there may still exist a mixed polynomial that is invariant under the spinor representations of the local proper orthochronous Lorentz groups and takes a nonzero value for the state. Thus the mixed polynomials may indicate forms of entanglement that are not indicated by any homogeneous polynomial.
One can go one step further and note that in general there exist entangled states that are not indicated by any
mixed polynomial of bidegree ($k$,$l$) where $k\neq l$ that is invariant under the spinor representations of the local proper orthochronous Lorentz groups. But there may still exist a mixed polynomial of bidegree $(k,k)$ for some $k$ that is invariant under the spinor representations of the local proper orthochronous Lorentz groups and takes a nonzero value for the state.

To evaluate if a state can be indicated by a mixed polynomial of bidegree $(k,l)$ where $k\neq l$ we can consider a generalization of the concept of balancedness called {\it affine balancedness} \cite{top1,top2}. A state $\psi_{ABC\dots}$ is affinely balanced if the set of basis vectors $S_{\psi}$ is such that there exists an affine combination of the weight vectors that equals zero.
Thus the definition of affine balancedness replaces the convex combination in the definition of balancedness with an affine combination.

A mixed polynomial with bidegree $(k,l)$ where $k\neq l$ that is invariant under the spinor representations of the local proper orthochronous Lorentz groups can be nonzero for a given state only if the state is affinely balanced for every choice of local reference frames \cite{top1,top2}.
Thus if the state $\psi_{ABC\dots}$ is not affinely balanced there can only exist mixed polynomials with bidegree $(k,k)$, for some $k$, that take a nonzero value for the state. One way to see this is to note that if the zero vector is not in the affine hull of the weight vectors there exist a one-parameter family $g(\theta)\in G_z^{\otimes n}$ such that $g(\theta)\psi_{ABC\dots}=e^{i\theta}\psi_{ABC\dots}$ for all $\theta\in \mathbb{R}$. No polynomial with bidegree $(k,l)$ where $k\neq l$ that is nonzero for $\psi_{ABC\dots}$ is invariant under the action of such a subgroup.
One can see this property by taking the ansatz $\exp(\alpha \gamma^1\gamma^2)\otimes \exp(\beta \gamma^1\gamma^2)\otimes\dots \otimes \exp(\zeta \gamma^1\gamma^2)(\phi_{j_A}\otimes \phi_{k_B}\otimes \phi_{l_C}\otimes\dots)=\exp(i\theta)\phi_{j_A}\otimes \phi_{k_B}\otimes \phi_{l_C}\otimes\dots$ for all the basis vectors $\phi_{j_A}\otimes \phi_{k_B}\otimes \phi_{l_C}\otimes\dots$ in $S_{\psi}$. The corresponding system of linear equations in $\alpha,\beta,\dots,\zeta,\theta$ has only solutions with a discrete set of values for $\theta$ if the state is balanced or affinely balanced \cite{top1}. This is regardless of whether the system of equations has a singular matrix or non-singular matrix because the values of $\theta$ are constrained by the condition of balancedness or affine balancedness. However, if the state is not balanced or affinely balanced solutions exist for arbitrary $\theta\in \mathbb{R}$ because the system of equations has a singular matrix and $\theta$ is not constrained.
Note that a condition on the form $g(\theta)\psi_{ABC\dots}=e^{i\theta}\psi_{ABC\dots}$ for all $\theta\in \mathbb{R}$, where $g(\theta)\in G_z^{\otimes n}$ is some one-parameter family defines a continuous Abelian local unitary symmetry of the state $\psi_{ABC\dots}$.

We can consider an example of a spinor entangled two-particle state that is neither balanced nor affinely balanced. The state in Eq. (\ref{xccx})
\begin{eqnarray}
\frac{1}{\sqrt{2}}({\phi_0^A}\otimes{\phi_1^B}+{\phi_1^A}\otimes{\phi_3^B}),
\end{eqnarray}
is of this kind.
The set of weight vectors corresponding to this state is $(-i,i)$ and $(i,i)$. There is no $a,b\in \mathbb{R}$ such that $a+b\neq 0$ and $a(-i,i)+b(i,i)=0$. Thus this state is not balanced and also not affinely balanced. Consequently all polynomials with bidegree $(k,l)$ where $k\neq l$ that are invariant under the spinor representations of the local proper orthochronous Lorentz groups are zero for this state. In particular the bidegree (2,0) polynomials $I_1,I_2,I_{2A},I_{2B}$ and the bidegree (3,1) polynomials $Q_1,Q_2,Q_{3},Q_{4}$ are identically zero. Only polynomials with bidegree $(k,k)$ that are invariant under the spinor representations of the local proper orthochronous Lorentz groups can be nonzero. In particular we have seen in Section \ref{exxx} that the bidegree (2,2) polynomials $R_1$ and $T_1$ are nonzero for this state. 

The three states in Eq. (\ref{xccx2}), Eq. (\ref{xccx3}) and Eq. (\ref{xccx4}) respectively are also neither balanced nor affinely balanced. The state in Eq. (\ref{xccx2}) has weight vectors $(-i,-i)$ and $(i,-i)$ and the states in  Eq. (\ref{xccx3}) and Eq. (\ref{xccx4}) both have the weight vectors $(-i,-i)$ and $(-i,i)$. The spinor entanglement of these three states is indicated by the three pairs of bidegree (2,2) polynomials $R_3,T_1$, and $R_4,T_1$ and $R_6,T_1$ respectively.

The state in Eq. (\ref{utoy})
\begin{eqnarray}
\frac{1}{\sqrt{3}}({\phi_0^A}\otimes{\phi_2^B}+{\phi_1^A}\otimes{\phi_0^B}+{\phi_2^A}\otimes{\phi_2^B}),
\end{eqnarray}
has a set of weight vectors $(-i,-i)$, $(i,-i)$ and $(-i,-i)$ and is another example of a state that is neither balanced nor affinely balanced.
For this state the bidegree (2,2) polynomials $R_1$, $R_2$ and $R_3$ are nonzero.
Similarly the state in Eq. (\ref{utoya}) has a set of weight vectors $(-i,-i)$, $(-i,i)$ and $(-i,-i)$
and is also neither balanced nor affinely balanced. For this state the bidegree (2,2) polynomials $R_4$, $R_5$ and $R_6$ are nonzero.

An example of a three-particle state that is neither balanced nor affinely balanced is the W-state \cite{dur}
\begin{eqnarray}
\frac{1}{{\sqrt{3}}}({\phi_0^A}\otimes{\phi_0^B}\otimes{\phi_1^C}+{\phi_0^A}\otimes{\phi_1^B}\otimes{\phi_0^C}+{\phi_1^A}\otimes{\phi_0^B}\otimes{\phi_0^C}),
\end{eqnarray}
with weight vectors $(-i,-i,i)$, $(-i,i,-i)$ and $(i,-i,-i)$. For this state the bidegree (3,3) polynomial $W_1$ is nonzero. States that are neither balanced nor affinely balanced can be constructed on a form that generalizes the three-particle W-state for any number of particles greater than three \cite{dur}.

Next we consider an example of a three-particle state that is not balanced but affinely balanced. The three-particle state in Eq. (\ref{req1})
\begin{eqnarray}
\frac{1}{{2}}&&({\phi_0^A}\otimes{\phi_0^B}\otimes{\phi_1^C}+{\phi_0^A}\otimes{\phi_1^B}\otimes{\phi_0^C}\nonumber\\&&+{\phi_1^A}\otimes{\phi_0^B}\otimes{\phi_0^C}+{\phi_0^A}\otimes{\phi_0^B}\otimes{\phi_0^C}),
\end{eqnarray}
is of this kind.
This state has the weight vectors $(-i,-i,i)$, $(-i,i,-i)$ $(i,-i,-i)$ and $(-i,-i,-i)$.
The zero vector is not contained in the weight polytope and thus the state is not balanced. Consequently there is no homogeneous polynomial invariant under the spinor representations of the local proper orthochronous Lorentz groups that takes a nonzero value for this state. 
However, the weight vectors satisfy $(-i,-i,i)+(-i,i,-i)+(i,-i,-i)-(-i,-i,-i)=(0,0,0)$, i.e., the zero vector is in the affine hull of the weight vectors, and thus the state is affinely balanced. 
As described in Section \ref{bideg3} the bidegree (3,1) polynomials $B_1$, $Z_1$, and $D_1$ and the bidegree (3,3) polynomial $W_1$ are nonzero for this state.
The states in Eq. (\ref{req2}), and Eq. (\ref{req3}) have the same weight vectors as the state in Eq. (\ref{req1}) and their spinor entanglement is indicated by the bidegree (3,1) polynomials $B_8$, $Z_8$, and $D_8$ and the bidegree (3,1) polynomials $B_6$, $Z_6$ and $D_6$, respectively.

\section{The bidegree (2,2) polynomial $T_2$}\label{cameron}
\begin{widetext}
Here we give the pure imaginary valued bidegree (2,2) polynomial $T_2$ written out
\begin{eqnarray}
T_2&=&(|\psi_{02}|^2 + |\psi_{12}|^2 - |\psi_{22}|^2 -|\psi_{32}|^2+ |\psi_{00}|^2 + |\psi_{10}|^2 - |\psi_{20}|^2 - 
    |\psi_{30}|^2)\nonumber\\&&\times     
    (\psi_{20}\psi^*_{02}-\psi^*_{20}\psi_{02} + \psi_{30}\psi^*_{12} - \psi^*_{30}\psi_{12} - \psi_{00}\psi^*_{22}  + \psi^*_{00}\psi_{22}-\psi_{10}\psi^*_{32}  +\psi^*_{10}\psi_{32})\nonumber\\&& 
  -(  |\psi_{03}|^2 + 
    |\psi_{13}|^2 - |\psi_{23}|^2 - |\psi_{33}|^2 +|\psi_{01}|^2 + |\psi_{11}|^2 - |\psi_{21}|^2 - 
    |\psi_{31}|^2 )\nonumber\\&& \times                  
    (\psi^*_{23}\psi_{01}-\psi_{23}\psi^*_{01}+ \psi^*_{33}\psi_{11} - \psi_{33}\psi^*_{11} + \psi_{03}\psi^*_{21}   - \psi^*_{03}\psi_{21} +\psi_{13}\psi^*_{31}   - \psi^*_{13}\psi_{31} )\nonumber\\&&                             
+( \psi_{02}\psi^*_{22}-\psi_{22}\psi^*_{02} - \psi_{32}\psi^*_{12}  + \psi_{12}\psi^*_{32} + \psi_{00}\psi^*_{20}-\psi_{20}\psi^*_{00} - \psi_{30}\psi^*_{10}  + \psi_{10}\psi^*_{30}) \nonumber\\&&\times
(\psi_{00}\psi^*_{02}+\psi^*_{00}\psi_{02} + 
    \psi_{10}\psi^*_{12} + 
    \psi^*_{10}\psi_{12}- \psi_{20}\psi^*_{22}- \psi^*_{20}\psi_{22}  - \psi_{30}\psi^*_{32}- \psi^*_{30}\psi_{32}  )\nonumber\\&&
 +( \psi_{01}\psi^*_{21}-\psi_{21}\psi^*_{01} - \psi_{31}\psi^*_{11}  + \psi_{11}\psi^*_{31} + \psi_{23}\psi^*_{03}  - \psi_{03}\psi^*_{23}+ \psi_{33}\psi^*_{13} - 
    \psi_{13}\psi^*_{33})\nonumber\\&&\times         
         (\psi_{03}\psi^*_{01}- \psi^*_{03}\psi_{01}  + \psi_{13}\psi^*_{11} - \psi^*_{13}\psi_{11}  - \psi_{23}\psi^*_{21}  + \psi^*_{23}\psi_{21}- \psi_{33}\psi^*_{31}
    + \psi^*_{33}\psi_{31})\nonumber\\&&    
 +(   \psi_{00}\psi^*_{21} -\psi_{20}\psi^*_{01} - \psi_{30}\psi^*_{11} + \psi_{10}\psi^*_{31}- \psi_{22}\psi^*_{03} - \psi_{32}\psi^*_{13} + \psi_{02}\psi^*_{23} + \psi_{12}\psi^*_{33})\nonumber\\&&\times   
   (\psi_{03}\psi^*_{00} + \psi_{13}\psi^*_{10} - \psi_{23}\psi^*_{20} - \psi_{33}\psi^*_{30}+\psi^{*}_{02}\psi_{01} +\psi^{*}_{12}\psi_{11} - \psi^{*}_{22}\psi_{21} - \psi_{32}^{*}\psi_{31})\nonumber\\&&         
+(\psi^*_{01}\psi_{02} + \psi^*_{11}\psi_{12} - \psi^*_{21}\psi_{22} - \psi^*_{31}\psi_{32}           + \psi^*_{03}\psi_{00} + \psi^*_{13}\psi_{10} - \psi^*_{23}\psi_{20} - \psi^*_{33}\psi_{30})\nonumber\\&&\times
            (\psi^*_{22}\psi_{03} + \psi^*_{32}\psi_{13} - \psi^*_{02}\psi_{23} - \psi^*_{12}\psi_{33} 
          -\psi_{21}\psi^*_{00} -  \psi_{31}\psi^*_{10} + \psi_{01} \psi^*_{20} + \psi_{11}\psi^*_{30}) \nonumber\\&&           
 -(\psi^*_{01}\psi_{00} + \psi^*_{11}\psi_{10} - \psi^*_{21}\psi_{20} - \psi^*_{31}\psi_{30}) (\psi^*_{20}\psi_{03} + \psi^*_{30}\psi_{13} - \psi^*_{00}\psi_{23} - \psi^*_{10}\psi_{33}        
     -  \psi_{21}\psi^*_{02} - \psi_{31}\psi^*_{12} + \psi_{01}\psi^*_{22} + \psi_{11}\psi^*_{32})\nonumber\\&&
   + (\psi_{01}\psi^*_{00} + \psi_{11}\psi^*_{10} - \psi_{21}\psi^*_{20} - \psi_{31}\psi^*_{30}) (\psi_{20}\psi^*_{03} + \psi_{30}\psi^*_{13} - \psi_{00}\psi^*_{23} - \psi_{10}\psi^*_{33}   
        - \psi^*_{21}\psi_{02} - \psi^*_{31}\psi_{12} + \psi^*_{01}\psi_{22} + \psi^*_{11}\psi_{32})\nonumber\\&&                                    
    -(\psi_{03}\psi^*_{02} + \psi_{13}\psi^*_{12} - \psi_{23}\psi^*_{22} - \psi_{33}\psi^*_{32})( \psi_{02}\psi^*_{21}  -\psi_{22}\psi^*_{01} - \psi_{32}\psi^*_{11} + \psi_{12}\psi^*_{31} 
    +\psi^*_{23}\psi_{00} + 
    \psi^*_{33}\psi_{10} -\psi^*_{03}\psi_{20} - \psi^*_{13}\psi_{30}  )\nonumber\\&& 
     +(\psi_{02}\psi^*_{03} + \psi_{12}\psi^*_{13} - \psi_{22}\psi^*_{23} - 
    \psi_{32}\psi^*_{33})(\psi^*_{02}\psi_{21} -\psi^*_{22}\psi_{01} - \psi^*_{32}\psi_{11} +\psi^*_{12}\psi_{31} +\psi_{23}\psi^*_{00}+ 
    \psi_{33}\psi^*_{10} -\psi_{03}\psi^*_{20}  - \psi_{13}\psi^*_{30}            
        ).\nonumber\\          
\end{eqnarray}

\section{The bidegree (2,2) polynomial $N_1N_4-N_2N_3$}\label{cameron2}

Here we give the real valued bidegree (2,2) polynomial $N_1N_4-N_2N_3$ written out
\begin{eqnarray}
N_1N_4-N_2N_3&=&(\psi_{20}\psi^*_{00}+ \psi_{21}\psi^*_{01}- \psi_{22}\psi^*_{02}- \psi_{23}\psi^*_{03}+ \psi_{30}\psi^*_{10}+ \psi_{31}\psi^*_{11}- \psi_{32}\psi^*_{12}-
     \psi_{33}\psi^*_{13}\nonumber\\&& - \psi_{00}\psi^*_{20}- \psi_{01}\psi^*_{21}+ \psi_{02}\psi^*_{22}+ \psi_{03}\psi^*_{23}- \psi_{10}\psi^*_{30}- 
    \psi_{11}\psi^*_{31}+ \psi_{12}\psi^*_{32}+ \psi_{13}\psi^*_{33})\nonumber\\&&\times (\psi_{02}\psi^*_{00}+ \psi_{03}\psi^*_{01}- \psi_{00}\psi^*_{02}- 
    \psi_{01}\psi^*_{03}+ \psi_{12}\psi^*_{10}+ \psi_{13}\psi^*_{11}- \psi_{10}\psi^*_{12}- \psi_{11}\psi^*_{13}\nonumber\\&& - \psi_{22}\psi^*_{20}- 
    \psi_{23}\psi^*_{21}+ \psi_{20}\psi^*_{22}+ \psi_{21}\psi^*_{23}- \psi_{32}\psi^*_{30}- \psi_{33}\psi^*_{31}+ \psi_{30}\psi^*_{32}+ 
    \psi_{31}\psi^*_{33})\nonumber\\&& + ( \psi_{20}\psi^*_{02}-\psi_{22}\psi^*_{00}- \psi_{23}\psi^*_{01}+ \psi_{21}\psi^*_{03}- \psi_{32}\psi^*_{10}- 
    \psi_{33}\psi^*_{11}+ \psi_{30}\psi^*_{12}+ \psi_{31}\psi^*_{13}\nonumber\\&& + \psi_{02}\psi^*_{20}+ \psi_{03}\psi^*_{21}- \psi_{00}\psi^*_{22}- 
    \psi_{01}\psi^*_{23}+ \psi_{12}\psi^*_{30}+ \psi_{13}\psi^*_{31}- \psi_{10}\psi^*_{32}- \psi_{11}\psi^*_{33}) \nonumber\\&&\times(|\psi_{00}|^2+ 
    |\psi_{01}|^2- |\psi_{02}|^2- |\psi_{03}|^2+ |\psi_{10}|^2+ |\psi_{11}|^2- |\psi_{12}|^2- 
    |\psi_{13}|^2\nonumber\\&& - |\psi_{20}|^2- |\psi_{21}|^2+ |\psi_{22}|^2+ |\psi_{23}|^2-|\psi_{30}|^2- 
    |\psi_{31}|^2+ |\psi_{32}|^2+ |\psi_{33}|^2).
\end{eqnarray}

\end{widetext}


\begin{thebibliography}{xx}
\bibitem{dirac2}P. A. M. Dirac, Proc. Royal Soc. A {\bf 117}, 610 (1928).
\bibitem{dirac} P. A. M. Dirac,
{\it Principles of Quantum Mechanics, Fourth edition} (Oxord University Press,
London, 1958).
\bibitem{bjorken} J. D. Bjorken and S. D. Drell,
{\it Relativistic Quantum Mechanics} (McGraw-Hill,
New York, 1964).
\bibitem{pykk}P. Pykk\"o, Chem. Rev. {\bf 88}, 563 (1988).
\bibitem{schwartz} M. D. Schwartz,
{\it Quantum Field Theory and the Standard Model} (Cambridge University Press,
Cambridge, 2014).
\bibitem{yukawa}H. Yukawa, Proc. Phys. Math. Soc. Japan {\bf 17}, 48 (1935).
\bibitem{thaller}B. Thaller, {\it The Dirac Equation} (Springer,
Berlin, 1992), Ch. 4.2.

\bibitem{weyl}H. Weyl, I. Z. Phys. {\bf 56}, 330 (1929).


\bibitem{epr} A. Einstein, B. Podolsky, and N. Rosen,  Phys. Rev. {\bf 47}, 777 (1935).
\bibitem{bell}J. S. Bell, Physics {\bf 1}, 195 (1964).
\bibitem{chsh}J. F. Clauser, M. A. Horne, A. Shimony, and R. A. Holt, Phys. Rev. Lett. {\bf 23}, 880 (1969).
\bibitem{bell2}J. S. Bell, Epistemol. Lett. {\bf 9}, 11 (1976).








\bibitem{svet}G. Svetlichny, Phys. Rev. D {\bf 35}, 3066 (1987).
\bibitem{steer}E. Schr\"odinger, Proc. Camb. Phil. Soc. {\bf 31}, 553 (1935).
\bibitem{wise}H. M. Wiseman, S. J. Jones, and A. C. Doherty,
Phys. Rev. Lett. {\bf 98}, 140402 (2007).

\bibitem{bennett}C. H. Bennett, G. Brassard, C. Cr{\'e}peau, R. Jozsa, A. Peres, and
W. K. Wootters, Phys. Rev. Lett. {\bf 70}, 1895 (1993).


\bibitem{oka}M. Oka, Kodai Math. J. {\bf 33}, 1 (2010).





\bibitem{wootters}S. A. Hill and W. K. Wootters, Phys. Rev. Lett. {\bf 78}, 5022 (1997).
\bibitem{wootters2}W. K. Wootters, Phys. Rev. Lett. {\bf 80}, 2245 (1998).
\bibitem{grassl}M. Grassl, M. R\"otteler, and T. Beth, Phys. Rev. A {\bf 58}, 1833 (1998).

\bibitem{popescu}N. Linden and S. Popescu, Fortsch. Phys.  {\bf 46 }, 567 (1998).
\bibitem{lindpop}N. Linden, S. Popescu, and A. Sudbery, Phys. Rev. Lett. {\bf 83}, 243 (1999).
\bibitem{carteret}H. A. Carteret, N. Linden, S. Popescu, and A. Sudbery, Foundations of Physics {\bf 29}, 527 (1999).
\bibitem{kempe}J. Kempe, Phys. Rev. A {\bf 60}, 910 (1999).
\bibitem{coffman}V. Coffman, J. Kundu, and W. K. Wootters, Phys. Rev. A {\bf 61}, 052306 (2000).
\bibitem{toni}A. Ac{\'i}n, A. Andrianov, L. Costa, E. Jan{\'e}, J. I. Latorre, and R. Tarrach,
Phys. Rev. Lett. {\bf 85}, 1560 (2000).
\bibitem{carsud}H. A. Carteret and A Sudbery, J. Phys. A {\bf 33}, 4981  (2000).
\bibitem{sud} A. Sudbery, J. Phys. A: Math. Gen. {\bf 34}, 643 (2001).
\bibitem{wong}A. Wong and N. Christensen, Phys. Rev. A {\bf 63}, 044301 (2001).
\bibitem{tarrach} A. Ac{\'i}n, A. Andrianov, E. Jan{\'e} and R. Tarrach, J. Phys. A: Math. Gen. {\bf 34}, 6725 (2001).
\bibitem{moor}F. Verstraete, J. Dehaene, and B. De Moor,
Phys. Rev. A {\bf 68}, 012103 (2003).
\bibitem{luque}J.-G. Luque and J.-Y. Thibon, Phys. Rev. A {\bf 67}, 042303 (2003).
\bibitem{toumazet}J.-G. Luque, J.-Y. Thibon, and F. Toumazet, Math. Struct. Comp. Sci. {\bf 17}, 1133 (2007).









\bibitem{czachor}M. Czachor, Phys. Rev. A {\bf 55}, 72 (1997).
\bibitem{alsing}P. M. Alsing and G. J. Milburn, Quantum Inf. Comput. {\bf 2}, 487 (2002).
\bibitem{terno} A. Peres, P. F. Scudo, and D. R. Terno,
Phys. Rev. Lett. {\bf 88}, 230402 (2002).

\bibitem{adami}R. M. Gingrich and C. Adami,
Phys. Rev. Lett. {\bf 89}, 270402 (2002).

\bibitem{pachos}J. Pachos and E. Solano, Quantum Inf. Comput. {\bf 3}, 115 (2003).
\bibitem{ahn}D. Ahn, H.-j. Lee, Y. H. Moon, and S. W. Hwang,
Phys. Rev. A {\bf 67}, 012103 (2003).
\bibitem{terno2}D. R. Terno,
Phys. Rev. A {\bf 67}, 014102 (2003).
\bibitem{tera}H. Terashima and M. Ueda, Quantum Inf. Comput. {\bf 3}, 224 (2003).

\bibitem{tera2}H. Terashima and M. Ueda, Int. J. Quantum Inform. {\bf 1},  93 (2003).
\bibitem{mano}E. B. Manoukian and N. Yongram, Eur. Phys. J. D {\bf 31}, 137 (2004).
\bibitem{won}W. T. Kim and E. J. Son,
Phys. Rev. A {\bf  71}, 014102 (2005).
\bibitem{caban3}P. Caban and J. Rembieli{\'n}ski, Phys. Rev. A {\bf 72}, 012103 (2005).
\bibitem{leon}L. Lamata, J. Le{\'o}n, and E. Solano,
Phys. Rev. A {\bf 73}, 012335 (2006).

\bibitem{caban}P. Caban and J. Rembieli{\'n}ski, Phys. Rev. A {\bf 74}, 042103 (2006).
\bibitem{tessier}P. M. Alsing, I. Fuentes-Schuller, R. B. Mann, and T. E. Tessier,
Phys. Rev. A {\bf 74}, 032326 (2006).
\bibitem{geng}H-J. Wang and W. T. Geng,  J. Phys. A: Math. Theor. {\bf 40}, 11617 (2007).
\bibitem{delgado}A. Bermudez and M. A. Martin-Delgado, J. Phys. A: Math. Theor. {\bf 41}, 485302 (2008).
\bibitem{moradi}S. Moradi, Jetp Lett. {\bf 89}, 50 (2009).
\bibitem{caban2}P. Caban, J. Rembieli{\'n}ski, and M. W\l odarczyk, Phys. Rev. A {\bf 79}, 014102 (2009).



\bibitem{spinorent}M. Johansson, Phys. Rev. A {\bf 105}, 032402 (2022).
\bibitem{multispinor}M. Johansson, Ann. Phys. (N. Y.) {\bf 457}, 169410 (2023).
\bibitem{dur}W. D\"ur, G. Vidal, and J. I. Cirac, Phys. Rev. A {\bf 62}, 062314  (2000).
\bibitem{peskin} M. E. Peskin and D. V. Schroeder,
{\it An Introduction to Quantum Field Theory} (Perseus Books,
Reading, 1995).



\bibitem{navascues}M. Navascu{\'e}s, S. Pironio, and A. Ac{\'i}n,
Phys. Rev. Lett. {\bf 98}, 010401 (2007).
\bibitem{tsirelson}B. S. Tsirelson, {\it Bell inequalities and operator algebras:
http://www.imaph.tu-bs.de/qi/problems/33.html}, (2006).
\bibitem{werner}V. B. Scholz and R. F. Werner, arXiv:0812.4305 (2008).

\bibitem{wald}R. M. Wald, {\it General Relativity} (The University of Chicago Press,
Chicago, 1984).
\bibitem{fock}V. Fock, Z. Phys. {\bf 57}, 261 (1929).



\bibitem{zuber} C. Itzykson and J-B. Zuber,
{\it Quantum Field Theory} (Dover,
New York, 2006), Ch. 2-1-3.

\bibitem{pauli}W. Pauli, Ann. de l'Inst. Henri Poincar\'{e} {\bf 6}, 109 (1936).









\bibitem{uhlmann}A. Uhlmann,
Phys. Rev. A {\bf 62}, 032307 (2000).
\bibitem{rungta}P. Rungta, V. Bu\v zek, C. M. Caves, M. Hillery, and G. J. Milburn,
Phys. Rev. A {\bf 64}, 042315  (2001).

\bibitem{omega}A. Cayley, Cambridge and Dublin Mathematical Journal {\bf 1}, 104 (1846).
\bibitem{cayley}A. Cayley, Cambridge Math. J. {\bf 4},  16 (1845).





\bibitem{lima}\AA . Lima, Proc. London Math. Soc. {\bf s3-25}, 27 (1972).

\bibitem{wakker} H. J. M. Peters, and P. P. Wakker, Econ. Lett. {\bf 22}, 251 (1986).

\bibitem{uhlmannn}A. Uhlmann, Open Syst. Inf. Dyn. {\bf 5}, 209 (1998).



\bibitem{birkhoff}G. Birkhoff and J. Von Neumann, Ann. Math. {\bf 37}, 823 (1936).

\bibitem{neumann}J. Von Neumann, {\it Mathematical Foundations of Quantum Mechanics} (Princeton University Press,
Princeton, 1955), Ch. III.5.


\bibitem{hall}B. C. Hall, {\it Lie Groups, Lie Algebras, and Representations: An Elementary Introduction, Second Edition} (Springer,
Cham, 2015).

\bibitem{wall}R. Goodman and N. R. Wallach, {\it Symmetry, Representations and Invariants} (Springer, New York, 2009).


\bibitem{mumford}D. Mumford, J. Fogarty, and F. Kirwan, {\it Geometric Invariant Theory} (Springer-Verlag, Berlin, 1994).
\bibitem{dolgachev}I. Dolgachev, {\it Lectures on Invariant Theory} (Cambridge University Press, Cambridge, 2003), Ch. 9.4.

\bibitem{osterloh}A. Osterloh and J. Siewert, New J. Phys. {\bf 12}, 075025 (2010).

\bibitem{top1}M. Johansson, M. Ericsson, K. Singh, E. Sj\"oqvist, and M. S. Williamson, Phys. Rev. A {\bf 85}, 032112 (2012).


\bibitem{top2}M. Johansson, M. Ericsson, E. Sj\"oqvist, and A. Osterloh, Phys. Rev. A {\bf 89}, 012320 (2014).



























\end{thebibliography}
\end{document}